\documentclass{aa}  

\usepackage{graphicx}
\usepackage{txfonts}
\usepackage{natbib}
\usepackage{multirow}
\bibpunct{(}{)}{;}{a}{}{,} 
\bibpunct[; ]{(}{)}{,}{a}{}{;}

\begin{document} 

\newcommand{\vexp}{\mbox{$V_{\rm exp}$}}
\newcommand{\tmb}{\mbox{$T_{\rm mb}$}}
\newcommand{\feff}{\mbox{$F_{\rm eff}$}}
\newcommand{\beff}{\mbox{$B_{\rm eff}$}}
\newcommand{\rco}{\mbox{$R_{\rm CO}$}}
\newcommand{\tex}{\mbox{$T_{\rm ex}$}}                                        

   \title{Circumstellar envelopes of semi-regular long-period variables: mass-loss rate estimates and general model fitting of the molecular gas}


   \author{J. J. D\'{\i}az-Luis
          \inst{1},
          J. Alcolea
          \inst{1},
          V. Bujarrabal
          \inst{2},
          M. Santander-Garc\'ia
          \inst{1},
          A. Castro-Carrizo
          \inst{3},
          M. G\'omez-Garrido
          \inst{2},
          \and
          J.\,-F.\,Desmurs
          \inst{1}}

   \institute{Observatorio Astron\'omico Nacional (OAN-IGN), 
              Alfonso XII, 3 y 5, 28014 Madrid, Spain\\
              \email{jjairo@oan.es} 
         \and     
              Observatorio Astron\'omico Nacional (OAN-IGN), Apartado 112, 28803 Alcal\'a de Henares, Spain
         \and
              Institut de Radioastronomie Millim\'etrique, 300 rue de la Piscine, 38406 Saint-Martin-d’H\`{e}res, France \\
              }

   \date{Received 13/06/2019; accepted 19/07/2019}

   \abstract{}
   {We aim to study the main properties of a volume-limited unbiased sample of well-characterized semi-regular variables (SRs) in order to clarify important issues that need to be further explained, such as the formation of axially symmetric planetary nebulae (PNe) from spherical circumstellar envelopes (CSEs), which takes place during the mass-loss process along the asymptotic giant branch (AGB) phase.}
   {We present new high-S/N IRAM 30m observations of the $^{12}$CO \textit{J}=2-1, $^{12}$CO \textit{J}=1-0, and $^{13}$CO \textit{J}=1-0 lines, in a volume-limited sample of SRs for which the Hipparcos distances are between 100 and 500 pc and the declinations are above $-$25$^{\circ}$. We analyzed the data by characterizing the main properties of the CSEs. The $^{12}$CO \textit{J}=2-1 data were used to study the profiles, while the $^{12}$CO \textit{J}=1-0 data were used to estimate mass-loss rates for the complete sample. Moreover, the $^{12}$CO \textit{J}=2-1 line has been used to determine the possible structures responsible for such profiles.}
   {We have classified the sources into four groups according to the different profiles and final gas expansion velocities. Type 1 and 2 profiles are broad and narrow symmetric lines, respectively. Furthermore, type 1 profiles  are more related to previously studied, standard, spherically symmetric envelopes. Type 3 profiles on the contrary are strange profiles with very pronounced asymmetries. Finally, type 4 profiles are those showing two different components: a narrow line profile superimposed on a broad pedestal component. We find that for sources with this latter kind of profile, the variation amplitude is very low, which means that these SRs do not have a well-developed inner envelope differentiated from the outer one. Interestingly, we report a moderate correlation between mass-loss rates and $^{12}$CO \textit{J}=1-0/$^{12}$CO \textit{J}=2-1 line intensity ratios for O-rich SRs, suggesting a different behaviour between C- and O-rich SRs. Using SHAPE+shapemol, we find a unified simple model based on an oblate spheroid placed in different orientations that may explain all the $^{12}$CO profiles in the sample, indicating that the gas expansion is in general predominantly equatorial. Moreover, in order to explain the type 4 profiles, we define an extra component which may somehow be a biconical structure or similar according to the structures already found in this kind of source. Type 1 and 2 profiles, curiously, may also be explained by standard spherically symmetric envelopes, but often requiring anomalously low velocities. Type 3 and 4 profiles however, need axial symmetry to be explained. We conclude that most circumstellar shells around SRs show axial, strongly nonspherical symmetry. More interferometric observations are needed in order to make firm conclusions about mass-loss processes and possible morphologies of SRs.}
   {}
 
               
   \titlerunning{Circumstellar envelopes of SRs}
   \authorrunning{J. J. D\'{\i}az-Luis et al.}            
               
   \maketitle
   
%

\section{Introduction}

For low- and intermediate-mass stars ($\sim$1-8 M$_{\odot}$), the latest stages of their evolution are controlled by the mechanisms of mass loss that occur on the asymptotic giant branch (AGB; and in general in the red giant or super-giant phases). In addition, these mass losses, with rates up to 10$^{-5}$ M$_{\odot}$\,yr$^{-1}$, are crucial for the formation of circumstellar envelopes (CSEs), which in the post-AGB phase will evolve into planetary nebulae (PNe); see, for example, \citet{olof99}. Although this scenario is well understood in general terms, there are still important issues that need to be further explained; in particular, the formation of axially symmetric PNe from spherical CSEs. Circumstellar envelopes around AGB stars are in general relatively spherical, and expand isotropically at moderate speeds (10-25 km\,s$^{-1}$), at least at large scales \citep{neri98,castro10}. However, PNe display a large variety of shapes with a high degree of symmetry; about four fifths of the cases are far from showing spherically symmetric envelopes \citep{parker06}. This transformation from spherical CSEs to axially symmetric PNe has been a matter of debate for more than two decades \citep{balick02}. Many theoretical, numerical, and observational studies have been conducted in order to shed light on the problem. In recent years, there is growing consensus that the presence of binary systems is the most likely explanation; see, for example, \citet{jones17}. This would be supported by the large prevalence of multiple systems in solar-type stars in general, and in the AGB population in particular. Some studies reveal the presence of spiral patterns and bipolar outflows at smaller scales, providing observational support for the binary system scenario as the best explanation for such structures \citep[e.g.,][]{morris06,mauron06,maercker12,kim15}. Moreover, recent studies involve searches for UV excess emission ---which is believed to be due to a main sequence companion--- from AGB stars using the GALEX archive \citep[e.g.,][]{Sahai08,ortiz16}. Almost 60$\%$ of the stars in the sample have a main sequence companion of spectral type earlier than K0. It is believed that more than 50$\%$ of PNe harbor binary systems \citep{douchin15} and almost 50$\%$ of solar-type stars in the solar neighborhood seem to form in a binary or a multiple system \citep{raghavan10}. However, in order to firmly establish the connection between binarity and nonspherical shape, more observations of complete and unbiased samples are necessary. 

During recent years, the results from the increasing number of good interferometric maps of the CSEs of a particular type of evolved star have attracted our attention. Semi-regular variables (SRs) are red giant stars showing quasi-regular or some irregular variations in the optical range (SRa and SRb variables, respectively), in contrast to the regular pulsators, the Mira-type variables (referred to here as `Miras'; we note that both SRs and Miras are classified as long-period variable stars, LPVs). In principle, the differences in their pulsation mode should not affect the envelope geometry and kinematics. However, a clear axial symmetry has been found in almost all CSEs around SRa and SRb variables that have been well mapped (using mm-wave interferometric observations of CO), in strong contrast to the CSEs of regular pulsators (Miras). This is a most fascinating result \citep{alcolea11}, and the number of well-mapped SRs exhibiting axially symmetric structures is becoming significant: RV Boo \citep{bergman00}, $\pi$$^{1}$ Gru \citep{chiu06,doan17}, EP Aqr \citep{winters07,homan18,nhung19,hoai19}, RX Boo and X Her \citep{castro10}, V Hya \citep{hirano04}, L$_{2}$ Pup \citep{homan17}, IRC+50049 (V370 And), and BM\,Gem (Castro-Carrizo et al., priv. comm.). The only exception is a small group of C-rich SRs that show detached and relatively spherical shells \citep{olof92}. In spite of a strong prevalence of nonspherical CSEs in SRs, we should take into account that the sample is limited and strongly biased. Hence, the question of whether or not the CSEs of SRs are really  different from those of regular variables remains open.

In this paper we present high-S/N IRAM 30m observations of $^{12}$CO \textit{J}=1-0 and \textit{J}=2-1, and $^{13}$CO \textit{J}=1-0 in a volume-limited unbiased sample of well-characterized SRs. We characterized the main properties of the CSEs, such as expansion velocity, mass-loss rate, wind density, and so on. The $^{12}$CO \textit{J}=2-1 data were used primarily to study the profiles of the lines (since the S/N is higher). The $^{12}$CO \textit{J}=1-0 data, on the contrary, were used to estimate the mass-loss rates for all sources in the sample; this line suffers less from opacity and extended source effects, and so it is easier to use it for this purpose. Moreover, the $^{12}$CO \textit{J}=2-1 line has been used to unveil the possible structures responsible for such profiles. Due to the low S/N of the $^{13}$CO \textit{J}=1-0 data, these are not used for any purpose other than to compare profiles and double-check our assumption on the thickness of the $^{12}$CO lines.


\section{Sample, observations, and previous studies}

\subsection{Sample selection}

We selected all SRa/SRb variables in the General Catalog of Variable Stars (GCVS; \citealt{samus09}) with declinations above $-$25$^{\circ}$, with information on the spectral and chemical type and variability period, and for which the Hipparcos parallaxes \citep{van07} are larger than 2 mas (all sources are closer than 500 pc; we note that \textit{Gaia} parallaxes were not available at the time of sample selection). Moreover, we further refined our sample by selecting only targets with IRAS 60 $\mu$m fluxes larger than 4 Jy (and with 12, 25, and 60 $\mu$m fluxes of quality 3; \citealt{iras88}). This last criterion was introduced in view of the good correlation between IRAS 60 $\mu$m fluxes and CO intensities: nothing is gained from observing sources with less than 4 Jy, as the CO limit obtained in a reasonable time is meaningless. This resulted in a list of 49 sources\footnote{We note that we were able to detect CO emission in all but three sources in the sample (see Figs. A.1-A.6).}. The main parameters of the observed sample of SRs are summarized in Table 1, where we give the observed coordinates, variability type, period of variability, spectral type, variation amplitude, the IRAS 12, 25, 60, and 100 $\mu$m fluxes, and Hipparcos and \textit{Gaia} distances, respectively. We also made observations of CW Leo, NGC 7027, and CRL 618 as calibrators.

Figure 1 shows a comparison between Hipparcos and \textit{Gaia} parallaxes (taken from the \textit{Gaia} DR2 archive; \citealt{gaia18}), which have been converted to distances in parsecs for the full sample. As we see, Hipparcos parallax uncertainties are higher than those from \textit{Gaia}. Moreover, we find that for Hipparcos distances between $\sim$250 and 500 pc, \textit{Gaia} distances continue to increase up to values of $\sim$1000 pc. Hipparcos was only able to accurately measure distances up to values of $\sim$250 pc (see the 1:1 relationship marked by a dashed line in Fig. 1). Therefore, although the first selected sample would now be different, we use the same sample but only taking into consideration the new distances (see Table 1). For S Dra, the only exception in our sample, we take the same parallax ($\sim$411 pc), since the \textit{Gaia} parallax seems to be negative.

\begin{figure}
\centering
\includegraphics[angle=0,width=8.6cm]{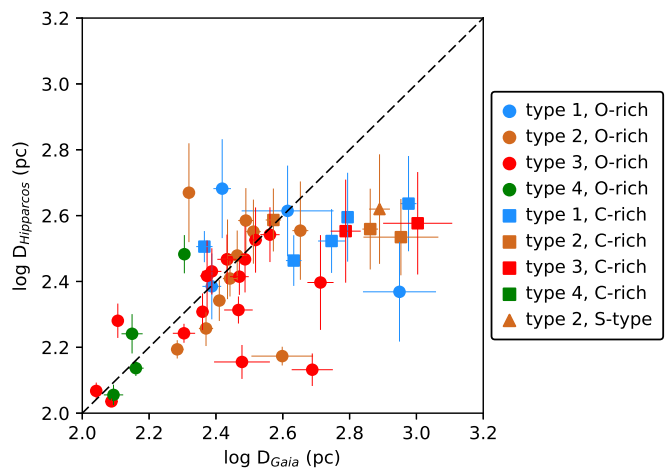}
\caption{Comparison between the Hipparcos and \textit{Gaia} distances derived from parallax measurements (see Table 1) for the 49 SRs in our sample. Circles, squares, and triangles are O-rich, C-rich, and S-type stars, respectively. Blue, brown, red, and green colors are type 1, 2, 3, and 4 profiles, respectively (see Sect. 3.1). The dashed line marks the 1:1 relationship.}
\label{figmassod_deltamv}%
\end{figure}

\subsection{Observations and data processing}

The observations were obtained at the 30m MRT of the Institut de Radioastronomie Millimétrique (IRAM) during the period from June to November, 2012. Precipitable water vapor during this observing run was between 3.5 and 8.5 mm. We used the broadband EMIR receivers E090 (3 mm) and E230 (1.3 mm) in Wobbler Switching observing mode in order to get very flat baselines and dual band polarization. We observed $^{12}$CO \textit{J}=1-0 in the E090 band at a rest frequency\footnote{Frequencies are from the Cologne Database of Molecular Spectroscopy (CDMS, \citealt{muller01}).} of 115.271 GHz, and $^{12}$CO \textit{J}=2-1 in the E230 band at 230.538 GHz. Moreover, we were able to observe $^{13}$CO \textit{J}=1-0 at 110.201 GHz. We reached a sensitivity of 23 and 18 mK (in \tmb~scale) for 230 and 115 GHz, respectively, by using a typical elevation of 45$^{\circ}$ and a spectrometer resolution of 1 km\,s$^{-1}$ in 1.5 hr. Thus, we invested 1.5 hr of telescope time on each source of the 23 objects in the sample with IRAS 60 $\mu$m fluxes between 4 and 10 Jy. For the 19 sources with IRAS 60 $\mu$m fluxes between 10 and 30 Jy, we invested 1 hr of time on each one. Finally, for the 7 sources with IRAS 60 $\mu$m fluxes between 30 and 70 Jy, we invested 0.5 hr of time on each one. In total, we needed 57 hr to complete the sample (including ON and OFF positions, pointing, and calibration). 

The observations were calibrated by observing hot (ambient) and cold (liquid nitrogen) loads every 20 minutes. During these calibration scans, we also observed the blank sky to derive its opacity from its emissivity and the values provided by the weather station (using ATM atmospheric model; see \citealt{pardo01}). Pointing was also checked every 2 hours and before moving to a new target by observing nearby strong continuum sources (quasars, planets, and compact HII regions). Uncertainties in the absolute calibration are of the order of 5 -- 10\% at 3\,mm, and 10 -- 15\% at 1.3\,mm. The calibration was also checked several times per day by observing strong line emitters of well-known properties.

Data were reduced with the CLASS software of the GILDAS\footnote{The GILDAS software package is available for download on the IRAM web pages.} package. For each source, scans were individually inspected. After the removal of bad scans and the flagging of bad channels, they were averaged (including horizontal and vertical polarizations) and a flat baseline was removed. Data were also converted into the \tmb~scale using the recommended values of \beff~and \feff~for EMIR at the frequencies of the observed lines. The spectra are displayed in main-beam temperature (\tmb; see Appendix A, Figs. A.1-A.6). 

\begin{table*}
\caption{Main properties of the observed sample of SRs.}
\label{table:1}   
\centering          
\begin{tabular}{l c c c c c c c c c c c c}     
\hline\hline
        Star & RA  &  Dec  & Var. & Period & Spectral &  $\Delta$m$_{V}$ & \multicolumn{4}{c}{IRAS fluxes (Jy)}           & Hip. dist. & \textit{Gaia} dist. \\
             &     &       &   type   &  (d)   & type     &   (mag)          & 12 & 25 & 60              & 100            & (pc)      & (pc)      \\
\hline  
      VX And & 00:19:54  &  +44:42:34  &       SRa& 375   &    C   &  2.2&  53     &  15  &  4   &   2       & 394     &  622        \\
    V465 Cas & 01:18:14  &  +57:48:11  &       SRb& 60    &    M   &  1.8&  72     &  32  & 5  &  2          & 348     &  364        \\
    V370 And & 01:58:44  &  +45:26:07  &       SRb& 228   &    M   &  1.0&  499    &  292   & 42  &  16      & 174     &  141        \\
       Z Eri & 02:47:56  &  -12:27:38  &       SRb& 80    &    M   &  1.6&  63     &  27    & 4  &  1        & 256     &  277        \\
      RZ Ari & 02:55:48  &  +18:19:54  &       SRb& 57    &    M   &  0.6&  147     &  37  &  6 &   2        & 108     &  98         \\
      SS Cep & 03:49:30  &  +80:19:21  &       SRb& 90    &    M   &  1.1&  114    &  55  & 11  &  4         & 260     &  295        \\
      TT Tau & 04:51:31  &  +28:31:37  &       SRb& 167   &    C   &  2.0&  33     &  11  & 7  &  4          & 362     &  727        \\
       W Ori & 05:05:24  &  +01:10:39  &       SRb& 212   &    C   &  4.2&  184    &  52  & 14  &  6         & 377     &  1009       \\
      RX Lep & 05:11:23  &  -11:50:57  &       SRb& 60    &    M   &  2.4&  280    &  124   & 19    &  6     & 149     &  397        \\
       Y Tau & 05:45:39  &  +20:41:42  &       SRb& 242   &    C   &  2.7&  144    &  51  & 13  &  4         & 357     &  614        \\
       S Lep & 06:05:45  &  -24:11:44  &       SRb& 89    &    M   &  1.6&  258    &  141   & 17  &  5       & 203     &  229        \\
      RY Mon & 07:06:56  &  -07:33:27  &       SRa& 456   &    C   &  1.7&  59     &  17  & 5  &  19         & 433     &  947        \\
      RT Hya & 08:29:41  &  -06:19:08  &       SRb& 290   &    M   &  3.2&  106    &  41  & 6   &  2         & 301     &  291        \\
      RT Cnc & 08:58:16  &  +10:50:43  &       SRb& 60    &    M   &  1.8&  73     &  29  & 4   &  1         & 243     &  244        \\
       R Crt & 11:00:34  &  -18:19:30  &       SRb& 160   &    M   &  1.4&  638    &  308   & 50 &  20       & 261     &  236        \\
      Z UMa  & 11:56:30  &  +57:52:18  &       SRb& 196   &    M   &  3.2&  64     &  26    & 4  &   2       & 356     &  325        \\
      BK Vir & 12:30:21  &  +04:24:59  &       SRb& 150   &    M   &  1.5&  249    &  102   & 19    &  8     & 181     &  235        \\
       Y UMa & 12:40:21  &  +55:50:48  &       SRb& 168   &    M   &  2.1&  193    &  91  & 16    &  7       & 385     &  309        \\
       Y CVn & 12:45:08  &  +45:26:25  &       SRb& 157   &    C   &  2.5&  276    &  70  & 17  &  8         & 320     &  232        \\
      RT Vir & 13:02:38  &  +05:11:08  &       SRb& 155   &    M   &  1.3&  462    &  226   & 39  &  15      & 135     &  488        \\
      SW Vir & 13:14:04  &  -02:48:25  &       SRb& 150   &    M   &  1.5&  681    &  340   & 49  &  15      & 143     &  300        \\
      RX Boo & 14:24:12  &  +25:42:13  &       SRb& 162   &    M   &  2.7&  846    &  419   & 69  &  26      & 191     &  128        \\
      RR UMi & 14:57:35  &  +65:55:57  &       SRb& 43    &    M   &  0.2&  124    &  33    & 5   &  1       & 141     &  100        \\
    $\tau$$^{4}$ Ser & 15:36:28  &  +15:06:05  &  SRb& 100   & M   &  1.2&  188    &  91  & 16  &  6         & 206     &  293        \\
      ST Her & 15:50:47  &  +48:28:59  &       SRb& 148   &    M   &  1.5&  199    &  97  & 17  &  6         & 293     &  271        \\
       X Her & 16:02:39  &  +47:14:25  &       SRb& 95    &    M   &  1.1&  484    &  241   & 39  &  18      & 137     &  145        \\
       g Her & 16:28:38  &  +41:52:54  &       SRb& 89    &    M   &  2.0&  438    &  149   & 24  &  7       & 109     &  122        \\
       S Dra & 16:42:56  &  +54:54:14  &       SRb& 136   &    M   &  1.0&  131    &  73  & 12  &  4         & 411     &  411        \\
      FI Lyr & 18:42:05  &  +28:57:30  &       SRb& 146   &    M   &  0.8&  93   &  55  & 7  &  2            & 358     &  449        \\
       S Sct & 18:50:20  &  -07:54:27  &       SRb& 148   &    C   &  1.3&  65   &  17  & 9  &  14           & 386     &  373        \\
      AF Cyg & 19:30:13  &  +46:08:52  &       SRb& 93    &    M   &  2.0&  78   &  34  & 4  &  1            & 220     &  257        \\
      AQ Sgr & 19:34:19  &  -16:22:27  &       SRb& 200   &    C   &  2.3&  57   &  19  & 6  &  6            & 333     &  557        \\
      RT Cap & 20:17:06  &  -21:19:04  &       SRb& 393   &    C   &  1.2&  73   &  21  & 4  &  4            & 291     &  429        \\
      UU Dra & 20:24:04  &  +75:15:14  &       SRb& 120   &    M   &  1.5&  132    &  90  & 13  &  5         & 336     &  330        \\
      EU Del & 20:37:55  &  +18:16:07  &       SRb& 60    &    M   &  1.1&  175    &  49  & 7 &  2           & 117     &  110        \\
       U Del & 20:45:28  &  +18:05:24  &       SRb& 110   &    M   &  1.3&  142    &  78  & 11 &  4          & 481     &  262        \\
      IQ Aqr & 20:49:17  &  -00:33:48  &       SRb& 385   &    M   &  0.6&  61   &  23  & 4   &  2           & 304     &  202        \\
       W Cyg & 21:36:02  &  +45:22:29  &       SRb& 131   &    M   &  2.1&  349    &  142   & 22 &  8        & 175     &  201        \\
      RU Cyg & 21:40:39  &  +54:19:29  &       SRa& 233   &    M   &  2.4&  190    &  105   & 18 &  19       & 249     &  516        \\
      EP Aqr & 21:46:32  &  -02:12:46  &       SRb& 55    &    M   &  0.4&  637    &  321   & 47 &  16       & 114     &  124        \\
      TW Peg & 22:03:59  &  +28:20:54  &       SRb& 930   &    M   &  0.9&  262    &  152   & 22 &  8        & 269     &  244        \\
    $\textit{o}$$^{1}$ Ori & 04:52:32  & +14:15:02 & SRb & 30& M   &  0.2& 85  & 21 & 5 & 4                  & 200     &  162        \\
      SV Lyn & 08:03:40  &  +36:20:42  &       SRb& 70    &    M   &  0.9&  52   &  29  & 5   &  1           & 293     &  272        \\
      SV Peg & 22:05:42  &  +35:20:55  &       SRb& 145   &    M   &  1.8&  265    &  146   & 24 &  10       & 234     &  890        \\
      AK Hya & 08:39:53  &  -17:18:11  &       SRb& 75    &    M   &  0.6&  192    &  93  & 15 &  6          & 156     &  192        \\
      RW Boo & 14:41:13  &  +31:34:20  &       SRb& 209   &    M   &  1.5&  61   &  30  & 5 &  3             & 293     &  307        \\
      IQ Her & 18:17:55  &  +17:58:53  &       SRb& 75    &    M   &  0.5&  64   &  21  & 6 &  3             & 467     &  208        \\
      AA Cyg & 20:04:28  &  +36:49:00  &       SRb& 213   &    S   &  3.0&  40   &  15  & 5  &  9            & 417     &  775        \\
       X Cnc & 08:55:23  &  +17:13:53  &       SRb& 195   &    C   &  2.0&  90     &  25  & 7 &  3           & 343     &  898        \\
\hline                                                                                                                                          
\end{tabular}
\tablefoot{Description of the columns from left to right. (1) GCVS designation \citep{samus09}; (2-3) R.A. and Dec. (J2000); (4-7) Variability type, period of variability, spectral type, and variation amplitude from the GCVS \citep{samus09}; (8-11) IRAS fluxes \citep{iras88}. (12) Distance derived from Hipparcos parallax \citep{van07}; (13) Distance derived from \textit{Gaia} parallax (\textit{Gaia} DR2; \citealt{gaia18}).\\
}
\end{table*}

\subsection{Previous studies}

One of the pioneering studies in this direction was made by \citet{olof93}, who performed a survey of circumstellar CO $\textit{J}=2-1$ and $\textit{J}=1-0$ emissions on a sample of 120 bright C-rich stars within 600 pc of the Sun, finding that most of the stars have circumstellar envelopes. The sample contained Mira (regular), Lb (irregular), SRa, and SRb (semi-regular) variables. Moreover, they found good correlations between the different characteristic parameters of the envelopes, such as the correlation between IR excess and the estimated mass-loss rate. Later on, \citet{kers96} observed a sample of 48 O-rich SRs (SRa and SRb) in the $^{12}$CO $\textit{J}=2-1$ and $\textit{J}=1-0$ lines, obtaining weak intensities and finding no significant correlations between stellar and circumstellar properties.

\citet{kers98}, interestingly, made the first systematic study of a sample of 31 O-rich irregular variables of type Lb by observing the $^{12}$CO $\textit{J}=1-0$, $\textit{J}=2-1$, and $\textit{J}=3-2$ lines. They found that the detected objects were weak in CO and the mass-loss properties were not influenced by the pulsation of the stars. Furthermore, \citet{kers99} carried out a survey of the CO $\textit{J}=1-0$, $\textit{J}=2-1$, $\textit{J}=3-2$, and $\textit{J}=4-3$ emissions on a sample of 109 O-rich SRa, SRb, and Lb variables. The catalogue contains all observational data, as well as discussions on detection statistics, line profiles, gas expansion velocity distributions, and correlations between CO line and IR continuum fluxes (including a connection with the mass-loss mechanisms).

In spite of the significant results of these works, we would like to add that the S/Ns were relatively low and therefore it is difficult to clearly identify the profile shapes other than for the brightest sources. Our sample is more significative and the detection rate is higher (we note that we are considering sources with IRAS 60 $\mu$m fluxes larger than 4 Jy).

\section{Expansion velocities and $^{12}$CO line profiles}

Lines were characterized using the CLASS software. The profile parameters (1$\sigma$ noise in K, integrated line intensity in K\,km\,s$^{-1}$, peak main-beam brightness temperature in K, and full width at half power in km\,s$^{-1}$) are displayed in Tables A.1$-$A.3 for the $^{12}$CO $\textit{J}=2-1$, $^{12}$CO $\textit{J}=1-0$, and $^{13}$CO $\textit{J}=1-0$ lines, respectively. In Table A.1 we also give the gas final expansion velocity in km\,s$^{-1}$, the systemic velocity in km\,s$^{-1}$, the full width at zero power (FWZP) in km\,s$^{-1}$, and the line profile type (see Sect. 3.1), since the $^{12}$CO $\textit{J}=2-1$ line has been used to study the main properties of the CSEs. Table A.2 also gives the $^{12}$CO line intensity ratios (A[$^{12}$CO \textit{J}=1-0]/A[$^{12}$CO \textit{J}=2-1]) and their uncertainties. We also present 3$\sigma$ upper detection limits for those lines which are unsuccessfully detected by making use of the FWZP of the $^{12}$CO $\textit{J}=2-1$ line.

\begin{figure*}
\centering
\includegraphics[angle=0,width=8.0cm]{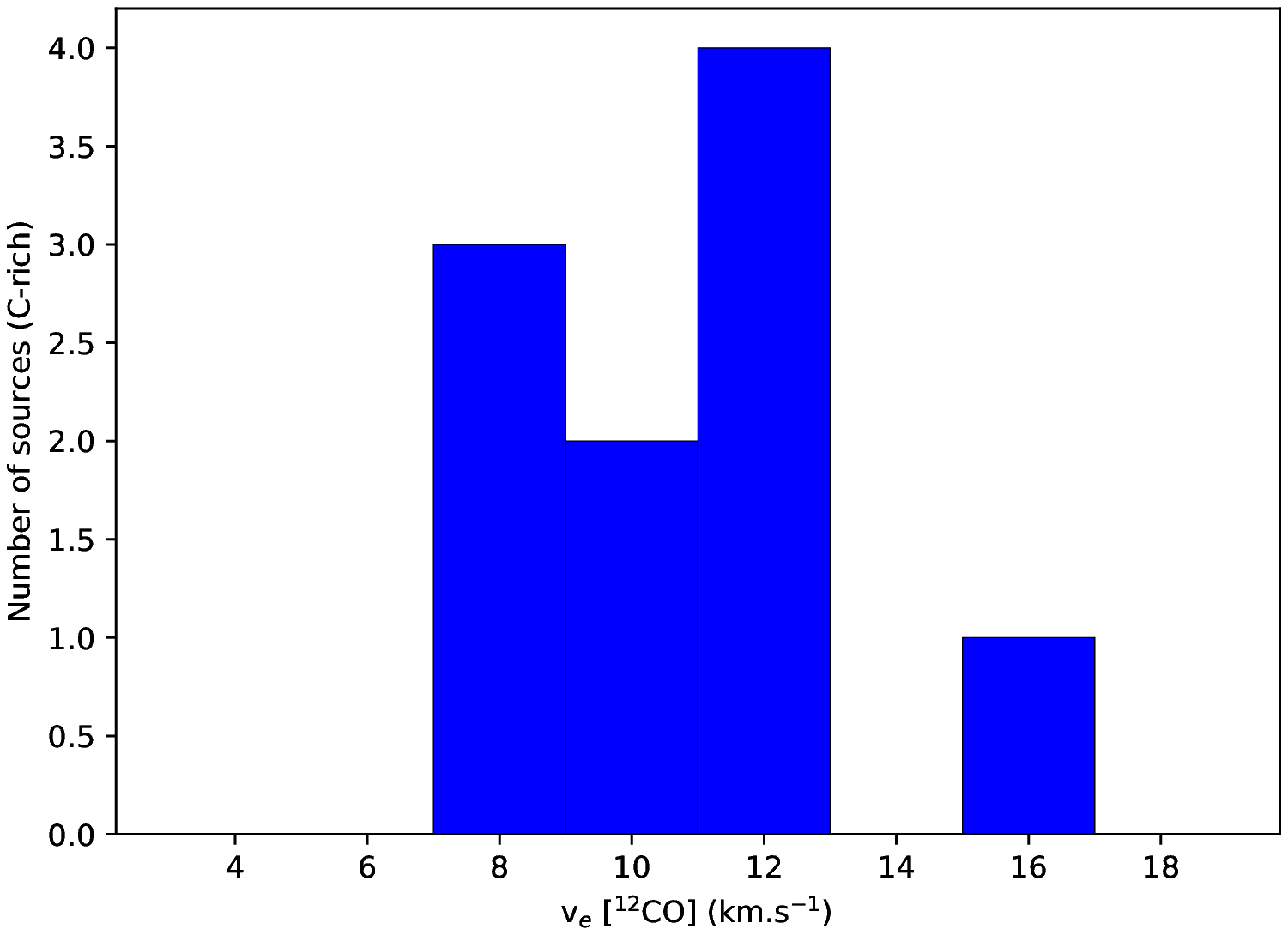}
\hspace{0.4cm}
\includegraphics[angle=0,width=8.0cm]{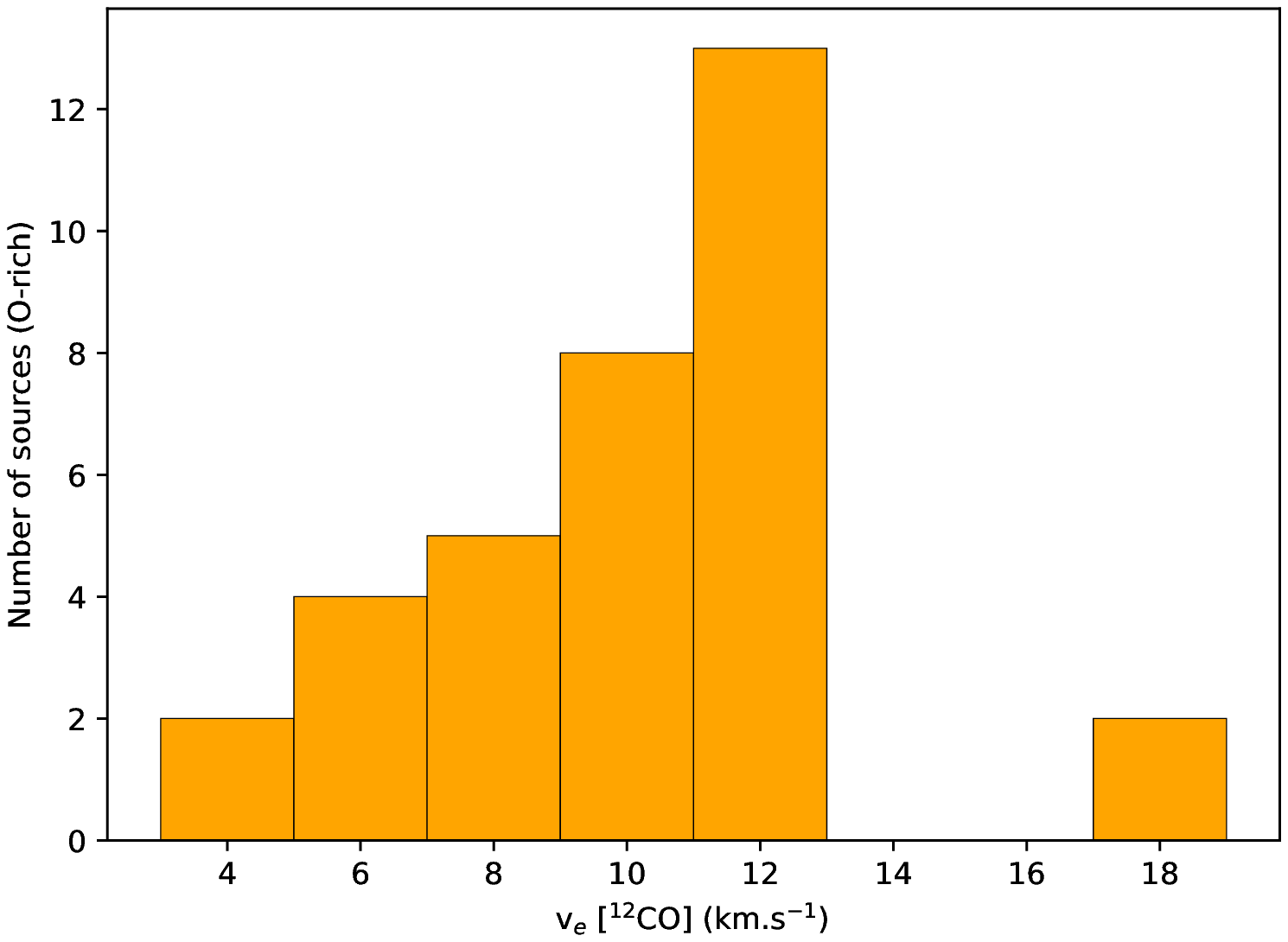}
\caption{Expansion velocity histograms for C-rich (left panel) and O-rich (right panel) SRs. We excluded the only S-type star in the sample (AA Cyg).}
\label{figcaroxi_vexp}%
\end{figure*}

Figure 2 presents expansion velocity histograms for C-rich and O-rich sources (AA Cyg, the only S-type star in the sample, is removed). Four C-rich and thirteen O-rich sources have $\vexp$ ranging from 11 to 13 km\,s$^{-1}$, where both distributions peak. Only one C-rich and two O-rich sources have $\vexp$ higher than 13 km\,s$^{-1}$, at $\sim$16 and 18 km\,s$^{-1}$, respectively. It seems anomalous that only a few O-rich SRs present unusually low $\vexp$ ranging from 3 to 7 km\,s$^{-1}$. The rest of the sample, either C-rich or O-rich SRs, has $\vexp$ ranging from 7 to 11 km\,s$^{-1}$. AA Cyg has $\vexp$ of only $\sim$6.1 km\,s$^{-1}$. 

\subsection{Line profiles}

\begin{figure*}
\centering
\includegraphics[angle=0,width=15cm]{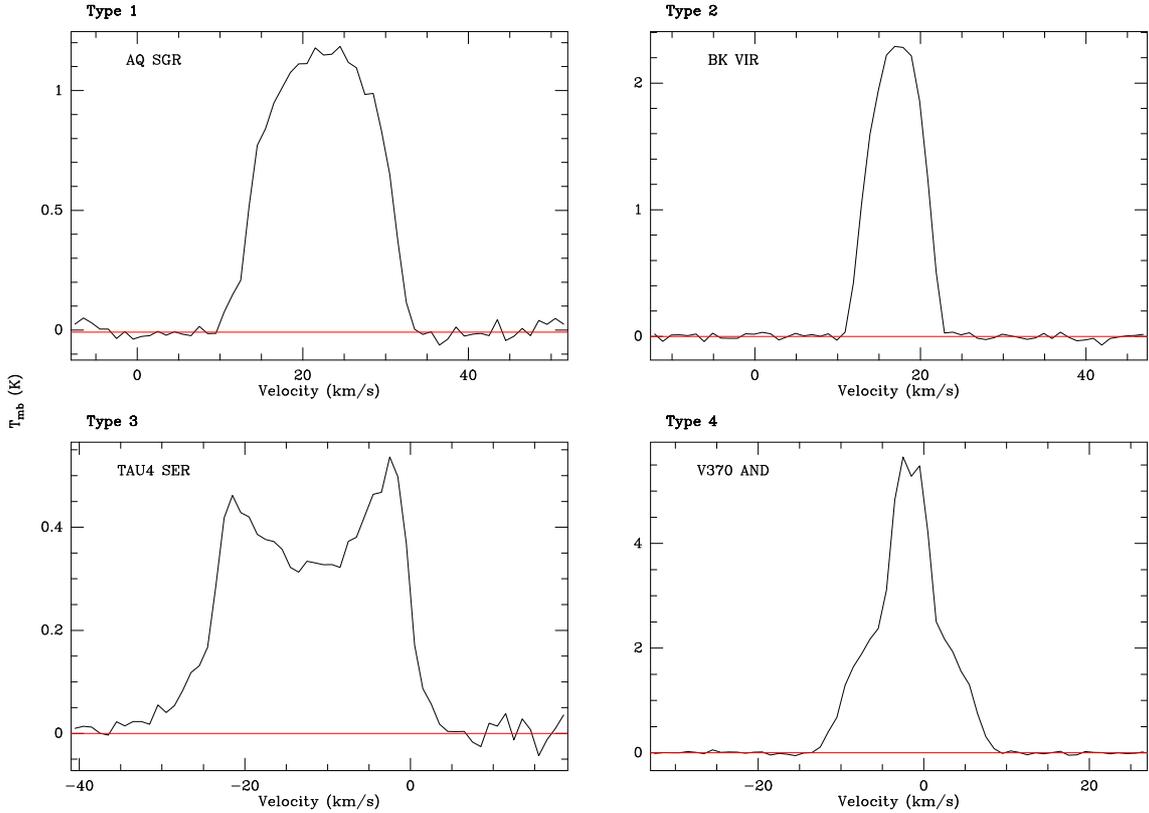}
\caption{Sample spectra of $^{12}$CO $\textit{J}=2-1$ for every profile type (see text). We selected the following sources: AQ Sgr (type 1), BK Vir (type 2), $\tau$$^{4}$ Ser (type 3), and V370 And (type 4).}
\label{figco21_types}%
\end{figure*}

\begin{figure*}
\centering
\includegraphics[angle=0,width=7cm]{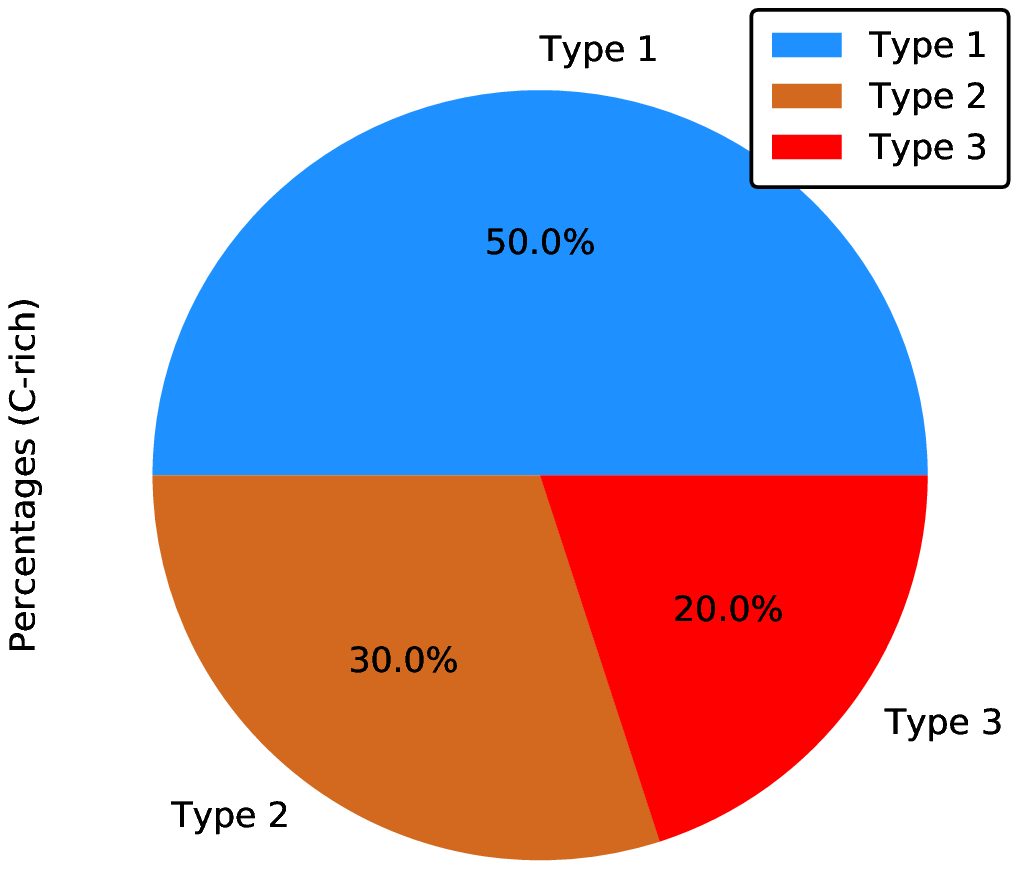}
\hspace{1.2cm}
\includegraphics[angle=0,width=7cm]{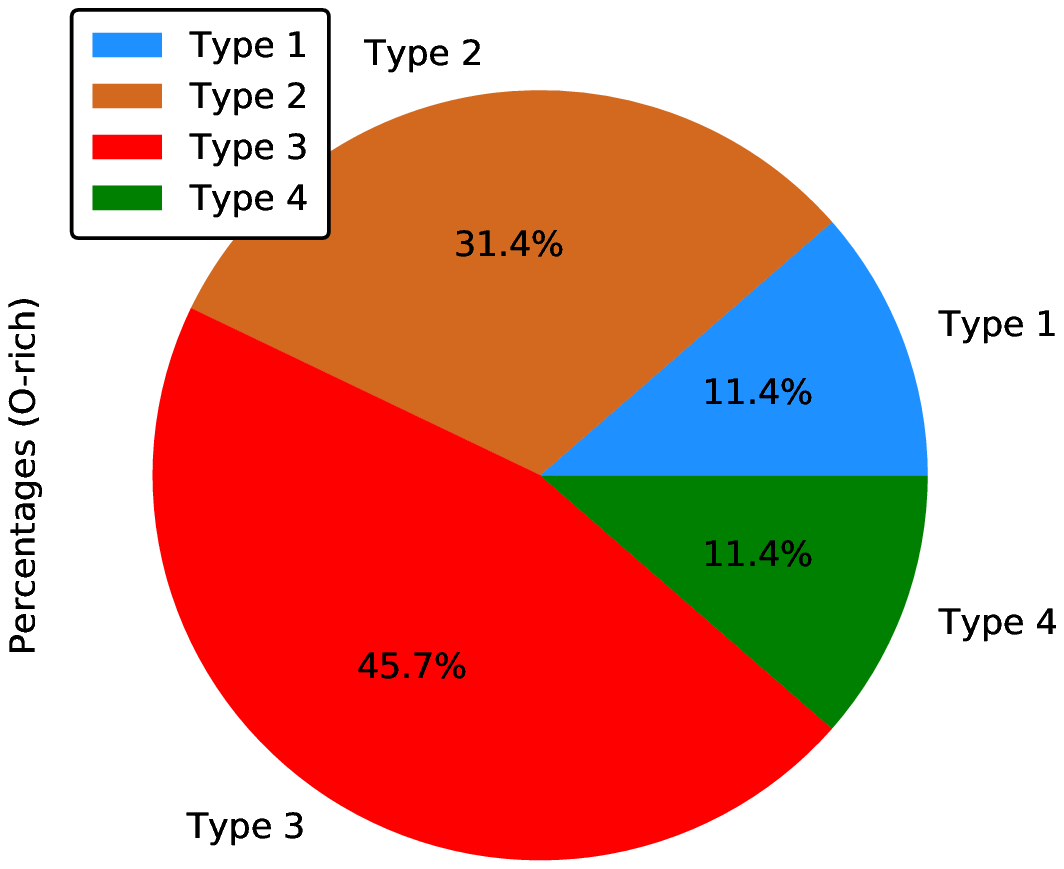}
\caption{Pie charts illustrating the percentage of C-rich (left panel) and O-rich (right panel) SRs for every profile type. Blue, brown, red, and green colors are type 1, 2, 3, and 4 profiles, respectively. We excluded the only S-type star in the sample (AA Cyg).}
\label{figcaroxi_prof}%
\end{figure*}

Sources have been classified into four groups based on the $^{12}$CO line profiles and $\vexp$ (see Fig. 3, Figs. A.1-A.4, and Table A.1). Type 1 profiles are broad and symmetric lines (with respect to the systemic velocity) with expansion velocities ($\vexp$) between 9 and 13 km\,s$^{-1}$. These profiles are in accordance with more standard spherically symmetric envelopes (e.g., parabolic, corresponding to optically thick and nonspatially resolved emissions; see, e.g., \citealt{olof82}). Type 2 profiles are unusually narrow and symmetric lines with $\vexp$ between 3 and 9 km\,s$^{-1}$. Type 3 profiles however are very strange profiles with very pronounced asymmetries; they are not compatible with previously studied, standard, optically thin lines (see, e.g., \citealt{olof82}), and have $\vexp$ between 9 and 17 km\,s$^{-1}$. Some of the U-shaped $^{12}$CO profiles may suggest that the envelopes, if standard, are optically thin and spatially resolved, but the difference between the blue and red horns suggests that the envelopes may be optically thick (e.g., the $^{12}$CO $\textit{J}=2-1$ line profile in SW Vir; see Fig. A.1). Moreover, by comparing these profiles with those obtained by \citet[e.g., SW Vir]{kers99}, we find that type 3 profiles do not change when using different telescopes and beams, demonstrating the nonstandard behavior. Finally, type 4 profiles are those lines which show two different components: one narrow component with low $\vexp$ ($\sim$2-3 km\,s$^{-1}$) and another broad component with higher $\vexp$ ($\sim$9-12 km\,s$^{-1}$), like the addition of a type 2 profile to a type 1 or type 3 profile (see Fig. 3). Although we expect that the $^{12}$CO $\textit{J}=2-1$, $^{12}$CO $\textit{J}=1-0$, and $^{13}$CO $\textit{J}=1-0$ lines have the same behavior for each source, there are some cases where this issue is not simple. There are some sources for which the $^{12}$CO $\textit{J}=2-1$ line seems to exhibit a type 1 profile, however the shape of the $^{12}$CO $\textit{J}=1-0$ line describes a type 3 profile, and so we consider that these lines are classified as type 3 (i.e., those towards SS Cep, RT Vir, RX Boo, EU Del, and TW Peg; see Figs. A.1-A.4). Other sources seem to show a specific profile type, but further analysis is hampered by overly high 1$\sigma$ noise (i.e., those towards Z Eri, S Lep, and RT Cnc; see Table A.1 and Figs. A.1, A.3, and A.5). Finally, it is necessary to mention S Sct, a particular exception which has already been studied thoroughly and in detail, and that we have classified as a type 2 profile. The circumstellar environment of this peculiar C-rich source was first studied by \citet{olof90} in CO line emission, who found a shell extended in the range of $\sim$4-5.5 $\times$ 10$^{17}$ cm from a pronounced double-peaked line profile. Nowadays, it is well known that this target shows spherically symmetric detached circumstellar emission (see, e.g., \citealt{mecina14}). In our data, we thus see three different components, and so we only take into account the central one (see Figs. A.2, A.4, and A.6, and Tables A.1-A.3).

This classification of SRs into four groups makes us wonder about the possible relation between profile types and spectral types. In Figure 4, two different pie charts show the percentage of C-rich and O-rich sources for each type of profile, respectively. Type 3 profiles form the largest group, with 20\% C-rich and 45.7\% O-rich SRs (18 sources of the total sample). This type of $^{12}$CO profile predominates and is representative of our complete sample of SRs. The second largest group is formed by sources displaying a type 2 profile, with 30\% C-rich and 31.4\% O-rich SRs (14 sources of the total sample). Curiously, we find that half of the C-rich SRs (five sources) and only four O-rich SRs in our sample present a type 1 profile. Finally, only 11.4\% O-rich SRs (four sources) present a type 4 profile. Therefore, it is inevitable to make a clear connection between the type 4 profile and O-rich SRs, noticing that the sample only contains 10 C-rich SRs, as compared with the dominant 38 O-rich SRs.

\citet{knapp98} found that double-component line profiles exist among O- and C-rich stars, however we find no evidence of double-component profiles in C-rich SRs of our volume-limited unbiased sample. These latter authors argued that this kind of profile may be due to the effect of episodic mass loss with varying expansion velocities or an effect of complicated geometries and kinematics. On the contrary, we find that only a few O-rich stars in our sample present very low $\vexp$, in agreement with \citet{olof02}, who found that the fraction of low-velocity sources is much higher in O-rich stars. It is clear that in this respect there is an interesting difference between the chemistries of C- and O-rich stars, and this may explain why we only see double-component profiles in O-rich SRs (referring to the narrow components with very low gas expansion velocities; see below), but of course the real nature of these enigmatic profiles is still unknown. The possible relation between mass-loss rates, profile types, spectral types, and variation amplitude is discussed in Sect. 4.1.

Recently, \citet{ortiz16} analyzed a volume-limited sample of AGB stars within 500 pc of the Sun with the aim of finding main sequence companions using a specific criterion. This criterion is based on searching for UV flux in excess of the companion star, which seems to prevail over that of the primary. Curiously, some of the SRs in our sample are confirmed and suspected binary AGB stars and most of them seem to show a type 3 profile in $^{12}$CO (i.e., SS Cep, S Lep, SW Vir, RW Boo, ST Her, g Her, and W Cyg). Only three of them show a type 2 profile (i.e., Z Eri, X Cnc, and Z UMa) and two are not detected in $^{12}$CO (i.e., $\textit{o}$$^{1}$ Ori and RR UMi). Therefore, we believe that sources that show a type 3 profile may be more likely to harbor binary systems.

\subsubsection{Double-component or type 4 profiles}

The first spatial information of this kind of profile was provided by \citet{kahane96}, in which they reveal that the broad plateau seen in X Her is a bipolar outflow, while a narrower feature seems not to be spatially resolved. Since then, these sources have become increasingly important over time. \citet{knapp98} formed a hypothesis based on a multiple-shell structure, in which each shell has a different gas expansion velocity that may be produced by episodic mass loss. The spatial-kinetic properties of EP Aqr seem to be an example of this multiple-shell model. Shortly after, \citet{kers99} confirmed this hypothesis by identifying a small group of O-rich semi-regular and irregular variables, including EP Aqr, that clearly show multi-component line profiles (i.e., EP Aqr, RV Boo, X Her, and SV\,Psc). All of them show a very narrow feature with a full width of less than $\sim$5 km\,s$^{-1}$ on a much broader plateau of $\sim$18 km\,s$^{-1}$. 

The latest results from good interferometric maps of X Her and EP Aqr, both of which are included in our sample, have provided significant insight into the problem of multi-component axisymmetric circumstellar envelopes. High-resolution CO data of X Her suggest that the broad emission plateau and the narrow emission feature arise from an expanding bipolar outflow and a disk, respectively \citep{naka05,castro10}. This disk, which is not resolved by \citet{castro10}, is interpreted as being in rotation (see \citealt{naka05} for more details). Moreover, these maps show, for the first time, an hourglass structure similar to that seen for the proto-planetary nebula (PPN) M 2-56 (we note that this structure has also been seen in the central parts of many other PPNe; see \citealt{castro10} for more details). EP\,Aqr, on the other hand, exhibits a bi-conical wind, also defined as polar jets, with a bright nearly face-on spiral feature, which resembles simulations of the spiral-inducing wind Roche-lobe overflow mechanism. This latter seems to be the first convincing detection of a spiral\footnote{We note that the symbiotic stellar system (SS) R Aqr also seems to show the formation of a double spiral extending significantly beyond the orbit (see \citealt{buja18}).} morphology, observed with ALMA, in an O-rich source (see \citealt{homan18}, \citealt{nhung19}, \citealt{hoai19}, and \citealt{tuan19} for more details).

Table A.5 gives the profile parameters (integrated line intensity, peak main-beam brightness temperature, systemic velocity, and gas expansion velocity) for each component in those sources that clearly show a $^{12}$CO $\textit{J}=2-1$ double-component profile in our sample (i.e., X Her, EP Aqr, V370 And, and IQ Aqr). We note that the same behavior is also seen in the $^{12}$CO $\textit{J}=1-0$ and $^{13}$CO $\textit{J}=1-0$ lines (see Figs. A.1-A.6). As we see, it was necessary to separate the two components in all the $^{12}$CO $\textit{J}=2-1$ profiles in order to measure the parameters. To do this in CLASS, we needed to delete the channels where the narrow component is located, and then interpolate between them using the channels of the broad plateau. The deleted channels were analyzed separately. Therefore, we have different values for each component. The peak main-beam brightness temperature of the narrow components is the same as that of the entire lines (see Table A.1), while for the broad plateaus it is always lower. The situation is less clear for the integrated line intensity, for which we assumed that the emissions are additive. For the broad plateau, its value provides a good approximation; however for the narrow component, we are missing part of the line, which is included in the broad one by default. Another option would be to fit the narrow line with a Gaussian profile, but we prefer to keep the procedure already used. On the other hand, we note that the centroid or systemic velocities for both components are slightly different in the four cases. Moreover, both components expand at different velocities (see Table A.5). For V370 And, this is the first time that all the parameters are given for both components, although this source has already been reported as having double-component winds (see, e.g., \citealt{dani15}). For X Her and EP Aqr, all parameters seem to be in accordance with \citet{kers99}. Interestingly, we report, for the first time, the tentative detection of IQ Aqr as presenting two components (or more) in the circumstellar gas envelope. The broad plateau seems to be much broader than in the other three cases, with a peak main-beam brightness temperature at the 3$\sigma$ noise level (see Fig. A.2, and Tables A.1 and A.5). The narrow component on the contrary is somewhat different and displays a double-peak profile, which is unusual in other narrow profiles. The gas velocity is also higher ($\sim$6 km\,s$^{-1}$; see Table A.5). Nevertheless, we prefer to be cautious until more observations confirm this detection, since the $^{12}$CO $\textit{J}=1-0$ and $^{13}$CO $\textit{J}=1-0$ lines do not show this behavior (see Figs. A.4 and A.6).

\section{Gas mass-loss rates and photodissociation radius}

According to the literature, mass-loss rates can be derived from CO observations by following two procedures: i) considering that the envelope is optically thick; and ii) considering that the envelope is optically thin. These two procedures do not use models that fit the $^{12}$CO emission; instead they calculate $\dot{M}$ as stated in \citet{knapp85} simply by using some equations. For optically thick cases, we used equation (5) from \citet{knapp85}, but we made some minor changes:

\begin{equation}
\dot{M} = {\mbox{$T^{\ast}_{\rm mb}$}}\frac{\vexp^{2}d^{2}}{{2.0 \times 10^{9}}f^{0.85}}{F^{-1} (R_{CO})}\;{M_{\odot}\,yr^{-1}},
\end{equation}
where $\mbox{$T^{\ast}_{\rm mb}$}$ is the peak $^{12}$CO $\textit{J}=1-0$ main-beam brightness temperature in K for a 7m telescope\footnote{We note that we are using a 30m telescope and we need to scale the diameter, thus, \mbox{$T^{\ast}_{\rm mb}$} $=$ $\tmb$/18, since (30/7)$^{2}$ $=$ 18.}, $\vexp$ is the expansion velocity in km\,s$^{-1}$, d is the distance in kilo parsecs (by using \textit{Gaia} parallaxes; \textit{Gaia} DR2; \citealt{gaia18}), and f is the abundance of CO relative to H$_{2}$. We use f = 3$\times$10$^{-4}$ for O-rich stars, 6$\times$10$^{-4}$ for S stars, and 8$\times$10$^{-4}$ for C-rich stars \citep{knapp85}. This equation should also take into account the variation of $\mbox{$T^{\ast}_{\rm mb}$}$ as a function of \rco,~called F (\rco), as shown in Fig. 16 of \citet{knapp85}. This function is normalized to 1 for~\rco = 3 $\times$ 10$^{17}$ cm (see \citealt{knapp85} for more details). 

For optically thin cases, we used equation (8) from \citet{knapp85}:

\begin{equation}
\begin{split}
\dot{M} = 10^{-7}{\mbox{$T^{\ast}_{\rm mb}$}}\left(\frac{2.0 \times 10^{-4}}{4.3 \times 10^{-4}}\right)\left(\frac{1500}{D}\right)^{-2}\left(\frac{15}{\vexp}\right)^{-2}\left(\frac{1}{f}\right)\frac{1}{\log\left(\frac{W}{0.04}\right)} \\
\times\;{F^{-1} (R_{CO})}\;{M_{\odot}\,yr^{-1}},
\end{split}
\end{equation}
where D is the distance in parsecs. For optically thin envelopes, the effects of IR excitation are included in W, which measures the ratio of the 4.6 $\mu$m flux to that of a standard 2000 K, 5 $\times$ 10$^{13}$ cm blackbody (see \citealt{knapp85} for more details). We also rescaled the original function to make $\mbox{$T^{\ast}_{\rm mb}$}$ coincide with 1.8 $\times$ 10$^{-3}$ K for an envelope radius of 10$^{17}$ cm (see Fig. 20 of \citealt{knapp85} for optically thin envelopes). We used F (\rco) as defined in optically thick envelopes (see above). Therefore, this equation can be rearranged to give

\begin{equation}
\dot{M} = {\mbox{$T^{\ast}_{\rm mb}$}}\frac{\vexp^{2}D^{2}}{{1.52 \times 10^{12}}f}{F^{-1} (R_{CO})}\;{M_{\odot}\,yr^{-1}}.
\end{equation}

The photodissociation radius on the other hand has been estimated by using equation (12) of \citet{groene17}, which takes into account the strength of the interstellar radiation field (ISRF; see \citealt{groene17} for more details):

\begin{equation}
\begin{split}
R_{CO} = 42\left(\frac{\dot{M}}{10^{-6}}\right)^{0.57}\left(\frac{\vexp}{15}\right)^{-0.14}\left(\frac{f}{10^{-4}}\right)^{0.25}\chi^{-0.41}\left(\frac{T_{ex}}{10}\right)^{+0.02} \\
+ f_{sc}\,4.00\,\left(\frac{\vexp}{15}\right)\left(\frac{2.6 \times 10^{-10}}{k_{0}}\right)\frac{1}{\chi}\;{(10^{15} cm)},
\end{split}
\end{equation}
where $\chi$ is a scaling factor indicating the strength of the ISRF ($\sim$1 with respect to the Draine field; see \citealt{draine78} for more details), \tex is the CO excitation temperature ($\sim$5 K; \cite{li14} used a model for CW Leo in which the excitation temperature is constant in the CSE and equal to 5 K), f$_{sc}$ is a correction applied to the second term because of the finite size of the central star (f$_{sc}$ = 4.144/4 = 1.036), and k$_{0}$ an unshielded CO photodissociation rate of 2.6 $\times$ 10$^{-10}$ s$^{-1}$, from \citet[see also \citealt{draine78}]{visser09}. Therefore, equation (4) can also be rearranged to give

\begin{equation}
\begin{split}
R_{CO} = 42\left(\frac{\dot{M}}{10^{-6}}\right)^{0.57}\left(\frac{\vexp}{15}\right)^{-0.14}\left(\frac{f}{10^{-4}}\right)^{0.25} \\ 
+\,4.144\,\left(\frac{\vexp}{15}\right)\;{(10^{15} cm)}.
\end{split}
\end{equation}

We then consistently calculated $\dot{M}$ and \rco~for each of the stars in our sample by solving the system iteratively in both cases.

\subsection{Results}

Since a priori we do not know if the CO emission is optically thin or thick, we finally decided to calculate mass-loss rates and envelope sizes, ($\dot{M}$$_{thin}$, R$^{thin}_{CO}$) and ($\dot{M}$$_{thick}$, R$^{thick}_{CO}$), for each source (see Table A.4 in the Appendix). We assume that the envelopes are not resolved by the telescope, since the estimated R$^{thin}_{CO}$ and R$^{thick}_{CO}$ are smaller than the telescope half power beam widths for the $^{12}$CO \textit{J}=2-1 ($\sim$13'') and the $^{12}$CO \textit{J}=1-0 ($\sim$21'') observations, respectively. In the following section, we discuss, through a simple test using SHAPE+shapemol, whether the emission is optically thin or optically thick in $^{12}$CO \textit{J}=2-1 for each source. In Table 2 we show the main results of this work, giving a conclusive mass-loss rate estimate and a final photodissociation radius for each source, in conformity with the above-mentioned test. Observations for which the $^{12}$CO \textit{J}=2-1 and 1-0 lines are not detected were eliminated (i.e., RZ Ari, RR UMi, and $\textit{o}$$^{1}$ Ori). For those sources for which the $^{12}$CO \textit{J}=2-1 line is detected, but not the $^{12}$CO \textit{J}=1-0, we derived 3$\sigma$ upper limits (i.e., VX And, Z Eri, S Lep, RT Cnc, and Z UMa) using the FWZP of the $^{12}$CO \textit{J}=2-1 transition and measuring the 1$\sigma$ noise.

The results are presented for 35 O-rich, 10 C-rich, and 1 S-type SR in Tables 2 and A.4. In Table A.4 we give mass-loss rates, photodissociation radii, and wind densities (the ratio of mass-loss rate to expansion velocity) for the optically thick and optically thin cases, respectively, as well as corresponding uncertainties. The derived mass-loss rates have a minimum of 9.20 $\times$ 10$^{-10}$ M$_{\odot}$\,yr$^{-1}$ and a maximum of 1.07 $\times$ 10$^{-6}$ M$_{\odot}$\,yr$^{-1}$ for optically thick cases, and 4.34 $\times$ 10$^{-10}$ M$_{\odot}$\,yr$^{-1}$ and 4.75 $\times$ 10$^{-7}$ M$_{\odot}$\,yr$^{-1}$ for optically thin cases, respectively. Curiously, we find that for each star the optically thick estimate of mass-loss rate is approximately twice as high as the optically thin estimate. This means that we cannot conclude from these estimations whether the circumstellar envelopes surrounding the central stars of the complete sample are optically thin or optically thick in $^{12}$CO (see below for more details about this aspect), but the influence on the derived mass-loss rate values is small.

\begin{table*}                                                                         
\caption{\label{t2a} Main results of this work for 46 SRs of the sample.}              
\centering    
\scalebox{0.83}{
\begin{tabular}{lcccl}                                                                 
\hline\hline                                                                           
\\ [-1ex]                                                                              
Source & Spectral & $\dot{M}$                 &     \rco        &    Comments       \\                                      
       & type     &  (M$_{\odot}$\,yr$^{-1}$) &  (10$^{15}$ cm)     &                   \\                          
\hline   
\\ [-1ex]
             VX And       &   C         & $<$1.94$_{-0.68}$ $\times$ 10$^{-9}$      & $<$5.61$_{-0.48}$       & Optically thin emission. Overly low mass-loss rate.                     \\  [1ex]           
           V370 And       &   M         & 3.66$^{+0.89}_{-0.81}$ $\times$ 10$^{-8}$ & 11.80$^{+1.22}_{-1.11}$ & Optically thin emission. SHAPE+shapemol model (see below).           \\  [1ex]           
              Z Eri       &   M         & $<$8.88$_{-3.31}$ $\times$ 10$^{-10}$     & $<$3.26$_{-0.34}$       & Optically thin emission. Overly low mass-loss rate.                     \\  [1ex]          
             SS Cep       &   M         & 4.08$^{+0.97}_{-0.88}$ $\times$ 10$^{-8}$ & 12.40$^{+1.26}_{-1.15}$ & Optically thin emission. Low mass-loss rate.                         \\  [1ex]          
             TT Tau       &   C         & 5.25$^{+1.65}_{-1.33}$ $\times$ 10$^{-8}$ & 16.60$^{+2.55}_{-2.05}$ & Optically thin emission. Low mass-loss rate.                         \\  [1ex]            
              W Ori       &   C         & 5.91$^{+5.11}_{-2.67}$ $\times$ 10$^{-7}$ &57.30$^{+26.60}_{-13.91}$& Optically thick emission. Similar to the case of RX Boo (see below). \\  [1ex]           
             RX Lep       &   M         & 1.44$^{+0.84}_{-0.67}$ $\times$ 10$^{-8}$ & 7.20$^{+1.88}_{-1.51}$  & Optically thin emission. Low mass-loss rate.                         \\  [1ex]           
              Y Tau       &   C         & 5.79$^{+4.03}_{-1.33}$ $\times$ 10$^{-7}$ &55.50$^{+20.16}_{-6.66}$ & Optically thick emission. Similar to the case of RX Boo (see below). \\  [1ex]           
              S Lep       &   M         & $<$9.89$_{-3.47}$ $\times$ 10$^{-10}$     &  $<$3.97$_{-0.34}$      & Optically thin emission. Overly low mass-loss rate.                     \\  [1ex]           
             RY Mon       &   C         & 3.65$^{+2.63}_{-0.71}$ $\times$ 10$^{-7}$ &44.30$^{+16.81}_{-4.54}$ & Optically thick emission. Similar to the case of RX Boo (see below). \\  [1ex]           
             RT Hya       &   M         & 1.32$^{+0.62}_{-0.60}$ $\times$ 10$^{-9}$ & 2.58$^{+0.46}_{-0.45}$  & Optically thin emission. Overly low mass-loss rate.                     \\  [1ex]           
             RT Cnc       &   M     & $<$4.34$_{-1.50}$ $\times$ 10$^{-10}$     &  $<$3.13$_{-0.29}$      & Optically thin emission. Overly low mass-loss rate.                     \\  [1ex]           
              R Crt       &   M         & 2.28$^{+1.44}_{-0.42}$ $\times$ 10$^{-7}$ & 27.90$^{+8.88}_{-2.56}$ & Optically thick emission. Similar to the case of RX Boo (see below). \\  [1ex]           
              Z UMa       &   M         & $<$4.89$_{-2.03}$ $\times$ 10$^{-10}$     & $<$1.87$_{-0.31}$       & Optically thin emission. Overly low mass-loss rate.                     \\  [1ex]           
             BK Vir       &   M         & 1.12$^{+0.35}_{-0.34}$ $\times$ 10$^{-8}$ & 6.56$^{+0.88}_{-0.86}$  & Optically thin emission. SHAPE+shapemol model (see below).           \\  [1ex]           
              Y UMa       &   M         & 4.24$^{+1.30}_{-1.25}$ $\times$ 10$^{-8}$ & 12.10$^{+1.77}_{-1.70}$ & Optically thin emission. Low mass-loss rate.                         \\  [1ex]           
              Y CVn       &   C         & 1.92$^{+0.47}_{-0.42}$ $\times$ 10$^{-8}$ & 10.60$^{+1.11}_{-1.00}$ & Optically thin emission. Low mass-loss rate.                         \\  [1ex]           
             RT Vir       &   M         & 4.85$^{+3.42}_{-1.56}$ $\times$ 10$^{-7}$ & 41.40$^{+15.49}_{-7.09}$& Optically thick emission. Similar to the case of RX Boo (see below). \\  [1ex]           
             SW Vir       &   M         & 3.62$^{+2.79}_{-1.43}$ $\times$ 10$^{-7}$ & 35.70$^{+14.51}_{-7.44}$& Optically thick emission. Similar to the case of RX Boo (see below). \\  [1ex]           
             RX Boo       &   M         & 6.49$^{+4.12}_{-1.24}$ $\times$ 10$^{-8}$ & 15.20$^{+4.39}_{-1.32}$ & Optically thick emission. SHAPE+shapemol model (see below).          \\  [1ex]          
   $\tau$$^{4}$ Ser       &   M         & 2.65$^{+0.73}_{-0.68}$ $\times$ 10$^{-8}$ & 11.70$^{+1.10}_{-1.02}$ & Optically thin emission. SHAPE+shapemol model (see below).           \\  [1ex]           
             ST Her       &   M         & 1.09$^{+0.70}_{-0.22}$ $\times$ 10$^{-7}$ & 19.50$^{+5.84}_{-1.83}$ & Optically thick emission. Similar to the case of RX Boo (see below). \\  [1ex]
              X Her       &   M         & 3.03$^{+0.72}_{-0.64}$ $\times$ 10$^{-8}$ & 10.70$^{+1.09}_{-0.97}$ & Optically thin emission. Similar to the case of V370 And (see below).\\  [1ex]           
              g Her       &   M         & 6.91$^{+1.35}_{-1.13}$ $\times$ 10$^{-9}$ &  6.85$^{+0.43}_{-0.39}$ & Optically thin emission. Overly low mass-loss rate.                     \\  [1ex]           
              S Dra       &   M         & 1.26$^{+1.23}_{-0.71}$ $\times$ 10$^{-7}$ & 20.80$^{+9.90}_{-5.73}$ & Optically thick emission. Similar to the case of RX Boo (see below). \\  [1ex]           
             FI Lyr       &   M         & 6.15$^{+1.78}_{-1.44}$ $\times$ 10$^{-8}$ & 14.60$^{+2.01}_{-1.62}$ & Optically thin emission. Similar to the case of BK Vir (see below).  \\  [1ex]           
              S Sct       &   C         & 9.02$^{+8.16}_{-8.82}$ $\times$ 10$^{-9}$ &  7.41$^{+2.74}_{-2.96}$ & Optically thin emission. Overly low mass-loss rate.                     \\  [1ex]           
             AF Cyg       &   M         & 6.29$^{+2.30}_{-1.82}$ $\times$ 10$^{-9}$ &  5.13$^{+0.75}_{-0.60}$ & Optically thin emission. Overly low mass-loss rate.                     \\  [1ex]           
             AQ Sgr       &   C         & 7.28$^{+2.01}_{-1.76}$ $\times$ 10$^{-8}$ & 19.70$^{+2.58}_{-2.26}$ & Optically thin emission. SHAPE+shapemol model (see below).           \\  [1ex]           
             RT Cap       &   C         & 1.60$^{+0.42}_{-0.35}$ $\times$ 10$^{-8}$ &  9.67$^{+1.10}_{-0.91}$ & Optically thin emission. Low mass-loss rate.                         \\  [1ex]           
             UU Dra       &   M         & 1.90$^{+1.22}_{-0.37}$ $\times$ 10$^{-7}$ & 25.50$^{+8.11}_{-2.47}$ & Optically thick emission. Similar to the case of RX Boo (see below). \\  [1ex]
             EU Del       &   M         & 2.64$^{+0.60}_{-0.50}$ $\times$ 10$^{-9}$ &  5.25$^{+0.35}_{-0.32}$ & Optically thin emission. Overly low mass-loss rate.                     \\  [1ex]
              U Del       &   M         & 3.35$^{+0.88}_{-0.66}$ $\times$ 10$^{-8}$ & 11.40$^{+1.25}_{-0.95}$ & Optically thin emission. Low mass-loss rate.                         \\  [1ex]
             IQ Aqr       &   M         & 5.28$^{+1.13}_{-1.37}$ $\times$ 10$^{-9}$ &  6.38$^{+0.42}_{-0.48}$ & Optically thin emission. Overly low mass-loss rate.                     \\  [1ex]
              W Cyg       &   M         & 1.91$^{+0.52}_{-0.42}$ $\times$ 10$^{-8}$ &  8.78$^{+0.97}_{-0.80}$ & Optically thin emission. Low mass-loss rate.                         \\  [1ex]
             RU Cyg       &   M         & 2.79$^{+1.82}_{-0.64}$ $\times$ 10$^{-7}$ & 30.80$^{+10.17}_{-3.57}$& Optically thick emission. Similar to the case of RX Boo (see below). \\  [1ex]
             EP Aqr       &   M         & 5.31$^{+1.15}_{-1.10}$ $\times$ 10$^{-8}$ & 14.00$^{+1.33}_{-1.27}$ & Optically thin emission. Similar to the case of V370 And (see below).\\  [1ex]
             TW Peg       &   M         & 4.20$^{+0.93}_{-0.87}$ $\times$ 10$^{-8}$ & 12.70$^{+1.18}_{-1.11}$ & Optically thin emission. Low mass-loss rate.                         \\  [1ex]
             SV Lyn       &   M         & 2.35$^{+1.60}_{-1.03}$ $\times$ 10$^{-9}$ & 3.01$^{+0.87}_{-0.57}$  & Optically thin emission. Overly low mass-loss rate.                     \\  [1ex]
             SV Peg       &   M         & 1.07$^{+0.92}_{-0.52}$ $\times$ 10$^{-6}$ &64.10$^{+30.33}_{-17.26}$& Optically thick emission. Similar to the case of RX Boo (see below). \\  [1ex]
             AK Hya       &   M         & 7.82$^{+2.53}_{-2.02}$ $\times$ 10$^{-9}$ & 5.85$^{+0.74}_{-0.60}$  & Optically thin emission. Overly low mass-loss rate.                     \\  [1ex]
             RW Boo       &   M         & 4.40$^{+0.72}_{-0.80}$ $\times$ 10$^{-8}$ & 13.90$^{+0.87}_{-0.96}$ & Optically thin emission. Low mass-loss rate.                         \\  [1ex]
             IQ Her       &   M         & 7.73$^{+2.47}_{-2.07}$ $\times$ 10$^{-9}$ & 5.72$^{+0.73}_{-0.62}$  & Optically thin emission. Overly low mass-loss rate.                     \\  [1ex]
             AA Cyg       &   S         & 6.25$^{+2.44}_{-2.06}$ $\times$ 10$^{-8}$ & 17.00$^{+3.41}_{-2.89}$ & Optically thin emission. Low mass-loss rate.                         \\  [1ex]
              X Cnc       &   C         & 2.87$^{+2.63}_{-1.45}$ $\times$ 10$^{-7}$ &40.00$^{+19.71}_{-10.90}$& Optically thick emission. Similar to the case of RX Boo (see below). \\  [1ex]
           V465 Cas       &   M         & 2.11$^{+0.56}_{-0.46}$ $\times$ 10$^{-8}$ &  9.38$^{+0.99}_{-0.81}$ & Optically thin emission. Low mass-loss rate.                         \\  [1ex]
\hline                        
\end{tabular}}                
      \tablefoot{Description of the columns from left to right. (1) GCVS designation \citep{samus09}; (2) Spectral type from the GCVS \citep{samus09}; (3-4) Conclusive mass-loss rate estimate and final photodissociation radius for each source in conformity with a simple test using SHAPE+shapemol (see below); (5) Comments. \\ }
\end{table*}

\begin{figure}
\centering
\includegraphics[angle=0,width=8cm]{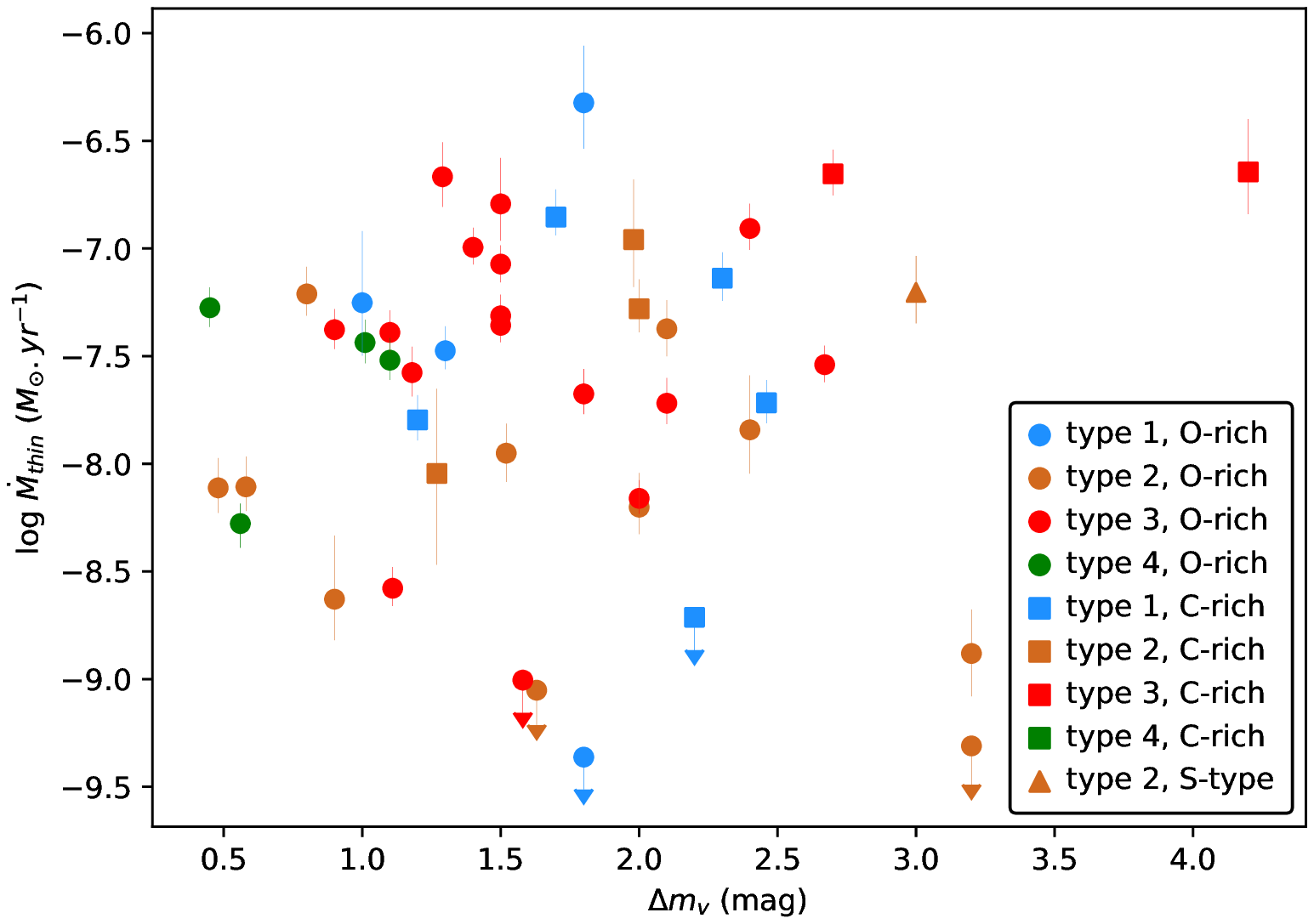}
\caption{Mass-loss rates for the optically thin case vs. $\Delta$m$_{V}$ for the SRs in our sample. Markers are as in Fig. 1. Upper limits are indicated with arrows.}
\label{figmassod_deltamv}%
\end{figure}

\begin{figure*}[!ht]
\centering
\includegraphics[angle=0,width=8cm]{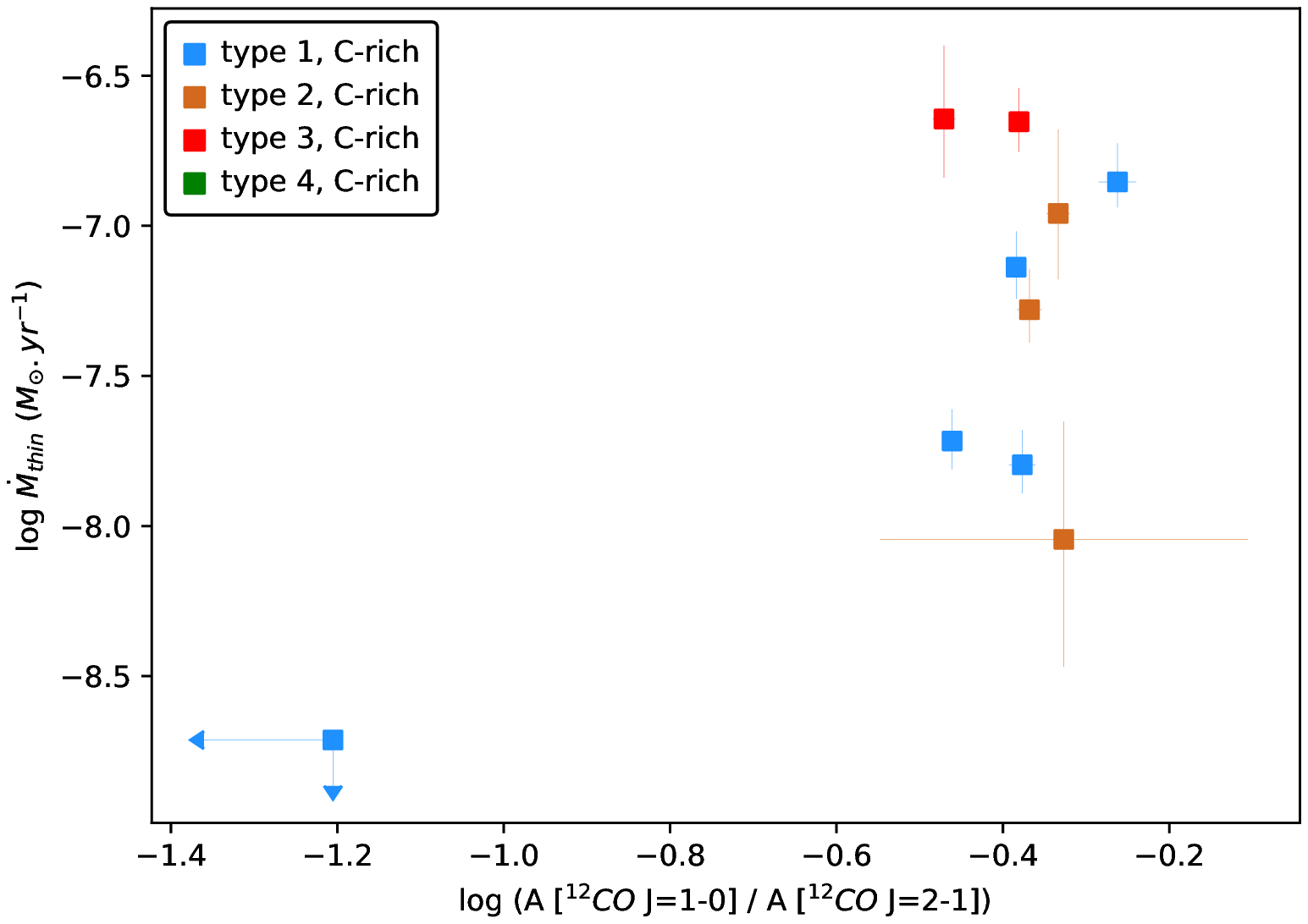}
\hspace{0.4cm}
\includegraphics[angle=0,width=8cm]{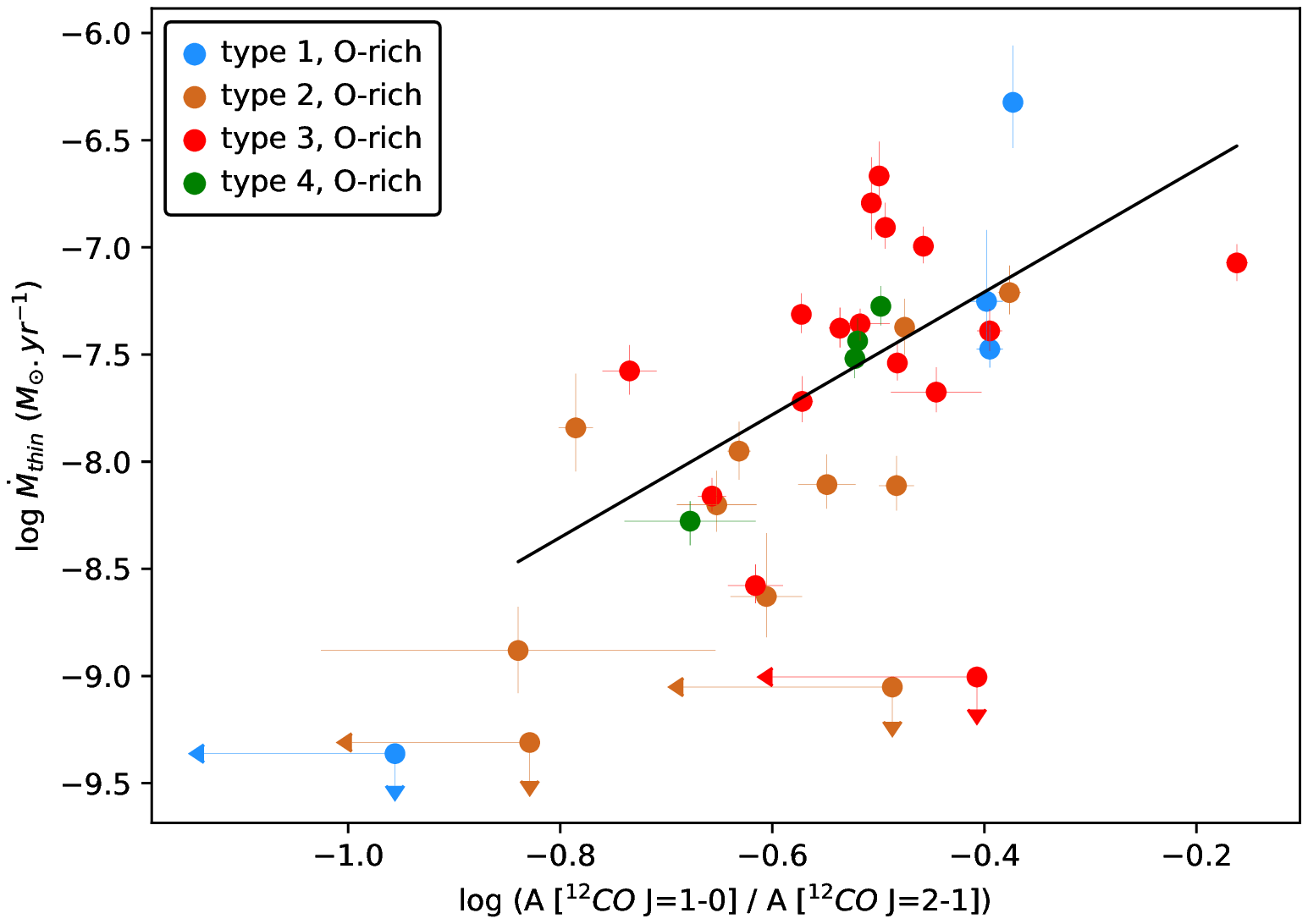}
\caption{Mass-loss rate vs. $^{12}$CO(1-0)-to-$^{12}$CO(2-1) intensity ratio in carbon-rich (left panel) and oxygen-rich (right panel) SRs, respectively. We note the moderate correlation between both parameters for O-rich SRs (r = 0.65 and RMSE = 0.44). Markers are as in Fig. 1. We excluded the only S-type star in the sample (AA Cyg). Upper limits are indicated with arrows.}
\label{figmassod_a10a21}%
\end{figure*}

When comparing mass-loss rates from O- and C-rich SRs, we find no significant difference except that the range is higher for O-rich SRs (from 10$^{-10}$ to 10$^{-6}$ M$_{\odot}$\,yr$^{-1}$; see Table A.4). Therefore, it seems obvious that mass-loss rates of SRs appear independent of chemistry. By making the same comparison for profile types, we do not find any relation either. Curiously, type 4 profiles seem to have rates in the range of 10$^{-8}$-10$^{-7}$ M$_{\odot}$\,yr$^{-1}$. Finally, the same comparison is made for the variation amplitude (see Tables 1, A.1, and A.4). In Figure 5, we present mass-loss rates for the optically thin case as a function of the variation amplitude for the complete sample. Similarly to the results of \citet{kers98}, we do not find any relation between mass-loss rates and pulsation activity, but interestingly, sources that show a type 4 profile have very low variation amplitude, and they are also O-rich SRs. \citet{alcolea90} argued that stars presenting high variation amplitude, and therefore a strong pulsation activity, should have a well-developed inner envelope where molecules such as SiO are expected to be found. In the case of sources presenting double-component winds, it is clear that there is no strong activity and the gas final expansion velocities are not particularly high (see above). We believe that these SRs do not show clearly differentiated inner and outer envelopes, but the mechanisms responsible for such a behavior are not yet well understood. Axisymmetric structures are expected to be more prominent in the inner regions of the envelope. New observations of the inner regions, for example using refractory species such as SiO, are needed in order to elucidate evidence of such a mystery. According to the photodissociation radius estimates, we find that the values range from 2 $\times$ 10$^{15}$ to 64 $\times$ 10$^{15}$ cm, being lower for the optically thin case (see Table A.4). Again, we do not find any relation between CO envelope size and profile type, nor between envelope size and spectral type (see Tables A.1 and A.4). The wind density estimates only provide information about the variation of mass-loss rate with respect to the gas expansion velocity on the surface (see Table A.4). As stated in \citet{loup93}, the gas final expansion velocity globally increases when $\dot{M}$ increases.

We believe that type 4 profiles, in addition to type 3 profiles as mentioned above, are related to binarity. It is known that low-mass AGB stars (M $<$ 1.5 M$_{\odot}$) are O-rich and that intermediate-mass AGB stars (1.5 M$_{\odot}$ $<$ M $<$ 4 M$_{\odot}$) are C-rich, and therefore the mass-loss rates of O-rich SRs (see Table A.4) may only be possible if a companion accretes material from the primary component. Thus, we believe that these few O-rich SRs showing a type 4 profile may very likely harbor binary systems, and this would support the different structures found in this kind of source (see above). If the jets are launched by stars, binarity may suggest two different scenarios in EP Aqr: (i) the jets may be launched by one mass-losing star, while the companion is simply focussing the wind and producing spiral patterns associated with its wake; or (ii) the jets may be launched by the companion and interact with the slow wind from the mass-losing star (see \citealt{tuan19} for more details). 

Interestingly, for O-rich SRs, we report a moderate correlation between mass-loss rates and $^{12}$CO line intensity ratio ($^{12}$CO \textit{J}=1-0/\textit{J}=2-1; see Fig. 6). Producing a fit using a linear regression model in PYTHON, we obtain the following relation: log $\dot{M}$$_{thin}$ = 2.86 $\times$ log ($^{12}$CO \textit{J}=1-0/\textit{J}=2-1) - 6.06, with a correlation coefficient of r = 0.65 and a scatter of RMSE = 0.44 for the optically thin case. Moreover, this correlation between mass-loss rate and $^{12}$CO line intensity ratio holds for the optically thick case, for which we obtain the following relation: log $\dot{M}$$_{thick}$ = 2.94 $\times$ log ($^{12}$CO \textit{J}=1-0/\textit{J}=2-1) - 5.68, with r = 0.66 and RMSE = 0.44. We can then determine a good estimate of mass-loss rate from these correlations, since the scatter provides uncertainties by a mean factor of $\sim$2.75 M$_{\odot}$\,yr$^{-1}$. These results are in good agreement with the correlation between $\dot{M}$ derived from CO observations and dust IR properties reported by \citet{loup93}, which was found for a sample of SRs and Miras. The IRAS$_{25}$ to IRAS$_{12}$ flux ratio ($\it{S}$$_{25}$/$\it{S}$$_{12}$) also seems to be an increasing function of the mass-loss rate and provides an approximate estimate of the dust mass-loss rate. Concerning C-rich SRs, the correlation between $\dot{M}$ and the $^{12}$CO line intensity ratio (see Fig. 6), as well as dust IR properties, is less clear, as found by \citet{loup93} and contrary to what was found by \citet[see above]{olof93}. The observed behavior of $\dot{M}$ may be more likely due to the effects of saturation of \tmb(1-0) for large mass-loss rates and low kinetic temperature. There could also be possible variations of the gas-to-dust ratio or changes in $\dot{M}$. We find that the $^{12}$CO \textit{J}=1-0/\textit{J}=2-1 ratio is, at the same time, moderately correlated with the wind density (r = 0.63) and the photodissociation radius (r = 0.64) for O-rich SRs. However, this is implicitly determined by the equations, since the photodissociation radius and the wind density both depend on the mass-loss rate (see above). In summary, the mechanisms behind the mass loss are still not well understood for our complete and unbiased sample of SRs, and therefore the question of whether these mechanisms are different from or similar to those of Miras is open for evaluation. Further CO observations are needed to obtain more clarity and comprehension, since the comparison for Miras from the literature is prevented by the lack of a similar unbiased sample.

\subsection{Uncertainties in the derived values of $\dot{M}$}

Mass-loss rates for the complete sample have been estimated using an equation similar to $\dot{M}$ = A{\mbox{$T^{\ast}_{\rm mb}$}}$\vexp$$^{2}$\:d$^{2}$\:{F$^{-1}$ (\rco)}, which depends on the measured input quantities {\mbox{$T^{\ast}_{\rm mb}$}}, $\vexp$, and d (assuming that A is a constant value; see above). By far the largest uncertainties are those on distance measurements taken from \citet{gaia18}. For {\mbox{$T^{\ast}_{\rm mb}$}}, we have taken the 1$\sigma$ noise values (see Table A.2), and for $\vexp$, uncertainties of $\pm$1 km\,s$^{-1}$ for the complete sample. In this way, we determined the variations in the result due to each input quantity, $\Delta$$\dot{M}$$_{d+}$ = $\dot{M}$(d + $\Delta$d) - $\dot{M}$(d) and $\Delta$$\dot{M}$$_{d-}$ = $\dot{M}$(d - $\Delta$d) - $\dot{M}$(d), by obtaining $\dot{M}$ in each case, and then adding the variations in quadrature, $\Delta$$\dot{M}$$_{+}$ = $\sqrt{\left({\Delta{\dot{M}_{d+}}}\right)^{2} + \left({\Delta{\dot{M}_{T^{\ast}_{mb}+}}}\right)^{2} + \left({\Delta{\dot{M}_{V_{exp}+}}}\right)^{2}}$ and $\Delta$$\dot{M}$$_{-}$ = $\sqrt{\left({\Delta{\dot{M}_{d-}}}\right)^{2} + \left({\Delta{\dot{M}_{T^{\ast}_{mb}-}}}\right)^{2} + \left({\Delta{\dot{M}_{V_{exp}-}}}\right)^{2}}$, for the upper and lower uncertainties, respectively. Once obtained, the same was done for the photodissociation radii and wind densities (see Table A.4). We note that the upper uncertainties are much higher in all cases.

\section{Modeling with SHAPE}

Although the mass-loss estimates have been computed assuming standard spherically symmetric CSEs, it can be deduced from the $^{12}$CO profiles that this may not be the case for most sources in the sample. In order to identify the kinds of structures that produce these types of $^{12}$CO profiles, we made use of SHAPE+shapemol \citep{steffen11,santander15}. This software is able to produce high-resolution synthetic spectral profiles which can be directly compared with observational data. While SHAPE models complex gaseous nebulae (e.g., PNe, supernova remnants, light echoes, high-energy phenomena, etc.), shapemol produces synthetic spectral profiles of $^{12}$CO. Our purpose is to compare our $^{12}$CO data with these synthetic profiles in order to shed light on the possible morphologies of a representative sample of SRs, and to make possible connections between them. To do this, we selected one SR of each profile type that we consider to be representative of that group (see Sect. 3.1 and Fig. 3) and then fit the best model to the observations. It is considered as a general simple model that only takes into account the $^{12}$CO \textit{J}=2-1 line and is able to produce most profiles in our sample. A more specific model would include all the possible available transitions. Therefore, we used the same abundance of $^{12}$CO ($\textit{X}$ = 3 $\times$ 10$^{-4}$) and the same logarithmic velocity gradient ($\epsilon$ = 1) for all the models (see \citealt{santander15} for more details). 

For type 1 profiles, we selected AQ Sgr. This star shows a broad and apparently symmetric profile for the $^{12}$CO \textit{J}=2-1 line (see Fig. A.2), and almost symmetric in the $^{12}$CO \textit{J}=1-0 line (see Fig. A.4), with an expansion velocity of $\sim$11.83 km\,s$^{-1}$. The best-fit model is defined by an oblate spheroid whose equatorial plane is inclined by $\sim$25$^{\circ}$ with respect to the plane of the sky (in a main Cartesian coordinate system\footnote{We note that in this system the XY plane is the plane of the sky and the Z axis is the line of sight from the observer.} with its origin in the center of the envelope), and equatorial and polar radii, r$_{1}$ and r$_{2}$, equal to \rco~(photodissociation radius taken from Table 2) and 0.2 $\times$ \rco,~respectively (see Appendix B, Fig. B.1). The parameters are indicated in Tables B.1 and B.2. We give the profile type, number of components, source distance in kiloparsecs (\citealt{gaia18}), and the radius of the envelope or photodissociation radius in centimeters (we note that this radius for an optically thin envelope was chosen after making a simple test using SHAPE+shapemol; see later). Moreover, in Table B.2 we give the physical conditions needed to find the best-model fitting according to a spherical coordinate system: gas density, gas temperature, and gas velocity laws, as well as the micro-turbulence velocity. The adopted density and temperature laws have already been used for standard spherically symmetric envelopes, which depend on the distance to the center of the envelope, r. The velocity law however depends on the azimuthal angle, $\varphi$, which indicates that the envelope expands equatorially outward, and not spherically outward, fulfilling $\mid$V$\mid$ $\propto$ $\mid$r$_{m}$$\mid$, with $\mid$r$_{m}$$\mid$ being the modulus of a maximum radius, r$_{m}$, to any point on the surface, defined as:

\begin{equation}
|r_{m}| = \sqrt{r^2_2 + (r^2_1-r^2_2)\sin^{2}\left(\varphi/57.3\right)}.
\end{equation}

In Figure B.2, we see that the resulting synthetic spectra (in red) and the $^{12}$CO \textit{J}=2-1 profile (in black) are very similar.

For type 2 profiles, we selected the O-rich SR BK Vir. This source presents a very narrow and symmetric profile in both transitions, with an expansion velocity of $\sim$6.27 km\,s$^{-1}$. We find that the best-fit model is defined by the same oblate spheroid (see Fig. B.3), whose equatorial plane is inclined by $\sim$25$^{\circ}$ with respect to the plane of the sky. The parameters are indicated in Tables B.1 and B.2. In Figure B.4, we see that the resulting synthetic spectra (in red) and the $^{12}$CO \textit{J}=2-1 profile (in black) are similar. Again, an equatorial expansion seems to predominate in sources that exhibit type 2 profiles. Moreover, it should be noted that the same model, with lower inclination than and similar expansion velocity to that of AQ Sgr, may also be possible for BK Vir.

In accordance with previously studied, standard, spherically symmetric envelopes (see, e.g., \citealt{olof82}), we find some similarity between CO profiles obtained for these sources (e.g., IRC\,+10216) and our type 1 profiles, and therefore we believe that this kind of profile can also be defined by a spherically symmetric envelope. This envelope would expand radially outward from the center. Curiously, we find that type 2 profiles, the narrowest, can also be defined by a sphere with lower expansion velocity. In this sphere, the radius is the photodissociation radius for each source (taken from Table 2); subsequently, r$_{1}$ = r$_{2}$ = \rco,~and the velocity field does not depend on the azimuthal angle. The parameters of the best-fit models are those given in Table C.1 of  Appendix C. In Figs. C.1 and C.2, we see the similarity between the profiles obtained for the oblate spheroid (in red), the sphere (in blue), and the $^{12}$CO \textit{J}=2-1 profile (in black) for AQ Sgr and BK Vir, respectively.

Type 3 profiles however are more difficult to characterize because they show significant asymmetries and bear little relation to the previous profiles. We selected the particular $^{12}$CO \textit{J}=2-1 profile of $\tau$$^{4}$ Ser, which displays two different horns (one of them blueshifted and another redshifted with respect to the systemic velocity; see Fig. A.1). The $^{12}$CO \textit{J}=1-0 profile on the contrary seems to have a lower S/N and is not clearly defined. We find that the best-fit model is defined by an edge-on oblate spheroid with the same dimensions and physical laws as the previous one (see Fig. B.7). This model is, at least for now, the only possible way of getting the profiles of the two different horns. Its parameters are indicated in Tables B.1 and B.2, and the best-fit spectrum is displayed in Figure B.8. Once again, the expansion of the molecular envelope seems to be predominantly equatorial.

When looking for the best-fit model for sources which display two different displaced horns, we find that some of them display the same horns located closer to the systemic velocity, but that these more difficult to visualize because they are unified as a single component (also classified as type 3 profiles; see Figs. A.1 and A.2). This is the case for RX Boo (see Fig. A.1), which seems to show only one component, but in reality is almost the same case as seen in $\tau$$^{4}$ Ser. Therefore, we define these cases as being intermediate between type 1 or 2 profiles and type 3 profiles exhibiting two clear horns. The best-fit model is defined by the same oblate spheroid with equatorial and polar radii, r$_{1}$ and r$_{2}$, equal to \rco~(photodissociation radius taken from Table 2) and 0.1 $\times$ \rco,~respectively (see Fig. B.5), and inclined by $\sim$50$^{\circ}$ with respect to the plane of the sky, whose parameters are indicated in Tables B.1 and B.2. The difference with respect to the other cases is the thickness, which is half the size, and the inclination, but the expansion remains equatorial.

Finally, for type 4 profiles we selected the source V370 And. This source shows a double-component $^{12}$CO \textit{J}=2-1 and $^{12}$CO \textit{J}=1-0 profile, for which we define two different components (see Fig. B.9): an oblate spheroid (similar to that of the first three cases), whose equatorial plane is inclined by $\sim$25$^{\circ}$ with respect to the plane of the sky, radii r$_{1}$ = 3 arcsec (which is not the photodissociation radius) and r$_{2}$ = 0.2 $\times$ r$_{1}$, attached to a biconical structure expanding with constant velocity. The parameters for both structures are given in Tables B.1 and B.2. The photodissociation radius, which is not easy to determine graphically, is the longitude of the cones (taken from Table 2). Moreover, the radius of the base of the cones is 0.7 $\times$ \rco~to simplify the model. Although we know that this source has already been mapped in CO (Castro-Carrizo, priv. comm.), this is a first general model to characterize the molecular envelope of V370 And. In Figure B.10, we see the strong similarity between the obtained synthetic spectra and the profile of the $^{12}$CO~\textit{J}=2-1~line. We should not discard the presence of an outflow and a disk, or a spiral morphology, as was found in X Her and EP Aqr, respectively (\citealt{winters07,castro10,homan18,nhung19}; see above). Moreover, it is worth noting that this model is very consistent with what was found in EP Aqr (see, e.g., \citealt{tuan19}). Furthermore, since type 4 profiles may be an addition of a type 2 profile to a type 1 or type 3 profile (see above), we find that the narrow component of V370 And can also be defined by a sphere (see Fig. C.3) whose paramenters are also given in Table C.1. In Figure C.3, we see the similarity between the synthetic profiles obtained for the biconical structure attached to the oblate spheroid (in red), the biconical structure attached to the sphere (in blue), and the observed $^{12}$CO \textit{J}=2-1 profile (in black).

To conclude, in Appendix D (Figs. D.1, D.2, and D.3) we show the velocity field behavior in AQ Sgr, $\tau$$^{4}$ Ser, and V370 And for a better understanding of the problem. Although the oblate spheroid has different inclinations in AQ Sgr and $\tau$$^{4}$ Ser, the expansion indicated by the arrows is predominantly equatorial, increasing from the poles to the equator. Moreover, we see clear symmetry about the equatorial plane and axial symmetry about the polar axis in both cases. In the case of V370 And, the expansion is a little different; the oblate spheroid expands equatorially while the biconical structure expands away from the equatorial plane of the spheroid at constant velocity, as jets. Again, a clear symmetry is seen about the equatorial plane of the spheroid and axial symmetry about the main axis of the cones (see Fig. D.3). We therefore conclude that these dust shells are not homogeneous or isotropic, and the expansion is predominantly equatorial, but this should be checked with the appropriate mapping (i.e., with NOEMA or ALMA). We believe that most SRs are axisymmetric with nonisotropic kinematics.

\subsection{Optically thin or optically thick emission}

The main problem when characterizing the circumstellar envelopes of these sources is that, a priori, we do not know if the $^{12}$CO emission is optically thin or optically thick. In order to take this crucial issue into consideration, we do a simple test using SHAPE+shapemol that consists of increasing the abundance of $^{12}$CO in each model and identifying the changes produced in the intensity of the $^{12}$CO \textit{J}=2-1 line. Curiously, only a 50$\%$ increase in the abundance (i.e., $\textit{X}$ = 4.5 $\times$ 10$^{-4}$), using the photodissociation radius for the optically thin case (see Table A.4), makes the intensity grow proportionally for the cases of AQ Sgr, BK Vir, $\tau$$^{4}$ Ser, and V370 And. For the case of RX Boo however, \tmb~for both lines saturates using the photodissociation radius for the optically thin case and the intensity remains unchanged. Thus, in this case it is clear that the emission may very likely be optically thick and the photodissociation radius would be that for the optically thick case. Moreover, the low or overly low mass-loss rates for the other cases favor this assumption (see Table 2). In cases for which the mass-loss rate estimates are lower than 10$^{-8}$ M$_{\odot}$\,yr$^{-1}$, the emission is probably optically thin. We note that these mass-loss rate estimates are calculated for reliable distances (\citealt{gaia18}) within the uncertainties given here.

In the recent study by \citet{saberi19} based on an updated treatment of the CO photodissociation, the authors conclude that the abundance of CO close to the star and the outflow density have a significant effect on the size of the molecular envelope. We therefore believe our test is a suitable way of defining the best photodissociation radius estimates for the molecular envelopes in our sample, in which an optically thin emission seems to predominate. Nonetheless, more theoretical and observational studies are needed for a better understanding of these questions.

\subsection{Model uncertainties}

As we consider a general model that involves only the $^{12}$CO \textit{J}=2-1 line, the sources of uncertainty can occur at any point in our study. For SHAPE+shapemol, these may be computed by varying a given parameter while leaving the others unchanged. However, we also have no real information about the geometry of these envelopes, making it more difficult for us to give real uncertainties. We prefer not to give any values and simply to provide a general description of the problem.

\section{Conclusions}

These observations of an unbiased sample of SRs, with high sensitivity and S/N ratio, provide a better understanding of the mass-loss mechanisms that take place during the AGB phase. Our analysis sheds light on the formation of axially symmetric PNe from spherical CSEs. The main results of this work can be summarized as follows.

   \begin{enumerate}
      \item We present high-S/N IRAM 30m observations of the $^{12}$CO \textit{J}=2-1, $^{12}$CO \textit{J}=1-0, and $^{13}$CO \textit{J}=1-0 lines, in a volume-limited unbiased sample of well-characterized SRs. The sample includes 38 O-rich, 10 C-rich, and 1 S-type SRs. We have characterized the main properties of the CSEs, and have analyzed the different profile types and modeled the possible structures that may produce them, making use of the $^{12}$CO \textit{J}=2-1 line. Moreover, we used the $^{12}$CO \textit{J}=1-0 line to estimate the mass-loss rates for all the sources in the sample. This work significantly improves on the pioneering studies by \citet{olof93}, \citet{kers96}, and \citet{kers98,kers99}.
      \vspace{0.1cm}    
      \item We classified sources into four groups according to the $^{12}$CO line profiles and final gas-expansion velocities. Type 1 profiles, which seem to be more standard, are broad and symmetric lines with $\vexp$ between 9 and 13 km\,s$^{-1}$. Type 2 profiles are unusually narrow and symmetric lines with $\vexp$ between 3 and 9 km\,s$^{-1}$. Type 3 profiles however are very peculiar and present very pronounced asymmetries which are not compatible with the standard optically thin lines that have long been studied. Finally, type 4 profiles are those presenting two different components: one narrow component with low $\vexp$ ($\sim$2-3 km\,s$^{-1}$) and another broad component with higher $\vexp$ ($\sim$9-12 km\,s$^{-1}$; resembling the addition of a type 2 profile to a type 1 or type 3 profile). Interestingly, we report the tentative detection of IQ Aqr as showing a type 4 profile with two different components (or more). On the other hand, according to a recent study by \citet{ortiz16} based on searching for prevailing UV flux excess from companion stars, SRs that seem to show a type 3 profile are more likely to harbor binary systems.
      \vspace{0.1cm}  
      \item It is remarkable that for SRs showing a type 4 profile, the variation amplitude is very low, meaning that these peculiar sources do not present a strong pulsation activity. We believe that there is a strong connection between their spectral type (O-rich), the variation amplitude, and the gas final expansion velocity. Moreover, we expect that these SRs do not show clearly differentiated inner and outer envelopes, but the mechanisms responsible are not yet known. The estimated mass-loss rates suggest the possible binarity of sources showing a type 4 profile, since a companion star may be accreting material from the primary one. More observations of the inner regions, for example of refractory species such as SiO, are needed in order to elucidate the possible causes of this particular profile type.
      \vspace{0.1cm}
      \item We estimated mass-loss rates, photodissociation radii, and surface wind densities for all the sources in our sample. Curiously, we find that for each star the optically thick estimate of mass-loss rate is approximately twice as high as the optically thin estimate. From these observations, we cannot conclude whether the circumstellar envelopes surrounding the central stars of the complete sample are optically thin or optically thick. Interestingly, we report a moderate correlation between mass-loss rate and $^{12}$CO \textit{J}=1-0/\textit{J}=2-1 line intensity ratio for O-rich SRs. No correlation is observed for C-rich SRs, contrary to what was found by \citet{olof93} in a sample of bright C-rich stars. Moreover, these results are in good agreement with the correlation between $\dot{M}$ derived from CO observations of a sample containing SRs and Miras, and dust infrared properties reported by \citet{loup93}. We cannot discern any new information about the mechanisms behind the mass-loss process, or find any differences with respect to Miras. We therefore propose to obtain more CO observations of a similar and unbiased sample of Miras in order to look for possible connections with SRs. 
      \vspace{0.1cm}
      \item Using SHAPE+shapemol, we find a general simple model based on an oblate spheroid placed in different orientations that may explain most $^{12}$CO profiles in our sample. An oblate spheroid whose equatorial plane is inclined by $\sim$25$^{\circ}$ with respect to the plane of the sky may explain type 1 and 2 profiles, whereas an edge-on oblate spheroid may explain type 3 profiles. We find an intermediate case between type 1 or 2 profiles and type 3 profiles, that may be defined by the same oblate spheroid inclined by $\sim$50$^{\circ}$ with respect to the plane of the sky. Finally, type 4 profiles may be defined by two different components: the same oblate spheroid and a biconical structure attached to it. We do not discard the presence of outflows, disks, or spiral morphologies on these structures, or a possible binarity. Curiously, type 1 and 2 profiles may also be explained by spherically symmetric envelopes, albeit requiring anomalously low velocities. Type 3 and 4 profiles can however only be explained with axial symmetry about an axis or a plane. Moreover, by doing a simple test using SHAPE+shapemol, we find that most SRs in the sample show optically thin emission in $^{12}$CO, which is also supported by the low or overly low mass-loss rates obtained for the complete sample. We conclude that circumstellar shells generally present axial and strongly nonspherical symmetry, and that the expansion is predominantly equatorial. This should be verified however by obtaining interferometric maps with NOEMA or ALMA. In summary, this study suggests that SRs may be the missing link between AGB stars and axial PNe.
   \end{enumerate}

\begin{acknowledgements}
The authors thank Dr. Kerschbaum, the referee, for his constructive comments. This  work  has  been  supported  by  the  Spanish  Ministry of Science, Innovation, and Universities grant AYA2016$-$78994$-$P. J.J.D.L. also acknowledges support provided by AYA2016$-$78994$-$P grant.
\end{acknowledgements}

\bibliographystyle{aa}
\bibliography{bib.bib}

%

\begin{appendix} 

\onecolumn

\section{Additional Tables and Figures}

We collect here additional material that has been used for the analysis.

\begin{figure*}[hbt!]
   \centering
   \includegraphics[width=14cm]{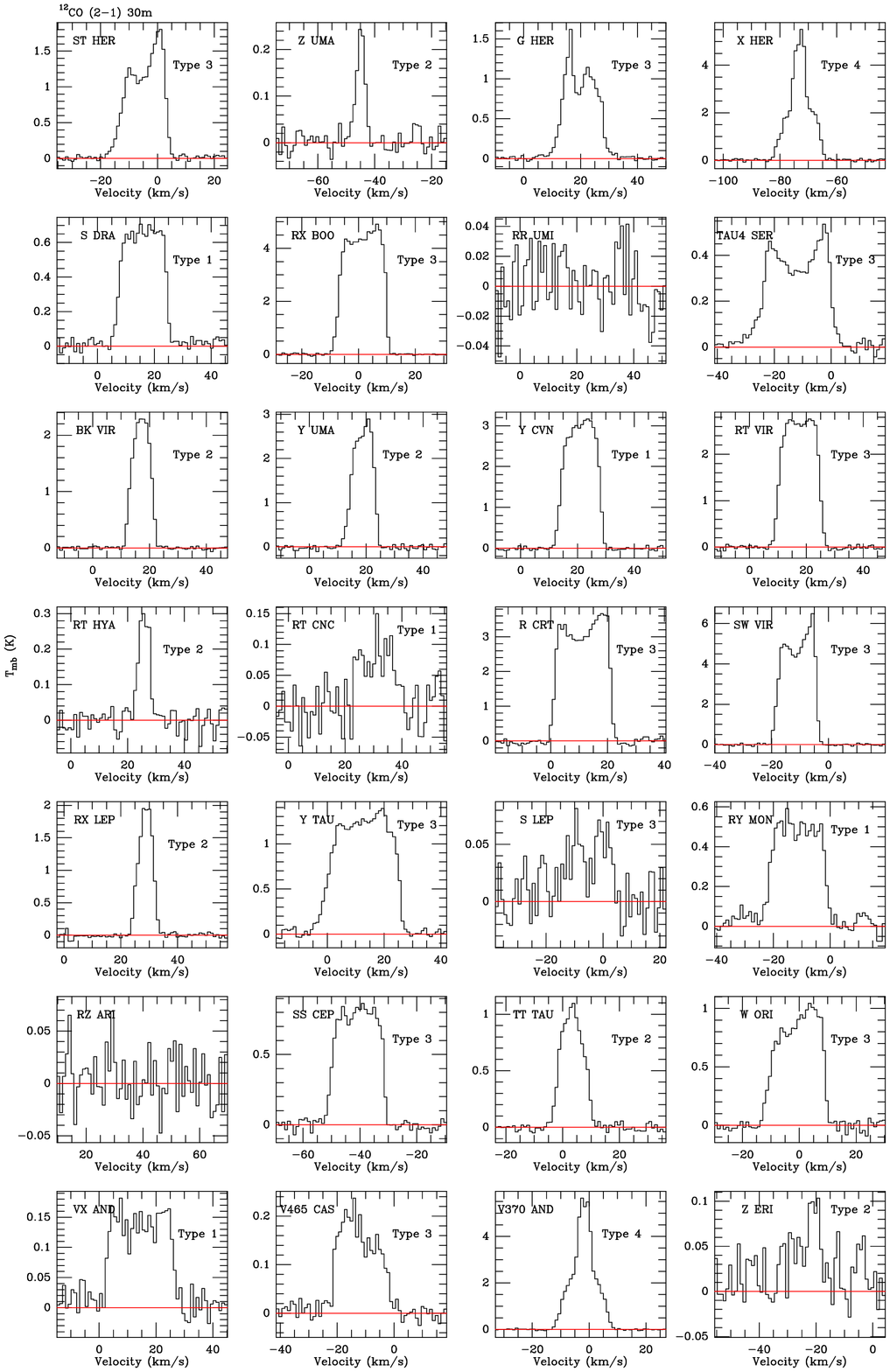}
      \caption{Overview of the 49 SRs of the sample detected with the IRAM 30m telescope in the $^{12}$CO \textit{J}=2-1 line (Part 1). Profile types are indicated.
              }
         \label{Figa1co21_1}
   \end{figure*}
   
\begin{figure*}
   \centering
   \includegraphics[width=14cm]{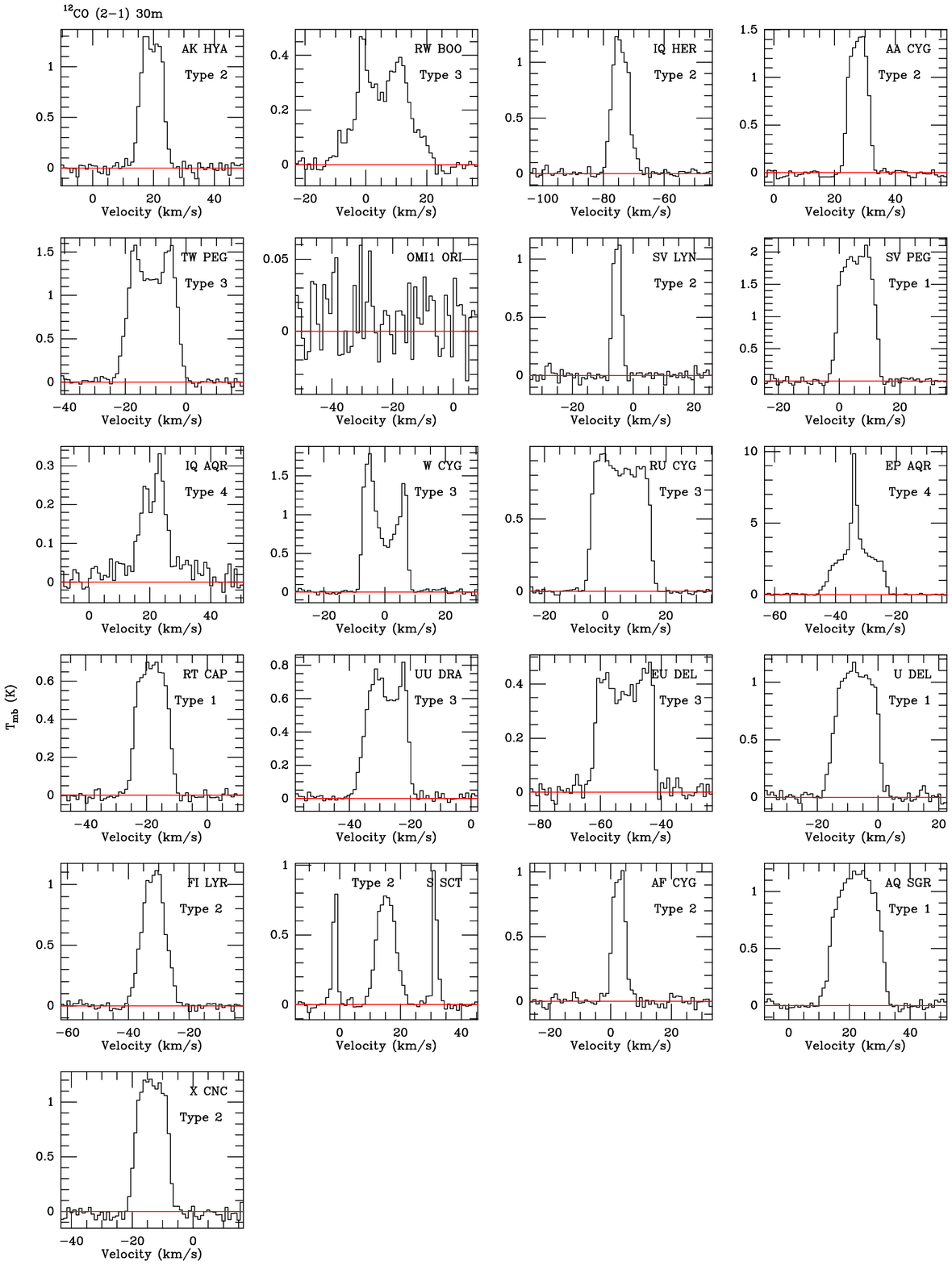}
      \caption{Overview of the 49 SRs of the sample detected with the IRAM 30m telescope in the $^{12}$CO \textit{J}=2-1 line (Part 2). Profile types are indicated.
              }
         \label{Figa2co21_2}
   \end{figure*}
   
\begin{figure*}
   \centering
   \includegraphics[width=14cm]{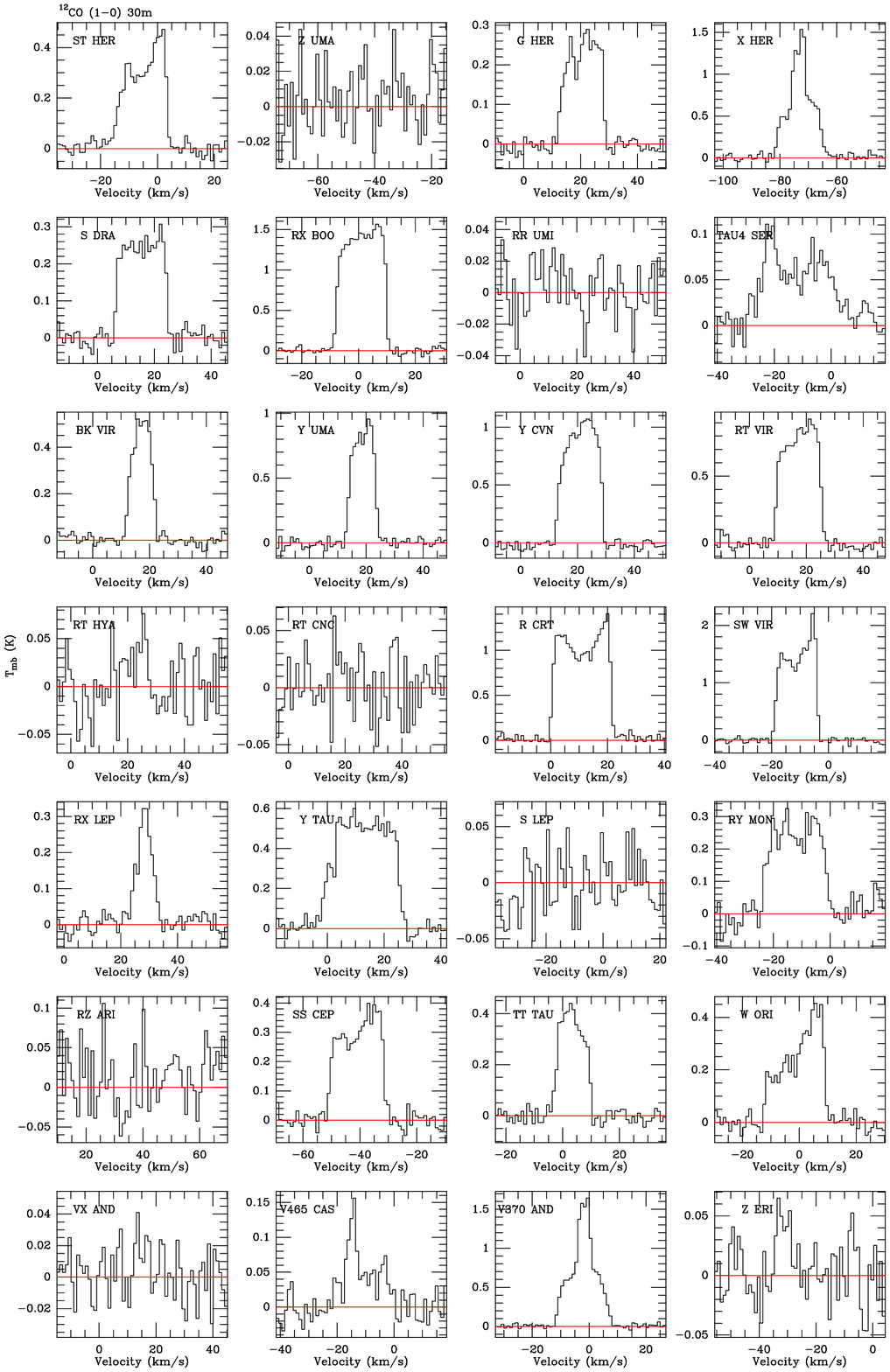}
      \caption{Overview of the 49 SRs of the sample detected with the IRAM 30m telescope in the $^{12}$CO \textit{J}=1-0 line (Part 1).
              }
         \label{Figa3co10_1}
   \end{figure*}

\begin{figure*}
   \centering
   \includegraphics[width=14cm]{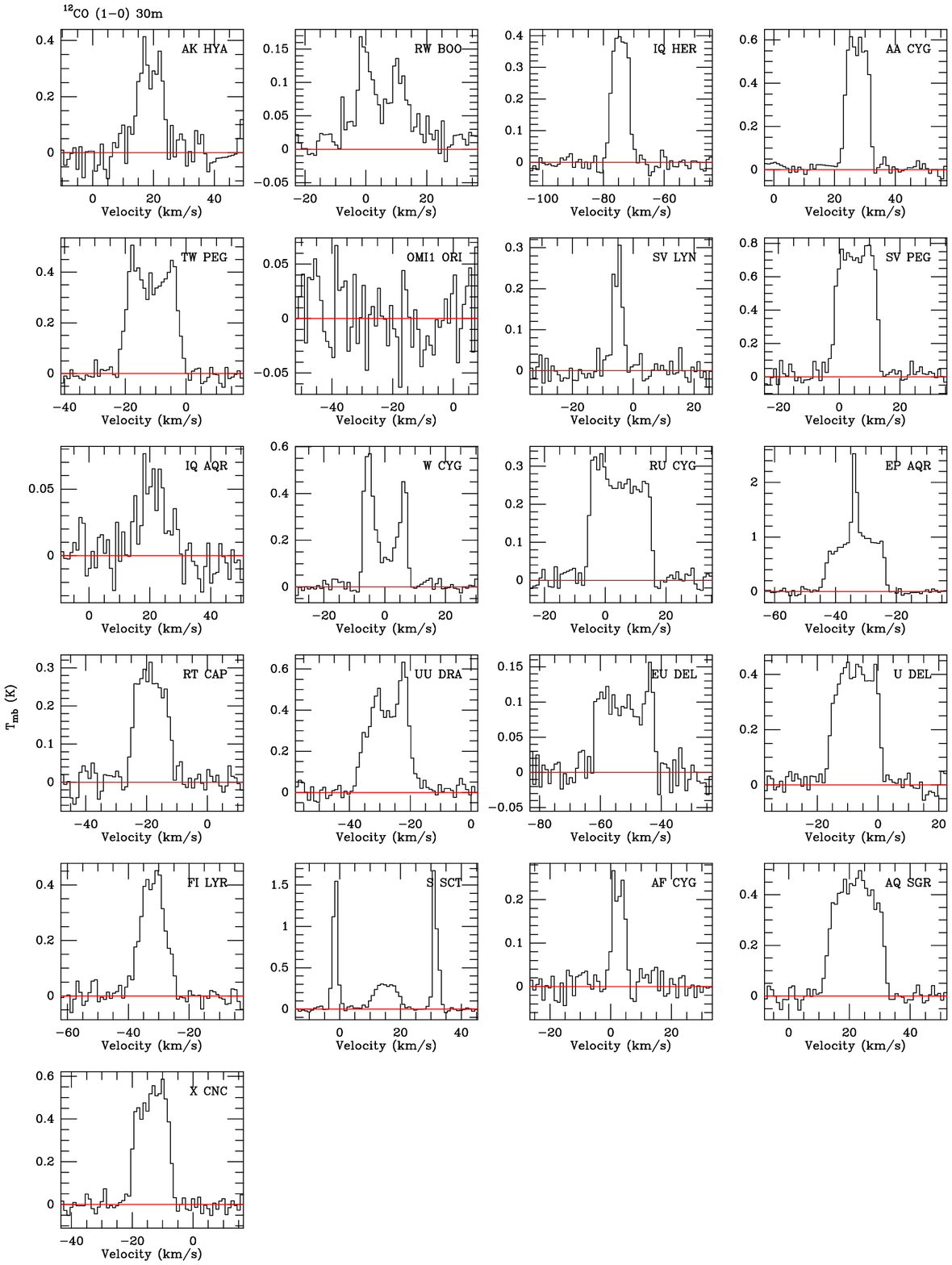}
      \caption{Overview of the 49 SRs of the sample detected with the IRAM 30m telescope in the $^{12}$CO \textit{J}=1-0 line (Part 2).
              }
         \label{Figa4co10_2}
   \end{figure*}

\begin{figure*}
   \centering
   \includegraphics[width=14cm]{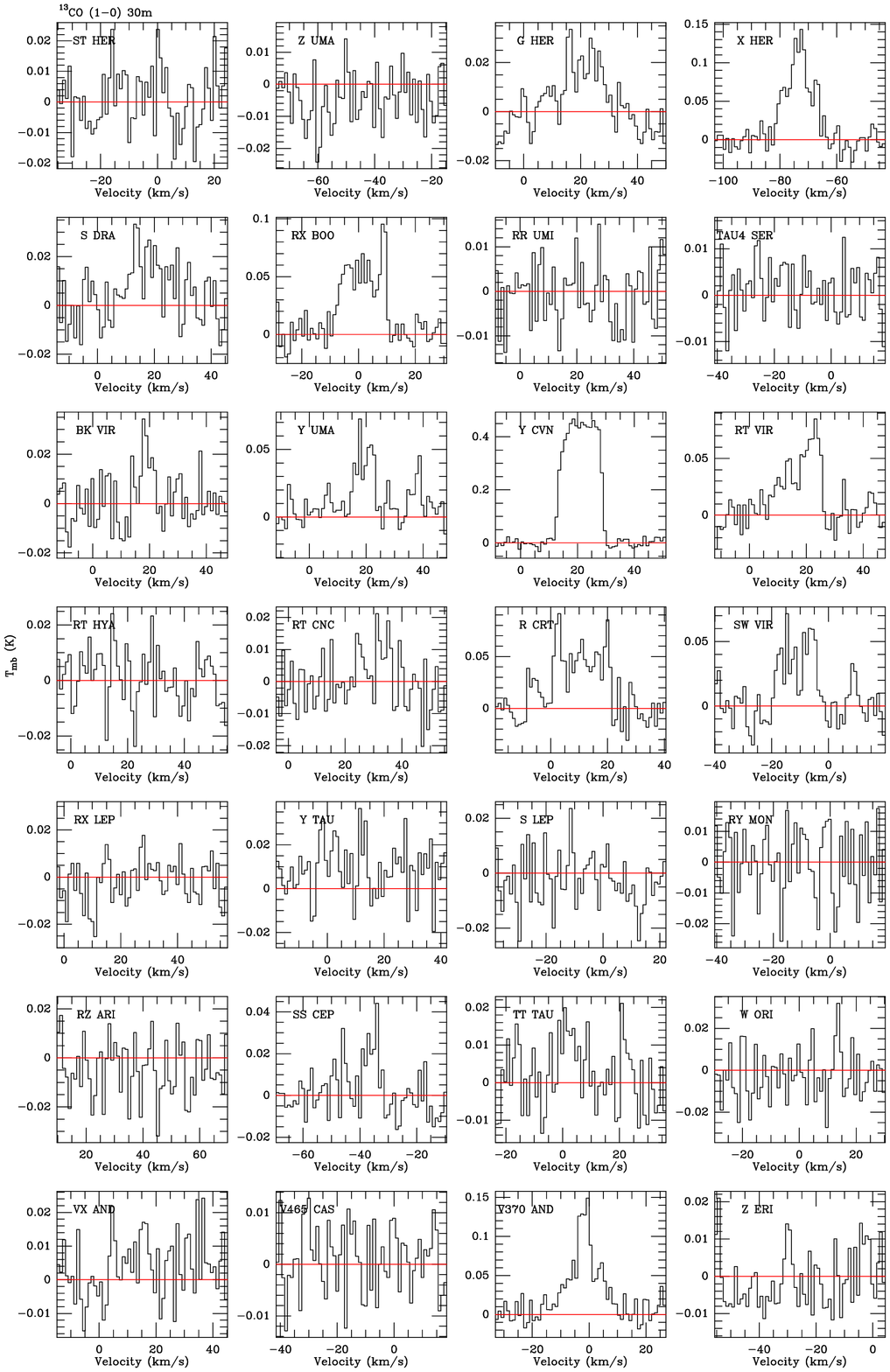}
      \caption{Overview of the 49 SRs of the sample detected with the IRAM 30m telescope in the $^{13}$CO \textit{J}=1-0 line (Part 1).
              }
         \label{Figa5co10_1}
   \end{figure*}
   
\begin{figure*}
   \centering
   \includegraphics[width=14cm]{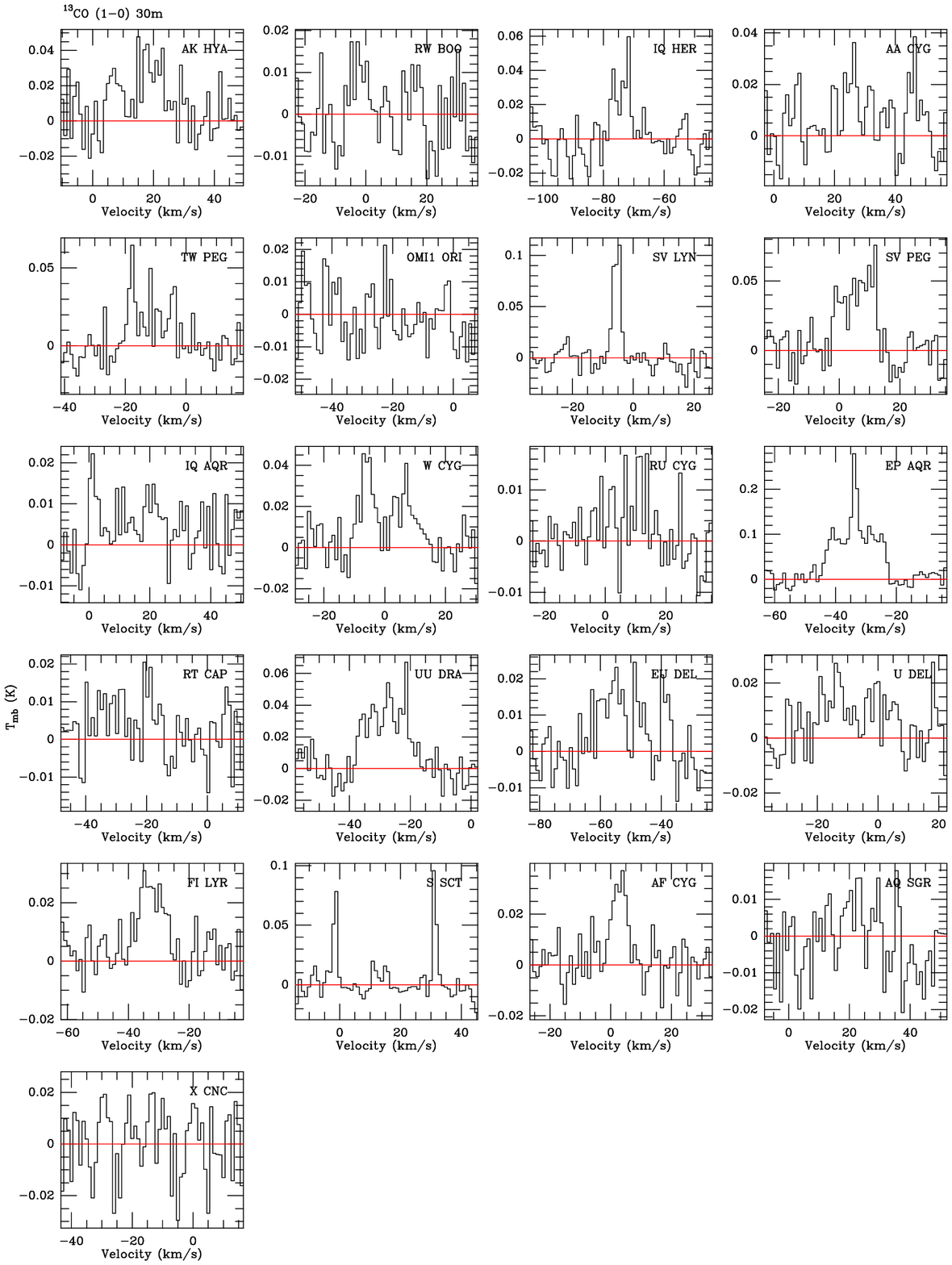}
      \caption{Overview of the 49 SRs of the sample detected with the IRAM 30m telescope in the $^{13}$CO \textit{J}=1-0 line (Part 2).
              }
         \label{Figa6co10_2}
   \end{figure*}   

\begin{table*}
\caption{\label{t1a}$^{12}$CO \textit{J}=2-1 observations in our sample.}
\centering
\begin{tabular}{lcccccccc}
\hline\hline
\\ [-1ex]
Source & 1$\sigma$ noise & A & \tmb & $\vexp$ & V$_{lsr}$ & FWZP & FWHP & Profile type \\
       & (K) & (K\,km\,s$^{-1}$) & (K) & (km\,s$^{-1}$) & (km\,s$^{-1}$) & (km\,s$^{-1}$) & (km\,s$^{-1}$) & \\
\hline
            VX And & 0.016  & 3.38  & 0.18 &  12.89  &  14.62     & 25.78  & 23.13   &   1       \\
          V370 And & 0.021  & 49.66 & 5.65 &  11.21  &  $-$1.98     & 22.42  & 6.18  &   4       \\ 
             Z Eri & 0.024  & 0.78  & 0.10 &  7.77   &  $-$25.48    & 15.54  & 10.95 &   2       \\
            SS Cep & 0.030  & 14.80 & 0.87 &  11.30  &  $-$41.39    & 22.60  & 18.36 &   3       \\
            TT Tau & 0.030  & 10.25 & 1.10 &  8.48   &  3.17      & 16.95  & 10.06   &   2       \\
             W Ori & 0.036      & 17.36 & 1.05 &  12.01  &  $-$1.34     & 24.01  & 18.89 &   3       \\
            RX Lep & 0.024      & 12.08 & 1.95 &  5.65   &  28.61     & 11.30  & 6.18    &   2       \\
             Y Tau & 0.031      & 31.46 & 1.39 &  16.95  &  11.16     & 33.90  & 24.36   &   3       \\
             S Lep & 0.021      & 0.86  & 0.08 &  10.33  &  $-$6.35     & 20.66  & 15.71 &   3       \\
            RY Mon & 0.039      & 10.20 & 0.60 &  12.27  &  $-$11.80    & 24.54  & 19.77 &   1       \\
            RT Hya & 0.029      & 1.43  & 0.30 &  3.80   &  25.76     & 7.59   & 4.77    &   2       \\
            RT Cnc & 0.026      & 1.33  & 0.15 &  8.74   &  31.66     & 17.48  & 13.60   &   1       \\
             R Crt & 0.078      & 65.26 & 3.67 &  11.92  &  11.16     & 23.84  & 19.60   &   3       \\
             Z UMa & 0.018      & 0.91  & 0.24 &  3.62   &  $-$45.45    & 7.24   & 3.36  &   2       \\
            BK Vir & 0.025      & 17.67 & 2.29 &  6.27   &  17.01     & 12.54  & 8.12    &   2       \\
             Y UMa & 0.042      & 23.64 & 2.91 &  7.06   &  17.99     & 14.12  & 8.48    &   2       \\
             Y CVn & 0.038      & 41.47 & 3.17 &  9.80   &  20.53     & 19.60  & 13.77   &   1       \\
            RT Vir & 0.036      & 39.60 & 2.77 &  10.33  &  17.13     & 20.66  & 14.65   &   3       \\
            SW Vir & 0.046      & 72.11 & 6.49 &  9.45   &  $-$11.20    & 18.89  & 13.42 &   3       \\
            RX Boo & 0.033      & 74.60 & 4.91 &  11.12  &  0.69      & 22.25  & 16.60   &   3       \\
  $\tau$$^{4}$ Ser & 0.019      & 10.11 & 0.53 &  17.48  &  $-$13.95    & 34.96  & 23.66 &   3       \\
            ST Her & 0.023      & 22.25 & 1.80 &  12.45  &  $-$6.32     & 24.89  & 14.65 &   3       \\
             X Her & 0.026      & 43.15 & 5.51 &  10.15  &  $-$73.39    & 20.30  & 5.47  &   4       \\
             g Her & 0.018      & 16.43 & 1.62 &  12.80  &  20.49     & 25.60  & 12.36   &   3       \\
             S Dra & 0.030      & 11.22 & 0.70 &  10.77  &  15.57     & 21.54  & 16.77   &   1       \\
            FI Lyr & 0.023      & 10.14 & 1.11 &  8.65   &  $-$31.98    & 17.30  & 8.48  &   2       \\
             S Sct & 0.163      & 6.03  & 0.78 &  7.58   &  15.47     & 15.15  & 7.49    &   2       \\
            AF Cyg & 0.031      & 5.16  & 1.01 &  5.915  &  2.30      & 11.83  & 4.77    &   2       \\
            AQ Sgr & 0.028      & 18.62 & 1.18 &  11.83  &  21.69     & 23.66  & 16.77   &   1       \\
            RT Cap & 0.018      & 8.14  & 0.70 &  9.00   &  $-$18.51  & 18.01  & 12.01   &   1       \\
            UU Dra & 0.017      & 10.51 & 0.82 &  11.56  &  $-$29.25  & 23.13  & 14.48   &   3       \\
            EU Del & 0.019      & 7.99  & 0.48 &  12.01  &  $-$52.98  & 24.01  & 19.42   &   3       \\
             U Del & 0.032      & 16.45 & 1.17 &  11.12  &  $-$7.35   & 22.25  & 15.18   &   1       \\
            IQ Aqr & 0.018      & 3.00  & 0.33 &  12.80  &  21.19     & 25.60  & 8.30    &   4       \\
             W Cyg & 0.022      & 16.69 & 1.78 &  9.45   &  0.03      & 18.89  & 15.01   &   3       \\
            RU Cyg & 0.016      & 17.26 & 0.95 &  12.45  &  5.16      & 24.89  & 19.95   &   3       \\
            EP Aqr & 0.042      & 64.13 & 9.85 &  12.18  &  $-$33.88  & 24.36  & 2.65    &   4       \\
            TW Peg & 0.029      & 24.04 & 1.57 &  12.36  &  $-$11.68  & 24.72  & 16.95   &   3       \\       
            SV Lyn & 0.042      & 4.06  & 1.12 &  2.92   &  $-$5.27   & 5.84   & 3.72    &   2       \\
            SV Peg & 0.045      & 25.01 & 2.11 &  9.07   &  5.29      & 18.14  & 12.81   &   1       \\
            AK Hya & 0.037      & 10.92 & 1.30 &  7.24   &  19.28     & 14.48  & 8.48    &   2       \\
            RW Boo & 0.024      & 7.34  & 0.47 &  17.29  &  $-$4.99   & 34.57  & 17.38   &   3       \\
            IQ Her & 0.020      & 8.06  & 1.23 &  6.70   &  $-$73.98  & 13.40  & 6.35    &   2       \\
            AA Cyg & 0.029      & 10.58 & 1.43 &  6.10   &     27.66  & 12.21  & 7.78    &   2       \\
             X Cnc & 0.033      & 13.54 & 1.21 &  8.20   &  $-$13.81  & 16.40  & 11.46   &   2       \\
          V465 Cas & 0.016      & 3.12  & 0.24 &  10.72  &  $-$11.27  & 21.43  & 17.36   &   3       \\
\hline                                                                                 
\end{tabular}    
      \tablefoot{Description of the columns from left to right. (1) GCVS designation \citep{samus09}; (2-9) 1$\sigma$ noise, integrated line intensity, peak main-beam brightness temperature, expansion velocity, systemic velocity, full width at zero power, full width at half power, and profile type.\\
      }
\end{table*}                                                                           
                                                                                       
\begin{table*}                                                                         
\caption{\label{t2a}$^{12}$CO \textit{J}=1-0 observations in our sample.}              
\centering                                                                             
\begin{tabular}{lccccc}                                                                 
\hline\hline                                                                           
\\ [-1ex]                                                                              
Source & 1$\sigma$ noise & A & \tmb  & FWHP & A[$^{12}$CO \textit{J}=1-0]/A[$^{12}$CO \textit{J}=2-1] \\                                         
       & (K) & (K\,km\,s$^{-1}$) & (K) & (km\,s$^{-1}$) &  \\                             
\hline                                                                                 
             VX And       & 0.003       & $<$0.21   &  $<$0.01   &    -  &  $<$0.062$-$0.022 \\             
           V370 And       & 0.024       & 15.02  &     1.64   &    5.83  &  0.302$\pm$0.003 \\             
              Z Eri       & 0.005       & $<$0.25   &  $<$0.02   &    -  &  $<$0.326$-$0.148 \\             
             SS Cep       & 0.022       & 5.96   &     0.40   &    18.36 &  0.403$\pm$0.011 \\             
             TT Tau       & 0.024       & 4.39   &     0.44   &    11.12 &  0.428$\pm$0.015 \\             
              W Ori       & 0.025       & 5.87   &     0.45   &    14.30 &  0.338$\pm$0.010 \\             
             RX Lep       & 0.018       & 1.98   &     0.32   &    6.53  &  0.164$\pm$0.006 \\             
              Y Tau       & 0.031       & 13.09  &     0.60   &    24.89 &  0.416$\pm$0.008 \\             
              S Lep       & 0.005       & $<$0.34   &     $<$0.02   & -   &  $<$0.392$-$0.174 \\             
             RY Mon       & 0.037       & 5.57   &     0.33   &    20.30 &  0.547$\pm$0.028 \\             
             RT Hya       & 0.028       & 0.21   &     0.08   &    3.53  &  0.145$\pm$0.062 \\             
             RT Cnc       & 0.003   & $<$0.15   &     $<$0.01   & -   &  $<$0.111$-$0.046 \\             
              R Crt       & 0.034       & 22.76  &     1.41   &    20.30 &  0.349$\pm$0.005 \\             
              Z UMa       & 0.006       & $<$0.14   &     $<$0.02   & -   &  $<$0.148$-$0.057 \\             
             BK Vir       & 0.023       & 4.13   &     0.52   &    8.48  &  0.234$\pm$0.006 \\             
              Y UMa       & 0.032       & 7.92   &     0.96   &    9.18  &  0.335$\pm$0.007 \\             
              Y CVn       & 0.033       & 14.35  &     1.08   &    14.83 &  0.346$\pm$0.005 \\             
             RT Vir       & 0.036       & 12.54  &     0.93   &    15.18 &  0.317$\pm$0.005 \\             
             SW Vir       & 0.037       & 22.46  &     2.21   &    13.77 &  0.311$\pm$0.003 \\             
             RX Boo       & 0.032       & 24.58  &     1.56   &    17.30 &  0.329$\pm$0.003 \\             
   $\tau$$^{4}$ Ser       & 0.015       & 1.86   &     0.11   &    23.13 &  0.184$\pm$0.011 \\             
             ST Her       & 0.019       & 5.95   &     0.47   &    15.18 &  0.268$\pm$0.006 \\
              X Her       & 0.024       & 12.97  &     1.54   &    5.47  &  0.300$\pm$0.003 \\             
              g Her       & 0.018       & 3.62   &     0.29   &    14.48 &  0.220$\pm$0.007 \\             
              S Dra       & 0.022       & 4.49   &     0.31   &    17.65 &  0.400$\pm$0.014 \\             
             FI Lyr       & 0.016       & 4.26   &     0.45   &    8.83  &  0.420$\pm$0.010 \\             
              S Sct       & 0.295       & 2.84   &     0.31   &    9.09  &  0.471$\pm$0.240 \\             
             AF Cyg       & 0.022       & 1.15   &     0.27   &    4.77  &  0.223$\pm$0.019 \\             
             AQ Sgr       & 0.028       & 7.69   &     0.49   &    17.65 &  0.413$\pm$0.010 \\             
             RT Cap       & 0.022       & 3.42   &     0.32   &    12.18 &  0.420$\pm$0.015 \\             
             UU Dra       & 0.024       & 7.24   &     0.63   &    12.89 &  0.689$\pm$0.016 \\
             EU Del       & 0.019       & 1.94   &     0.15   &    19.42 &  0.242$\pm$0.014 \\
              U Del       & 0.027       & 6.63   &     0.44   &    16.07 &  0.403$\pm$0.011 \\
             IQ Aqr       & 0.014       & 0.63   &     0.08   &    9.89  &  0.210$\pm$0.030 \\
              W Cyg       & 0.018       & 4.47   &     0.57   &    14.83 &  0.268$\pm$0.006 \\
             RU Cyg       & 0.015       & 5.54   &     0.33   &    20.48 &  0.321$\pm$0.006 \\
             EP Aqr       & 0.035       & 20.39  &     2.53   &    2.82  &  0.318$\pm$0.004 \\
             TW Peg       & 0.025       & 6.99   &     0.50   &    16.95 &  0.291$\pm$0.007 \\
             SV Lyn       & 0.022       & 1.01   &     0.31   &    3.36  &  0.248$\pm$0.019 \\
             SV Peg       & 0.034       & 10.59  &     0.79   &    14.23 &  0.424$\pm$0.009 \\
             AK Hya       & 0.040       & 3.09   &     0.42   &    8.65  &  0.283$\pm$0.018 \\
             RW Boo       & 0.017       & 2.23   &     0.17   &    15.78 &  0.304$\pm$0.019 \\
             IQ Her       & 0.021       & 2.65   &     0.40   &    6.70  &  0.329$\pm$0.012 \\
             AA Cyg       & 0.020       & 4.91   &     0.61   &    8.67  &  0.464$\pm$0.011 \\
              X Cnc       & 0.034       & 6.28   &     0.59   &    12.35 &  0.464$\pm$0.015 \\
           V465 Cas       & 0.018       & 1.12   &     0.15   &    3.54  &  0.359$\pm$0.035 \\
\hline 
\end{tabular}
      \tablefoot{Description of the columns from left to right. (1) GCVS designation \citep{samus09}; (2-6) 1$\sigma$ noise, integrated line intensity, peak main-beam brightness temperature, full width at half power, and $^{12}$CO line intensity ratio. 3$\sigma$ upper detection limits are indicated. \\ }
\end{table*}

\begin{table*}
\caption{\label{t3a}$^{13}$CO \textit{J}=1-0 observations in our sample.}
\centering
\begin{tabular}{lcccc}
\hline\hline
\\ [-1ex]
Source & 1$\sigma$ noise & A & \tmb  & FWHP \\
       & (K) & (K\,km\,s$^{-1}$) & (K) & (km\,s$^{-1}$) \\
\hline
             VX And          & 0.007 & $<$0.54 &   $<$0.02 &   -    \\
           V370 And          & 0.011 & 1.25 &   0.15 &   4.94       \\ 
              Z Eri          & 0.008 & $<$0.37 &   $<$0.02 &   -    \\
             SS Cep          & 0.009 & 0.24 &   0.04 &   12.71      \\
             TT Tau          & 0.008 & 0.13 &   0.02 &   10.77      \\
              W Ori          & 0.013 & $<$0.94 &   $<$0.04 &   -    \\
             RX Lep          & 0.012 & $<$0.41 &   $<$0.04 &   -    \\
              Y Tau          & 0.011 & 0.29 &   0.04 &   30.72      \\
              S Lep          & 0.006 & $<$0.37 &   $<$0.02 &   -    \\
             RY Mon          & 0.009 & $<$0.66 &   $<$0.03 &   -    \\
             RT Hya          & 0.009 & $<$0.21 &   $<$0.03 &   -    \\
             RT Cnc          & 0.009 & $<$0.47 &   $<$0.03 &   -    \\
              R Crt          & 0.017 & 1.03 &   0.09 &   19.95      \\
              Z UMa          & 0.008 & $<$0.17 &   $<$0.02 &   -    \\
             BK Vir          & 0.009 & 0.11 &   0.04 &   7.24       \\
              Y UMa          & 0.011 & 0.39 &   0.07 &   7.24       \\
              Y CVn          & 0.014 & 6.71 &   0.47 &   15.36      \\
             RT Vir          & 0.013 & 0.82 &   0.08 &   12.36      \\
             SW Vir          & 0.015 & 0.63 &   0.07 &   13.60      \\
             RX Boo          & 0.013 & 1.00 &   0.10 &   16.07      \\
   $\tau$$^{4}$ Ser          & 0.007 & $<$0.73 &   $<$0.02 &   -    \\
             ST Her          & 0.010 & $<$0.75 &   $<$0.03 &   -    \\
              X Her          & 0.014 & 1.30 &   0.14 &   5.65       \\
              g Her          & 0.008 & 0.42 &   0.04 &   12.89      \\
              S Dra          & 0.010 & 0.27 &   0.03 &   9.00       \\
             FI Lyr          & 0.008 & 0.24 &   0.03 &   8.65       \\
              S Sct          & 0.016 & 0.08 &   0.02 &   5.70       \\
             AF Cyg          & 0.010 & 0.20 &   0.04 &   5.83       \\
             AQ Sgr          & 0.008 & 0.12 &   0.02 &   17.65      \\
             RT Cap          & 0.008 & 0.06 &   0.02 &   3.00       \\
             UU Dra          & 0.009 & 0.63 &   0.07 &   12.54      \\
             EU Del          & 0.009 & 0.22 &   0.02 &   17.30      \\
              U Del          & 0.010 & 0.33 &   0.03 &   18.72      \\
             IQ Aqr          & 0.007 & $<$0.54 &   $<$0.02 &   -    \\
              W Cyg          & 0.009 & 0.40 &   0.04 &   15.36      \\
             RU Cyg          & 0.005 & 0.10 &   0.02 &   15.89      \\
             EP Aqr          & 0.015 & 2.20 &   0.28 &   2.65       \\
             TW Peg          & 0.010 & 0.40 &   0.06 &   15.36      \\
             SV Lyn          & 0.010 & 0.35 &   0.11 &   3.18       \\
             SV Peg          & 0.013 & 0.63 &   0.08 &   14.05      \\
             AK Hya          & 0.015 & 0.31 &   0.05 &   9.00       \\
             RW Boo          & 0.006 & 0.05 &   0.02 &   -          \\
             IQ Her          & 0.009 & 0.24 &   0.06 &   6.35       \\
             AA Cyg          & 0.012 & 0.15 &   0.04 &   4.07       \\
              X Cnc          & 0.012 & $<$0.59 &   $<$0.04 &   -    \\
           V465 Cas          & 0.006 & $<$0.39 &   $<$0.02 &   -    \\
\hline 
\end{tabular}
      \tablefoot{Description of the columns from left to right. (1) GCVS designation \citep{samus09}; (2-5) 1$\sigma$ noise, integrated line intensity, peak main-beam brightness temperature, and full width at half power. 3$\sigma$ upper detection limits are indicated.\\ }
\end{table*}

\begin{table*}   
\caption{\label{t7}Mass-loss rates, photodissociation radii, and wind densities of SRs in the sample for the optically thick and optically thin cases, respectively.}   
\centering
\scalebox{0.82}{
\begin{tabular}{lccccccc}
\hline\hline  
\\ [-0.5ex]
Source & Spectral type & $\dot{M}$$_{thick}$ & $\dot{M}$$_{thin}$ & R$^{thick}_{CO}$ & R$^{thin}_{CO}$ & ($\dot{M}$/$\vexp$)$_{thick}$ & ($\dot{M}$/$\vexp$)$_{thin}$ \\ 
     &            & (M$_{\odot}$\,yr$^{-1}$) & (M$_{\odot}$\,yr$^{-1}$) & (10$^{15}$ cm) & (10$^{15}$ cm) & (M$_{\odot}$\,yr$^{-1}$\,km$^{-1}$\,s) & (M$_{\odot}$\,yr$^{-1}$\,km$^{-1}$\,s) \\
\hline
\\ [-1ex]
        VX And        & C  & $<$4.90$_{-1.70}$ $\times$ 10$^{-9}$ & $<$1.94$_{-0.68}$ $\times$ 10$^{-9}$  & $<$7.04$_{-0.73}$ &    $<$5.61$_{-0.48}$  &   $<$3.80$_{-1.62}$ $\times$ 10$^{-10}$  &   $<$1.50$_{-0.64}$ $\times$ 10$^{-10}$ \\ [1ex]
      V370 And        & M  & 8.23$^{+5.35}_{-1.82}$ $\times$ 10$^{-8}$ & 3.66$^{+0.89}_{-0.81}$ $\times$ 10$^{-8}$  & 17.00$^{+5.14}_{-1.75}$ &   11.80$^{+1.22}_{-1.11}$  &   7.34$^{+5.42}_{-2.28}$ $\times$ 10$^{-9}$  &   3.26$^{+1.08}_{-1.01}$ $\times$ 10$^{-9}$ \\ [1ex]
         Z Eri        & M  & $<$1.89$_{-0.71}$ $\times$ 10$^{-9}$ & $<$8.88$_{-3.31}$ $\times$ 10$^{-10}$  &  $<$3.85$_{-0.43}$ &    $<$3.26$_{-0.34}$  &  $<$2.43$_{-1.23}$ $\times$ 10$^{-10}$  &   $<$1.14$_{-0.57}$ $\times$ 10$^{-10}$ \\ [1ex]
        SS Cep        & M  & 9.17$^{+5.93}_{-1.98}$ $\times$ 10$^{-8}$ & 4.08$^{+0.97}_{-0.88}$ $\times$ 10$^{-8}$  & 17.90$^{+5.43}_{-1.81}$ &   12.40$^{+1.26}_{-1.15}$  &   8.11$^{+5.97}_{-2.46}$ $\times$ 10$^{-9}$  &   3.61$^{+1.17}_{-1.10}$ $\times$ 10$^{-9}$ \\ [1ex]
        TT Tau        & C  & 1.37$^{+0.99}_{-0.35}$ $\times$ 10$^{-7}$ & 5.25$^{+1.65}_{-1.33}$ $\times$ 10$^{-8}$  & 27.00$^{+10.25}_{-3.55}$ &  16.60$^{+2.55}_{-2.05}$  &   1.61$^{+1.37}_{-0.60}$ $\times$ 10$^{-8}$  &   6.20$^{+2.68}_{-2.30}$ $\times$ 10$^{-9}$ \\ [1ex]
         W Ori        & C  & 5.91$^{+5.11}_{-2.67}$ $\times$ 10$^{-7}$ & 2.27$^{+1.28}_{-1.02}$ $\times$ 10$^{-7}$  &  57.30$^{+26.60}_{-13.91}$ & 34.60$^{+10.06}_{-8.06}$  &   4.92$^{+4.66}_{-2.63}$ $\times$ 10$^{-8}$  &   1.89$^{+1.22}_{-1.01}$ $\times$ 10$^{-8}$ \\ [1ex]
        RX Lep        & M  & 3.06$^{+2.53}_{-1.42}$ $\times$ 10$^{-8}$ & 1.44$^{+0.84}_{-0.67}$ $\times$ 10$^{-8}$  & 10.20$^{+4.09}_{-2.30}$  &    7.20$^{+1.88}_{-1.51}$  &   5.41$^{+5.43}_{-3.47}$ $\times$ 10$^{-9}$  &   2.54$^{+1.93}_{-1.64}$ $\times$ 10$^{-9}$ \\ [1ex]
         Y Tau        & C  & 5.79$^{+4.03}_{-1.33}$ $\times$ 10$^{-7}$ & 2.22$^{+0.58}_{-0.51}$ $\times$ 10$^{-7}$  & 55.50$^{+20.16}_{-6.66}$ &  34.10$^{+4.36}_{-3.86}$ &   3.42$^{+2.58}_{-0.99}$ $\times$ 10$^{-8}$  &   1.31$^{+0.42}_{-0.38}$ $\times$ 10$^{-8}$ \\ [1ex]
         S Lep        & M  & $<$2.11$_{-0.73}$ $\times$ 10$^{-9}$ & $<$9.89$_{-3.47}$ $\times$ 10$^{-10}$  &  $<$4.58$_{-0.42}$  &    $<$3.97$_{-0.34}$   &  $<$2.04$_{-0.90}$ $\times$ 10$^{-10}$  &   $<$9.58$_{-4.28}$ $\times$ 10$^{-11}$ \\ [1ex]
        RY Mon        & C  & 3.65$^{+2.63}_{-0.71}$ $\times$ 10$^{-7}$ & 1.40$^{+0.41}_{-0.27}$ $\times$ 10$^{-7}$  & 44.30$^{+16.81}_{-4.54}$   &  27.10$^{+4.00}_{-2.63}$   &   2.97$^{+2.39}_{-0.82}$ $\times$ 10$^{-8}$  &   1.14$^{+0.43}_{-0.31}$ $\times$ 10$^{-8}$ \\ [1ex]
        RT Hya        & M  & 2.52$^{+1.85}_{-1.24}$ $\times$ 10$^{-9}$ & 1.32$^{+0.62}_{-0.60}$ $\times$ 10$^{-9}$  & 3.26$^{+0.94}_{-0.65}$  &    2.58$^{+0.46}_{-0.45}$ &  6.63$^{+6.61}_{-5.00}$ $\times$ 10$^{-10}$  &   3.46$^{+2.54}_{-2.50}$ $\times$ 10$^{-10}$ \\ [1ex]
        RT Cnc        & M  & $<$9.39$_{-3.35}$ $\times$ 10$^{-10}$ & $<$4.34$_{-1.50}$ $\times$ 10$^{-10}$  & $<$3.53$_{-0.34}$  &    $<$3.13$_{-0.29}$ &  $<$1.08$_{-0.51}$ $\times$ 10$^{-10}$  &      $<$4.97$_{-2.29}$ $\times$ 10$^{-11}$ \\ [1ex]
         R Crt        & M  & 2.28$^{+1.44}_{-0.42}$ $\times$ 10$^{-7}$ & 1.01$^{+0.21}_{-0.18}$ $\times$ 10$^{-7}$  & 27.90$^{+8.88}_{-2.56}$  &   18.80$^{+1.86}_{-1.61}$ &   1.91$^{+1.37}_{-0.51}$ $\times$ 10$^{-8}$  &   8.50$^{+2.50}_{-2.26}$ $\times$ 10$^{-9}$ \\ [1ex]
         Z UMa        & M  & $<$9.20$_{-4.05}$ $\times$ 10$^{-10}$ & $<$4.89$_{-2.03}$ $\times$ 10$^{-10}$  & $<$2.25$_{-0.38}$  &    $<$1.87$_{-0.31}$ &  $<$2.54$_{-1.82}$ $\times$ 10$^{-10}$  &   $<$1.35$_{-0.93}$ $\times$ 10$^{-10}$ \\ [1ex]
        BK Vir        & M  & 2.39$^{+1.61}_{-0.72}$ $\times$ 10$^{-8}$ & 1.12$^{+0.35}_{-0.34}$ $\times$ 10$^{-8}$  &  9.17$^{+2.85}_{-1.28}$  &    6.56$^{+0.88}_{-0.86}$ &  3.81$^{+3.17}_{-1.76}$ $\times$ 10$^{-9}$  &   1.79$^{+0.85}_{-0.83}$ $\times$ 10$^{-9}$ \\ [1ex]
         Y UMa        & M  & 9.53$^{+6.53}_{-2.80}$ $\times$ 10$^{-8}$ & 4.24$^{+1.30}_{-1.25}$ $\times$ 10$^{-8}$  &  18.00$^{+6.28}_{-2.70}$  &   12.10$^{+1.77}_{-1.70}$ &   1.35$^{+1.12}_{-0.59}$ $\times$ 10$^{-8}$  &   6.00$^{+2.69}_{-2.61}$ $\times$ 10$^{-9}$ \\ [1ex]
         Y CVn        & C  & 5.00$^{+3.49}_{-1.13}$ $\times$ 10$^{-8}$ & 1.92$^{+0.47}_{-0.42}$ $\times$ 10$^{-8}$  & 16.30$^{+5.41}_{-1.75}$  &   10.60$^{+1.11}_{-1.00}$ &   5.10$^{+4.08}_{-1.67}$ $\times$ 10$^{-9}$  &    1.96$^{+0.68}_{-0.63}$ $\times$ 10$^{-9}$ \\ [1ex]
        RT Vir        & M  & 4.85$^{+3.42}_{-1.56}$ $\times$ 10$^{-7}$ & 2.15$^{+0.80}_{-0.69}$ $\times$ 10$^{-7}$  & 41.40$^{+15.49}_{-7.09}$ &  27.10$^{+5.12}_{-4.46}$  &   4.69$^{+3.76}_{-1.97}$ $\times$ 10$^{-8}$  &   2.09$^{+0.97}_{-0.87}$ $\times$ 10$^{-8}$ \\ [1ex]
        SW Vir        & M  & 3.62$^{+2.79}_{-1.43}$ $\times$ 10$^{-7}$ & 1.61$^{+0.79}_{-0.64}$ $\times$ 10$^{-7}$  & 35.70$^{+14.51}_{-7.44}$ &  23.40$^{+5.82}_{-4.69}$  &   3.84$^{+3.36}_{-1.92}$ $\times$ 10$^{-8}$  &   1.71$^{+1.02}_{-0.85}$ $\times$ 10$^{-8}$ \\ [1ex]
        RX Boo        & M  & 6.49$^{+4.12}_{-1.24}$ $\times$ 10$^{-8}$ & 2.89$^{+0.58}_{-0.54}$ $\times$ 10$^{-8}$  & 15.20$^{+4.39}_{-1.32}$ &   10.70$^{+0.90}_{-0.83}$  &  5.84$^{+4.23}_{-1.64}$ $\times$ 10$^{-9}$  &   2.60$^{+0.76}_{-0.72}$ $\times$ 10$^{-9}$ \\ [1ex]
    $\tau$$^{4}$ Ser  & M  & 5.96$^{+3.91}_{-1.52}$ $\times$ 10$^{-8}$ & 2.65$^{+0.73}_{-0.68}$ $\times$ 10$^{-8}$  & 15.70$^{+4.05}_{-1.59}$ &   11.70$^{+1.10}_{-1.02}$  &   3.41$^{+2.43}_{-1.06}$ $\times$ 10$^{-9}$  &   1.52$^{+0.51}_{-0.47}$ $\times$ 10$^{-9}$ \\ [1ex]
        ST Her        & M  & 1.09$^{+0.70}_{-0.22}$ $\times$ 10$^{-7}$ & 4.87$^{+1.10}_{-0.97}$ $\times$ 10$^{-8}$  & 19.50$^{+5.84}_{-1.83}$ &   13.60$^{+1.32}_{-1.16}$  &   8.79$^{+6.31}_{-2.46}$ $\times$ 10$^{-9}$  &   3.91$^{+1.20}_{-1.09}$ $\times$ 10$^{-9}$ \\ [1ex]
         X Her        & M  & 6.81$^{+4.38}_{-1.46}$ $\times$ 10$^{-8}$ & 3.03$^{+0.72}_{-0.64}$ $\times$ 10$^{-8}$  & 15.40$^{+4.63}_{-1.54}$ &   10.70$^{+1.09}_{-0.97}$  &  6.71$^{+4.98}_{-2.10}$ $\times$ 10$^{-9}$  &   2.98$^{+1.00}_{-0.92}$ $\times$ 10$^{-9}$ \\ [1ex]
         G Her        & M  & 1.50$^{+0.92}_{-0.24}$ $\times$ 10$^{-8}$ & 6.91$^{+1.35}_{-1.13}$ $\times$ 10$^{-9}$  &  8.69$^{+1.81}_{-0.52}$ &    6.85$^{+0.43}_{-0.39}$  &  1.17$^{+0.81}_{-0.28}$ $\times$ 10$^{-9}$  &   5.40$^{+1.47}_{-1.31}$ $\times$ 10$^{-10}$ \\ [1ex]
         S Dra        & M  & 1.26$^{+1.23}_{-0.71}$ $\times$ 10$^{-7}$ & 5.60$^{+4.29}_{-3.17}$ $\times$ 10$^{-8}$  &  20.80$^{+9.90}_{-5.73}$ &   14.20$^{+4.89}_{-3.61}$  &   1.17$^{+1.25}_{-0.77}$ $\times$ 10$^{-8}$  &   5.20$^{+4.47}_{-3.42}$ $\times$ 10$^{-9}$ \\ [1ex]
        FI Lyr        & M  & 1.38$^{+0.91}_{-0.32}$ $\times$ 10$^{-7}$ & 6.15$^{+1.78}_{-1.44}$ $\times$ 10$^{-8}$  & 21.70$^{+7.26}_{-2.58}$ &   14.60$^{+2.01}_{-1.62}$  &   1.60$^{+1.24}_{-0.56}$ $\times$ 10$^{-8}$  &    7.11$^{+2.88}_{-2.48}$ $\times$ 10$^{-9}$ \\ [1ex]
         S Sct        & C  & 2.23$^{+2.63}_{-2.19}$ $\times$ 10$^{-8}$ & 9.02$^{+8.16}_{-8.82}$ $\times$ 10$^{-9}$  &  11.00$^{+5.98}_{-4.97}$ &    7.41$^{+2.74}_{-2.96}$  &   2.95$^{+3.86}_{-3.28}$ $\times$ 10$^{-9}$  &   1.19$^{+1.23}_{-1.32}$ $\times$ 10$^{-9}$ \\ [1ex]
        AF Cyg        & M  & 1.36$^{+0.91}_{-0.42}$ $\times$ 10$^{-8}$ & 6.29$^{+2.30}_{-1.82}$ $\times$ 10$^{-9}$  &  7.06$^{+2.07}_{-0.97}$ &    5.13$^{+0.75}_{-0.60}$  &  2.30$^{+1.92}_{-1.10}$ $\times$ 10$^{-9}$  &   1.06$^{+0.57}_{-0.49}$ $\times$ 10$^{-9}$ \\ [1ex]
        AQ Sgr        & C  & 1.90$^{+1.34}_{-0.46}$ $\times$ 10$^{-7}$ & 7.28$^{+2.01}_{-1.76}$ $\times$ 10$^{-8}$  &  31.60$^{+11.41}_{-3.89}$  &  19.70$^{+2.58}_{-2.26}$   &   1.60$^{+1.27}_{-0.52}$ $\times$ 10$^{-8}$  &    6.15$^{+2.22}_{-2.00}$ $\times$ 10$^{-9}$ \\ [1ex]
        RT Cap        & C  & 4.14$^{+2.91}_{-0.98}$ $\times$ 10$^{-8}$ & 1.60$^{+0.42}_{-0.35}$ $\times$ 10$^{-8}$  & 14.80$^{+4.95}_{-1.68}$ &    9.67$^{+1.10}_{-0.91}$ &  4.60$^{+3.74}_{-1.60}$ $\times$ 10$^{-9}$  &   1.78$^{+0.67}_{-0.59}$ $\times$ 10$^{-9}$ \\ [1ex]
        UU Dra        & M  & 1.90$^{+1.22}_{-0.37}$ $\times$ 10$^{-7}$ & 8.46$^{+1.70}_{-1.65}$ $\times$ 10$^{-8}$  &  25.50$^{+8.11}_{-2.47}$ &   17.20$^{+1.61}_{-1.56}$ &   1.65$^{+1.19}_{-0.46}$ $\times$ 10$^{-8}$  &    7.32$^{+2.10}_{-2.06}$ $\times$ 10$^{-9}$ \\ [1ex]
        EU Del        & M  & 5.80$^{+3.64}_{-1.12}$ $\times$ 10$^{-9}$ & 2.64$^{+0.60}_{-0.50}$ $\times$ 10$^{-9}$  & 6.34$^{+1.11}_{-0.41}$ &    5.25$^{+0.35}_{-0.32}$ &  4.83$^{+3.44}_{-1.34}$ $\times$ 10$^{-10}$  &   2.20$^{+0.68}_{-0.60}$ $\times$ 10$^{-10}$ \\ [1ex]
         U Del        & M  & 7.54$^{+4.82}_{-1.48}$ $\times$ 10$^{-8}$ & 3.35$^{+0.88}_{-0.66}$ $\times$ 10$^{-8}$  & 16.30$^{+4.82}_{-1.49}$ &   11.40$^{+1.25}_{-0.95}$ &  6.77$^{+4.94}_{-1.94}$ $\times$ 10$^{-9}$  &   3.01$^{+1.06}_{-0.86}$ $\times$ 10$^{-9}$ \\ [1ex]
        IQ Aqr        & M  & 1.15$^{+0.72}_{-0.29}$ $\times$ 10$^{-8}$ & 5.28$^{+1.13}_{-1.37}$ $\times$ 10$^{-9}$  &  7.97$^{+1.59}_{-0.68}$ &    6.38$^{+0.42}_{-0.48}$ &  8.97$^{+6.30}_{-3.00}$ $\times$ 10$^{-10}$  &    4.12$^{+1.21}_{-1.39}$ $\times$ 10$^{-10}$ \\ [1ex]
         W Cyg        & M  & 4.20$^{+2.71}_{-1.00}$ $\times$ 10$^{-8}$ & 1.91$^{+0.52}_{-0.42}$ $\times$ 10$^{-8}$  & 12.30$^{+3.57}_{-1.32}$ &    8.78$^{+0.97}_{-0.80}$ &  4.44$^{+3.34}_{-1.53}$ $\times$ 10$^{-9}$  &   2.02$^{+0.76}_{-0.66}$ $\times$ 10$^{-9}$ \\ [1ex]
        RU Cyg        & M  & 2.79$^{+1.82}_{-0.64}$ $\times$ 10$^{-7}$ & 1.24$^{+0.33}_{-0.28}$ $\times$ 10$^{-7}$  & 30.80$^{+10.17}_{-3.57}$ &  20.70$^{+2.61}_{-2.25}$ &   2.24$^{+1.64}_{-0.69}$ $\times$ 10$^{-8}$  &   9.95$^{+3.44}_{-3.07}$ $\times$ 10$^{-9}$ \\ [1ex]
        EP Aqr        & M  & 1.19$^{+0.77}_{-0.25}$ $\times$ 10$^{-7}$ & 5.31$^{+1.15}_{-1.10}$ $\times$ 10$^{-8}$  & 20.30$^{+6.20}_{-2.00}$ &   14.00$^{+1.33}_{-1.27}$ &   9.80$^{+7.09}_{-2.83}$ $\times$ 10$^{-9}$  &   4.36$^{+1.30}_{-1.26}$ $\times$ 10$^{-9}$ \\ [1ex]
        TW Peg        & M  & 9.44$^{+6.08}_{-1.96}$ $\times$ 10$^{-8}$ & 4.20$^{+0.93}_{-0.87}$ $\times$ 10$^{-8}$  & 18.20$^{+5.43}_{-1.75}$ &   12.70$^{+1.18}_{-1.11}$ &   7.64$^{+5.53}_{-2.20}$ $\times$ 10$^{-9}$  &   3.39$^{+1.02}_{-0.98}$ $\times$ 10$^{-9}$ \\ [1ex]
        SV Lyn        & M  & 4.76$^{+3.45}_{-2.42}$ $\times$ 10$^{-9}$ & 2.35$^{+1.60}_{-1.03}$ $\times$ 10$^{-9}$  &   4.10$^{+1.37}_{-0.96}$ &    3.01$^{+0.87}_{-0.57}$ &  1.63$^{+1.74}_{-1.39}$ $\times$ 10$^{-9}$  &   8.05$^{+8.24}_{-6.29}$ $\times$ 10$^{-10}$ \\ [1ex]
        SV Peg        & M  & 1.07$^{+0.92}_{-0.52}$ $\times$ 10$^{-6}$ & 4.75$^{+2.90}_{-2.33}$ $\times$ 10$^{-7}$  & 64.10$^{+30.33}_{-17.26}$  & 41.30$^{+13.52}_{-10.87}$ &   1.18$^{+1.15}_{-0.71}$ $\times$ 10$^{-7}$  &   5.24$^{+3.78}_{-3.15}$ $\times$ 10$^{-8}$ \\ [1ex]
        AK Hya        & M  & 1.68$^{+1.11}_{-0.43}$ $\times$ 10$^{-8}$ & 7.82$^{+2.53}_{-2.02}$ $\times$ 10$^{-9}$  &   7.96$^{+2.24}_{-0.88}$  &    5.85$^{+0.74}_{-0.60}$ &  2.32$^{+1.85}_{-0.91}$ $\times$ 10$^{-9}$  &     1.08$^{+0.50}_{-0.43}$ $\times$ 10$^{-9}$ \\ [1ex]
        RW Boo        & M  & 9.89$^{+6.10}_{-1.79}$ $\times$ 10$^{-8}$ & 4.40$^{+0.72}_{-0.80}$ $\times$ 10$^{-8}$  & 19.30$^{+5.10}_{-1.50}$  &   13.90$^{+0.87}_{-0.96}$ &  5.72$^{+3.86}_{-1.37}$ $\times$ 10$^{-9}$  &   2.54$^{+0.56}_{-0.61}$ $\times$ 10$^{-9}$ \\ [1ex]
        IQ Her        & M  & 1.66$^{+1.10}_{-0.44}$ $\times$ 10$^{-8}$ & 7.73$^{+2.47}_{-2.07}$ $\times$ 10$^{-9}$  &   7.84$^{+2.26}_{-0.92}$  &    5.72$^{+0.73}_{-0.62}$ &  2.48$^{+2.01}_{-1.03}$ $\times$ 10$^{-9}$  &   1.15$^{+0.54}_{-0.48}$ $\times$ 10$^{-9}$ \\  [1ex]
        AA Cyg        & S  & 1.56$^{+1.15}_{-0.51}$ $\times$ 10$^{-7}$ & 6.25$^{+2.44}_{-2.06}$ $\times$ 10$^{-8}$  &  27.50$^{+10.84}_{-4.88}$  &  17.00$^{+3.41}_{-2.89}$ &    2.55$^{+2.30}_{-1.26}$ $\times$ 10$^{-8}$  &   1.02$^{+0.57}_{-0.51}$ $\times$ 10$^{-8}$ \\ [1ex]
         X Cnc        & C  & 2.87$^{+2.63}_{-1.45}$ $\times$ 10$^{-7}$ & 1.10$^{+0.71}_{-0.56}$ $\times$ 10$^{-7}$  & 40.00$^{+19.71}_{-10.90}$  & 24.10$^{+8.01}_{-6.31}$ &   3.49$^{+3.63}_{-2.20}$ $\times$ 10$^{-8}$  &   1.34$^{+1.03}_{-0.84}$ $\times$ 10$^{-8}$ \\ [1ex]
      V465 Cas        & M  & 4.69$^{+3.07}_{-1.09}$ $\times$ 10$^{-8}$ & 2.11$^{+0.56}_{-0.46}$ $\times$ 10$^{-8}$  & 13.10$^{+3.78}_{-1.35}$   &    9.38$^{+0.99}_{-0.81}$  &  4.38$^{+3.28}_{-1.43}$ $\times$ 10$^{-9}$  &   1.97$^{+0.71}_{-0.61}$ $\times$ 10$^{-9}$ \\  [1ex]
\hline
\end{tabular}}
\tablefoot{Description of the columns from left to right. (1-2) GCVS designation and spectral type \citep{samus09}; (3-8) Mass-loss rate, photodissociation radius, and surface wind density estimates with the corresponding asymmetric uncertainties for the optically thick and optically thin case, respectively. 3$\sigma$ upper detection limits are indicated.\\}
\end{table*}

\begin{table*}
\caption{\label{tabmod}Observational results for those sources which clearly show double-component in the $^{12}$CO $\textit{J}=2-1$ line (type 4 profiles).}
\centering
\begin{tabular}{lccccc}
\hline\hline
\\ [-1ex]
Source  & Components & A & \tmb & V$_{lsr}$ & $\vexp$ \\
        &            & (K\,km\,s$^{-1}$) & (K) & (km\,s$^{-1}$) & (km\,s$^{-1}$) \\
\hline
  V370 And   & broad  & 35.87 & 2.51 & -1.94  & 10.97  \\
             & narrow & 13.93 & 5.65 & -2.03  &  3.03  \\
  X Her      & broad  & 29.89 & 2.16 & -73.46 &  9.46  \\
             & narrow & 13.15 & 5.51 & -73.01 &  2.94  \\          
  EP Aqr     & broad  & 51.70 & 3.18 & -33.95 & 11.51  \\
             & narrow & 12.37 & 9.85 & -33.41 &  2.05  \\         
  IQ Aqr     & broad  &  1.20 & 0.06 &  21.24 & 12.48  \\         
             & narrow &  2.07 & 0.33 &  20.71 &  6.02  \\         
\hline                                                                                 
\end{tabular}      
\tablefoot{Description of the columns from left to right. (1) GCVS designation \citep{samus09}; (2) Component; (3-6) Integrated line intensity, peak main-beam brightness temperature, systemic velocity, and expansion velocity for each component. We note that the peak main-beam brightness temperature of the narrow components coincides with the peak main-beam brightness temperature of the entire emission lines (see Table A.1). For the integrated line intensity, the situation is more complex (see text). \\
}
\end{table*}   

\onecolumn

\newpage
\section{Oblate spheroid models}

In these models, we have considered a general and  simple model based on an oblate spheroid placed in different orientations.

\begin{table*}[htbp]
\caption{\label{tabmod} Structure of the molecular envelopes of AQ Sgr, BK Vir, RX Boo, $\tau$$^{4}$ Ser, and V370 And derived from our model fitting of the $^{12}$CO data with SHAPE+shapemol.}
\centering
\begin{tabular}{lcccc}
\hline\hline
\\ [-1ex]
 Source                   & Profile type & Components     &  D      &  \rco                   \\
                          &              &                &  (kpc)  &  (10$^{15}$ cm)             \\
\hline
  AQ Sgr                  & 1            & oblate spheroid              &  0.56   &  19.70        \\
  BK Vir                  & 2            & oblate spheroid              &  0.23   &   6.56        \\
  RX Boo                  & 3            & oblate spheroid              &  0.13   &  15.20        \\
 $\tau$$^{4}$ Ser         & 3            & oblate spheroid              &  0.29   &  11.70        \\
\hline
 V370 And                 & 4            & biconical structure &  0.14   & 11.80~\tablefootmark{a} \\  
                          &              & oblate spheroid     &  0.14   &  6.32                   \\
\hline                                                                                 
\end{tabular}   
\tablefoot{Description of the columns from left to right. (1) GCVS designation \citep{samus09}; (2-3) Profile type and components; (4) Source distance (\textit{Gaia} DR2; \citealt{gaia18}); (5) Photodissociation radius of the envelope or equatorial radius, r$_{1}$ (taken from Table 2). In the oblate spheroid of V370 And this radius is 3 arcsec ($\sim$6.32 $\times$ 10$^{15}$ cm), which is not the photodissociation radius. \\
\tablefoottext{a}{We note that in this model the photodissociation radius corresponds to the longitude of the cones, $\sim$5.6 arcsec ($\sim$11.80 $\times$ 10$^{15}$ cm). Moreover, the radius of the base of the cones is $\sim$3.9 arcsec ($\sim$8.22 $\times$ 10$^{15}$ cm).}
}
\end{table*}     

\begin{table*}[htbp]
\caption{\label{tabmod} Physical conditions in the molecular envelopes of AQ Sgr, BK Vir, RX Boo, $\tau$$^{4}$ Ser, and V370 And, derived from our model fitting of the $^{12}$CO data with SHAPE+shapemol.}
\centering
\scalebox{0.96}{
\begin{tabular}{lcccc}
\hline\hline
\\ [-1ex]
 Source                       & Gas density                             &  Temperature   &   Velocity     &  $\delta$$_{V}$   \\
                              & (cm$^{-3}$)                             & (K)            & (km\,s$^{-1}$) &  (km\,s$^{-1}$)   \\
\hline
\\ [-1ex] 
                              & \it{n(r)} = \it{n$_{o}$} \normalfont{$\times$ 1/(r$^{2}$+0.1)} & \it{T(r)} = \it{T$_{o}$} \normalfont{$\times$ 1/(r+0.1)}  & \it{V(r)} = \it{V$_{o}$}                              &    \\
                              & \it{n($\theta$)} = \it{n$_{o}$}                   & \it{T($\theta$)} = \it{T$_{o}$}              & \it{V($\theta$)} = \it{V$_{o}$}                       &    \\
                              & \it{n($\varphi$)} = \it{n$_{o}$}                  & \it{T($\varphi$)} = \it{T$_{o}$}             & \it{V($\varphi$)} = \it{V$_{o}$} $\times$ $\sqrt{r^2_2 + (r^2_1-r^2_2)\sin^{2}\left(\varphi/57.3\right)}$   &  \\
\hline
                              & \it{n$_{o}$}                                      & \it{T$_{o}$}                                 & \it{V$_{o}$}           &  \\
\hline
\\ [-1ex]  
  AQ Sgr                      & 1.70 $\times$ 10$^{5}$             & 50             & 5.5               &   2.9         \\
  BK Vir                      & 1.12 $\times$ 10$^{6}$             & 100            & 3.5               &   1.4      \\
  RX Boo                      & 1.70 $\times$ 10$^{6}$             & 110            & 1.2               &   1.6          \\
 $\tau$$^{4}$ Ser             & 1.80 $\times$ 10$^{5}$             & 120            & 4.4               &   2.0      \\  
\hline\hline
 V370 And                     &                                    &                                       &                         &  \\
\hline
\\ [-1ex] 
Oblate                        & \it{n(r)} = \it{n$_{o}$} \normalfont{$\times$ 1/(r+0.1)}                  & \it{T(r)} = \it{T$_{o}$} \normalfont{$\times$ 1/(r+0.1)} &  \it{V(r)} = \it{V$_{o}$}        &  1.1   \\
 spheroid                     & \it{n($\theta$)} = \it{n$_{o}$}                              & \it{T($\theta$)} = \it{T$_{o}$}             &  \it{V($\theta$)} = \it{V$_{o}$} &        \\
                              & \it{n($\varphi$)} = \it{n$_{o}$}                             & \it{T($\varphi$)} = \it{T$_{o}$}            &  \it{V($\varphi$)} = \it{V$_{o}$} $\times$ $\sqrt{r^2_2 + (r^2_1-r^2_2)\sin^{2}\left(\varphi/57.3\right)}$ &  \\    
                              & \it{n$_{o}$} = \normalfont{6.50 $\times$ 10$^{5}$} & \it{T$_{o}$} = \normalfont{100}                          &  \it{V$_{o}$} = \normalfont{0.9}              &        \\ 
\hline 
\\ [-1ex] 
 Biconical                    & \it{n(r)} = \it{n$_{o}$} \normalfont{$\times$ $\mid1/(r+0.1)\mid$} &  \it{T(r)} = \it{T$_{o}$} \normalfont{$\times$ 1/(r+0.1)}  &  \it{V(r)} = \it{V$_{o}$}        &  3.4 \\  
 structure                    & \it{n($\theta$)} = \it{n$_{o}$}                                   &  \it{T($\theta$)} = \it{T$_{o}$}              &  \it{V($\theta$)} = \it{V$_{o}$} &      \\
                              & \it{n}(\normalfont{z})~\tablefootmark{a} = \it{n$_{o}$}                        &  \it{T}(\normalfont{z}) = \it{T$_{o}$}                     &  \it{V}(\normalfont{z}) = \it{V$_{o}$}        &      \\
                              & \it{n$_{o}$} = \normalfont{1.30 $\times$ 10$^{5}$}      &  \it{T$_{o}$} = \normalfont{100}                           &  \it{V$_{o}$} = \normalfont{4.6}          &      \\    
\hline\hline                             
\end{tabular}}   
\tablefoot{Description of the columns from left to right. (1) GCVS designation \citep{samus09}; (2-5) Gas density, gas temperature, gas velocity, and micro-turbulence velocity. Note that r, r$_1$, and r$_2$ are given in arcsec; r$_1$ = \rco~(photodissociation radius taken from Table 2) and r$_2$ = 0.2 $\times$ \rco.~For RX Boo, the only exception, r$_2$ = 0.1 $\times$~\rco.~$\varphi$ is given in degrees. \\
\tablefoottext{a}{We note that in this model we use cylindrical coordinates for defining the physical conditions in the biconical structure.}
}
\end{table*}     
\clearpage

\begin{figure*}
   \centering
   \includegraphics[angle=0,width=9cm]{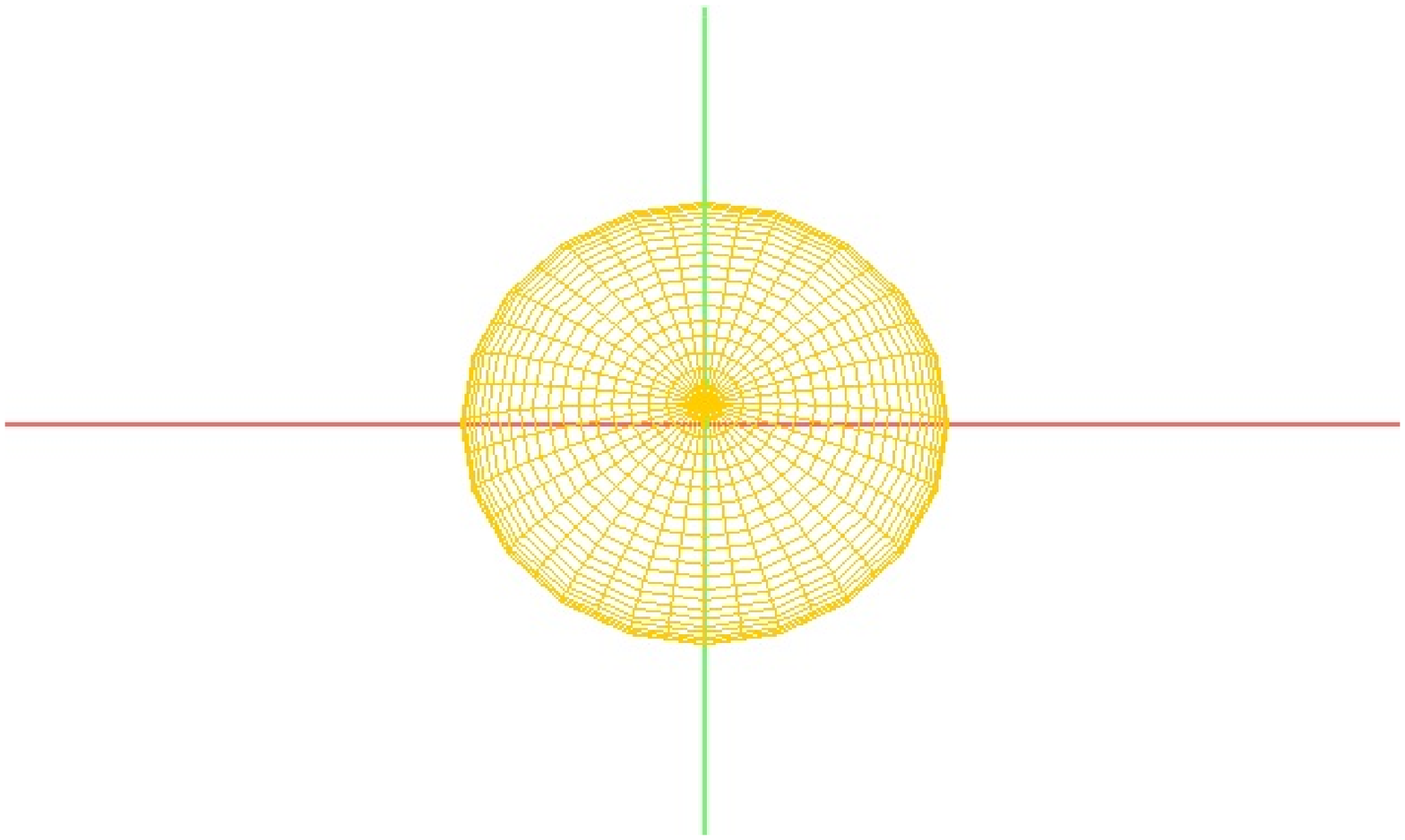}
   \includegraphics[angle=0,width=9cm]{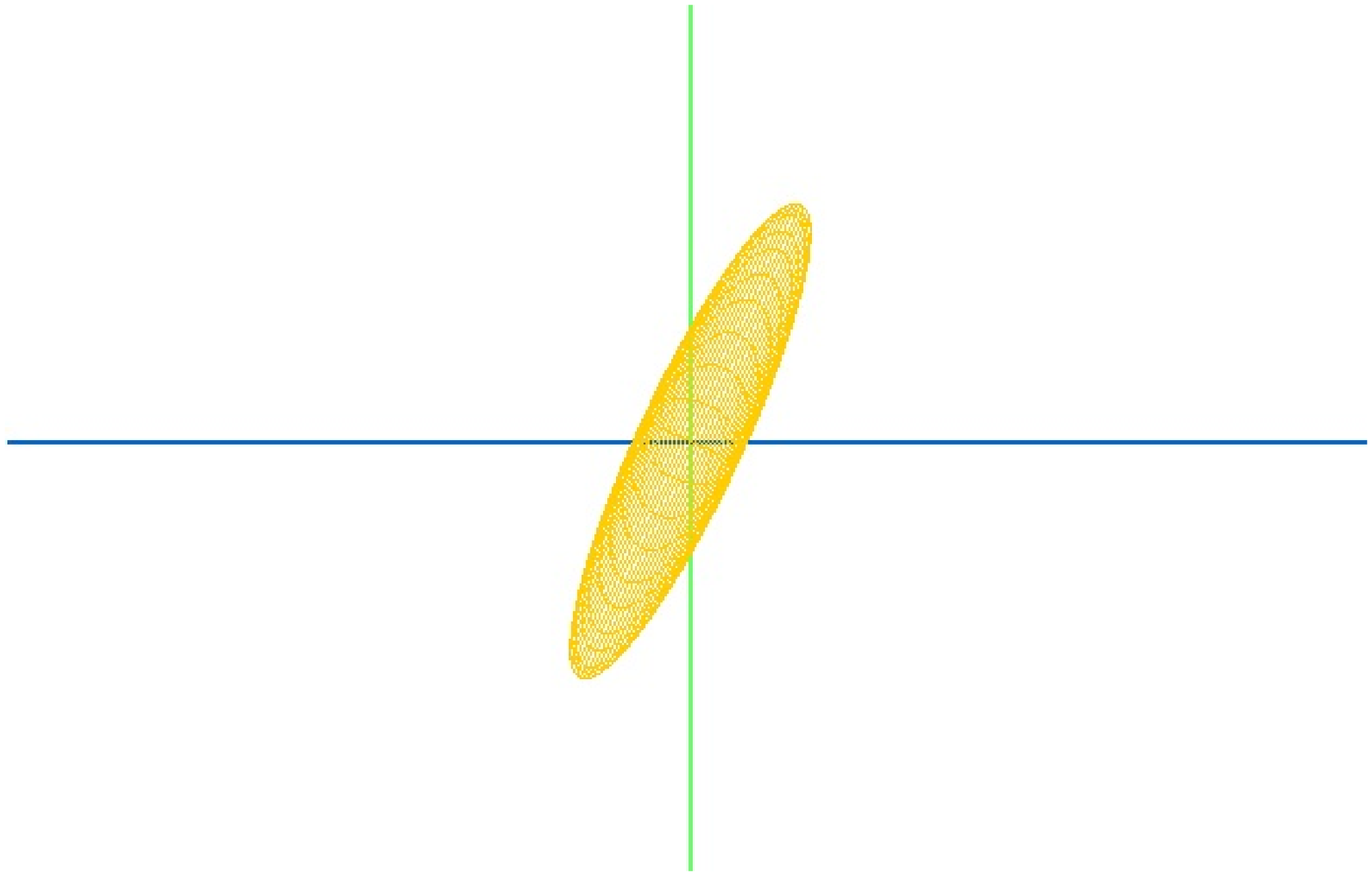}
      \caption{Three-dimensional mesh view of the model of the molecular envelope of AQ Sgr, which is defined by an oblate spheroid whose equatorial plane is inclined by $\sim$25$^{\circ}$ with respect to the plane of the sky. Left: View from Earth. Right: View from the direction defined by west in the plane of the sky (the observer is on the left). The X, Y, and Z axes are represented by red, green, and blue, respectively (positive directions are in the clockwise direction).}
      \label{mod1}
\end{figure*}

 \begin{figure*}
   \centering
   \includegraphics[angle=0,width=9cm]{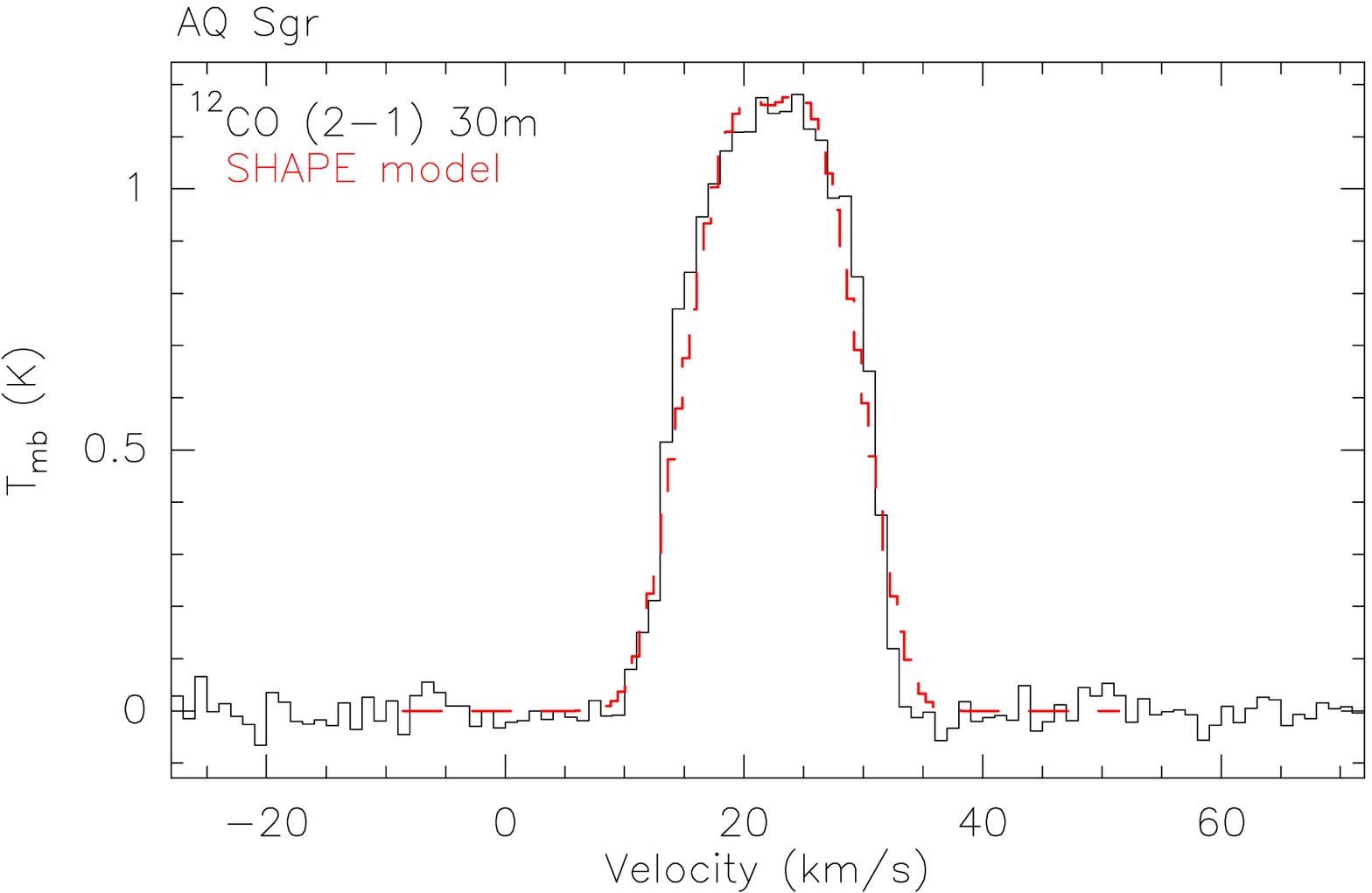}
      \caption{Resulting synthetic spectrum (red) and observation (black) for the $^{12}$CO \textit{J}=2-1 transition toward AQ Sgr.}
         \label{mod2}
   \end{figure*}  
   
\begin{figure*}
   \centering
   \includegraphics[angle=0,width=9cm]{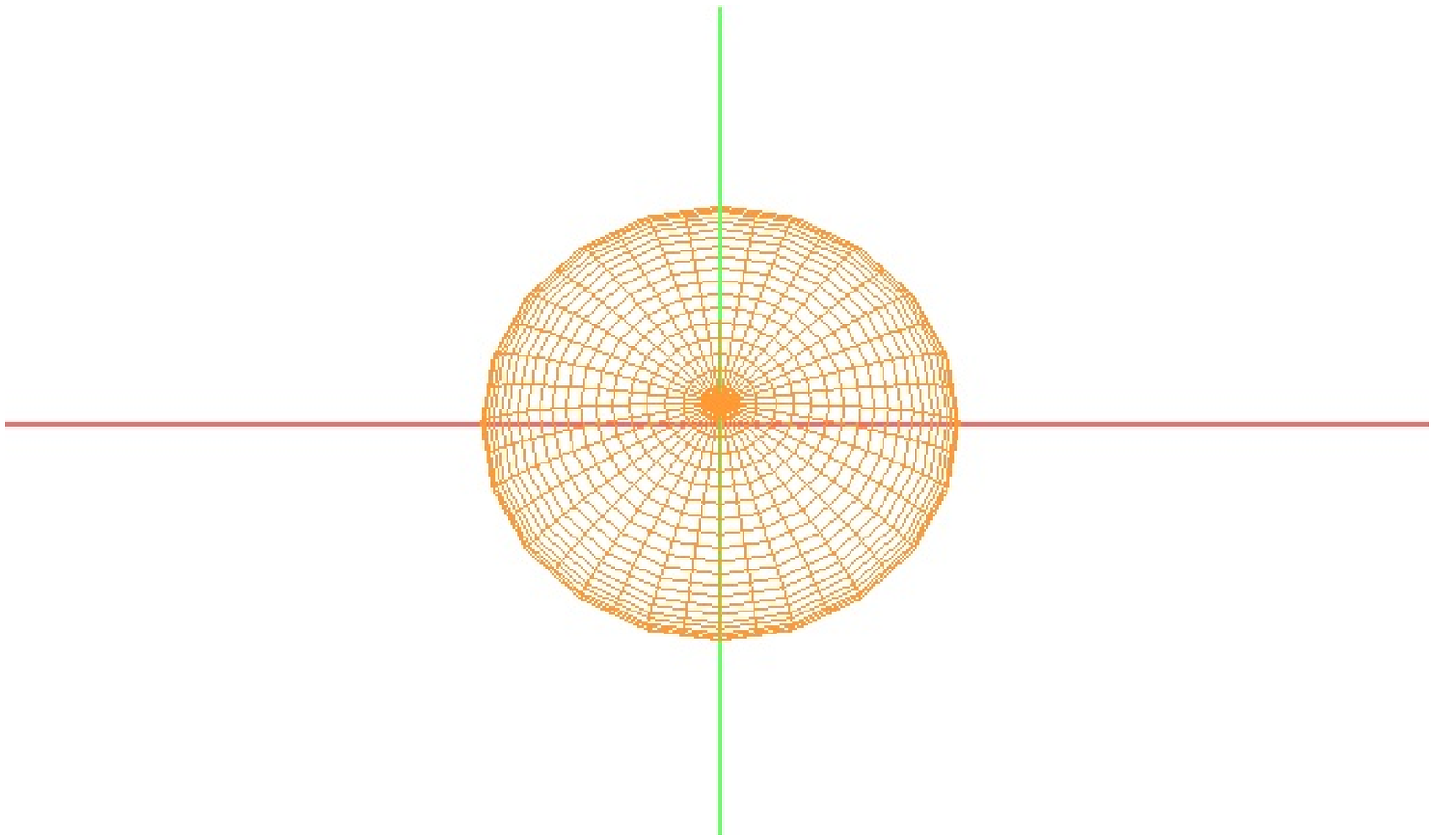}
   \includegraphics[angle=0,width=9cm]{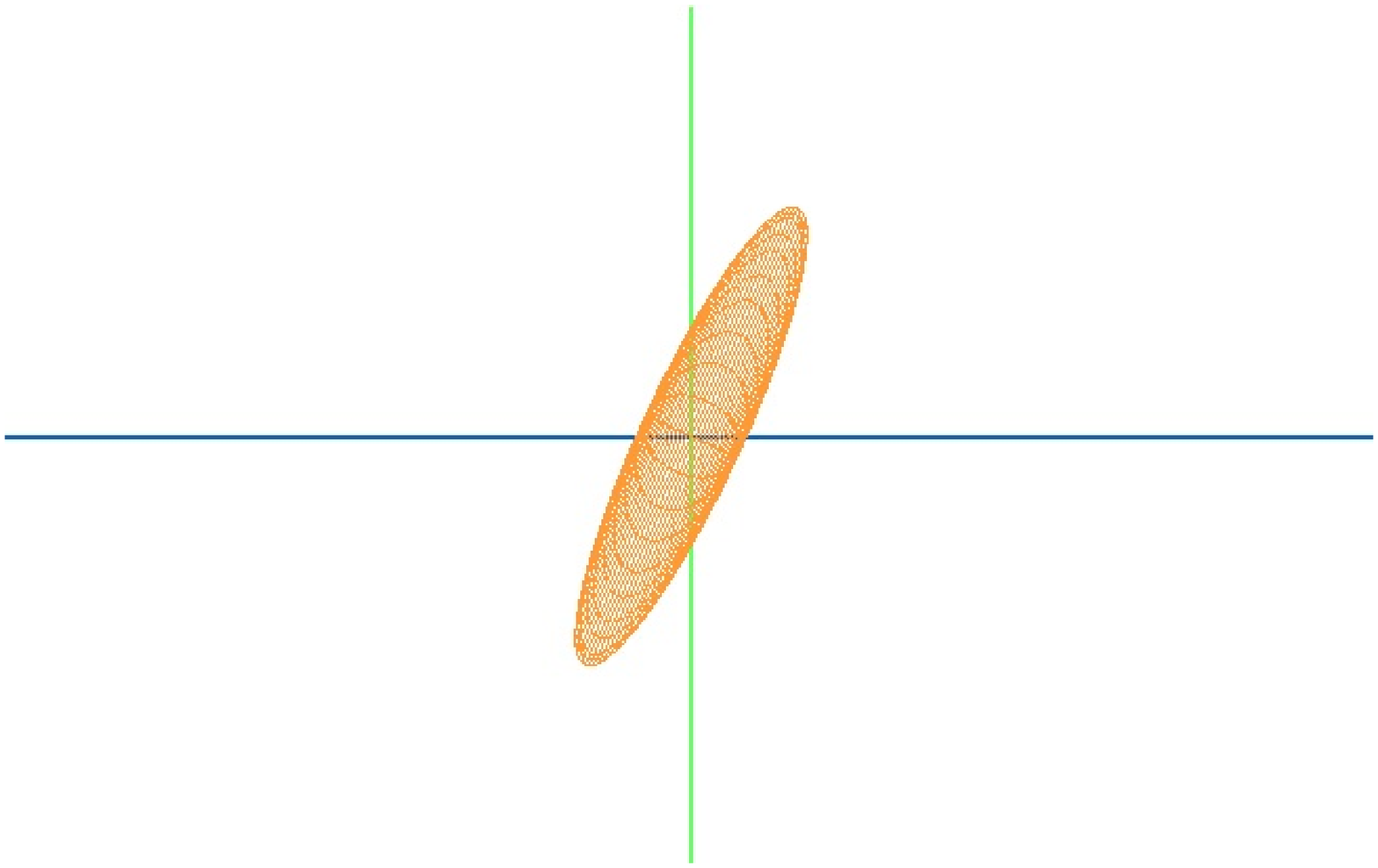}
      \caption{3D mesh view of the model of the molecular envelope of BK Vir, which is defined by an oblate spheroid whose equatorial plane is inclined by $\sim$25$^{\circ}$ with respect to the plane of the sky. Left: view from Earth. Right: view from the direction defined by west in the plane of the sky (the observer is on the left). X, Y, and Z axes are represented by red, green, and blue, respectively (positive directions are in the clockwise direction).}
         \label{mod3}
   \end{figure*}

 \begin{figure*}
   \centering
   \includegraphics[angle=0,width=9cm]{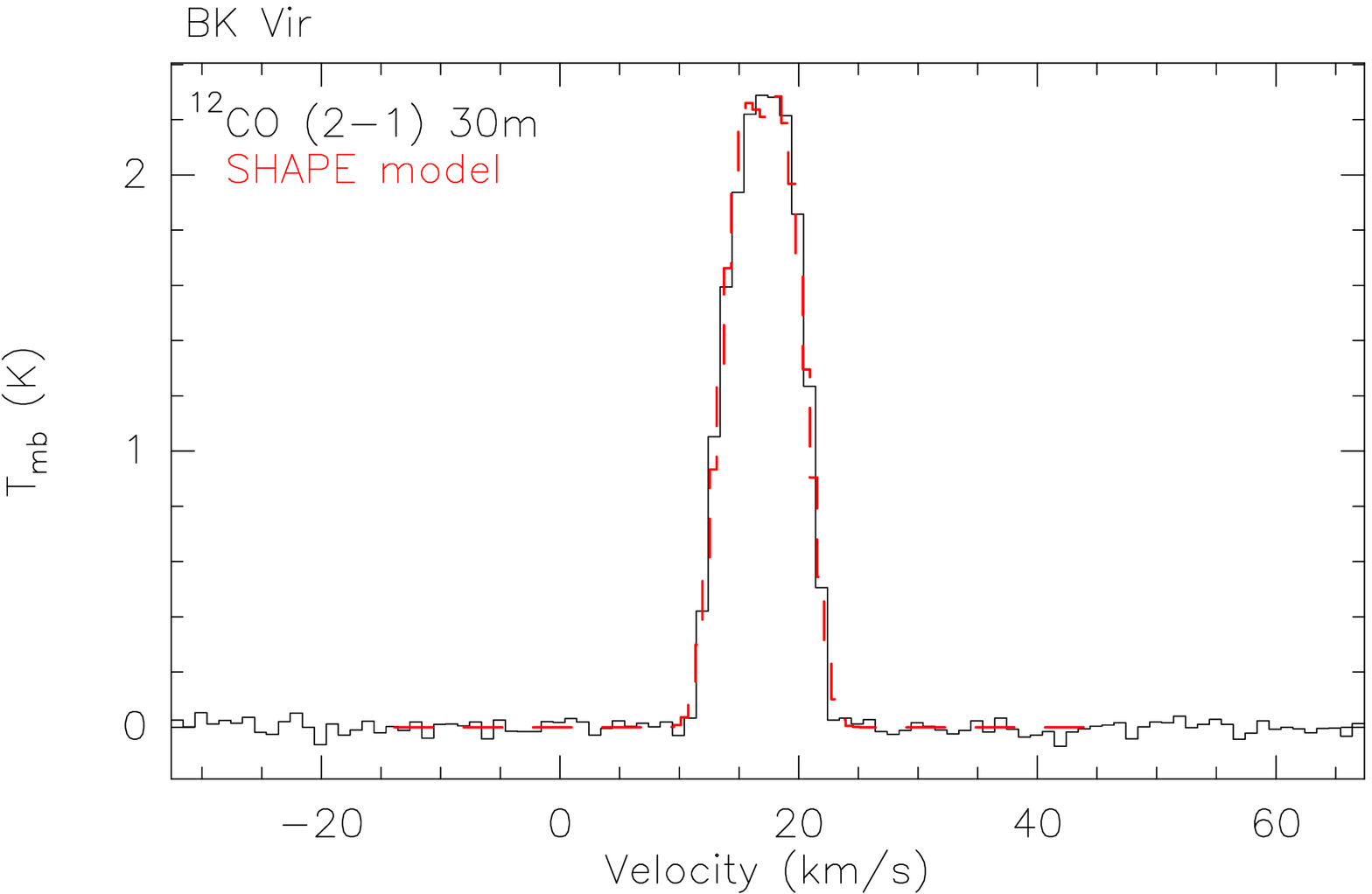}
      \caption{Resulting synthetic spectrum (red) and observation (black) for the $^{12}$CO \textit{J}=2-1 transition toward BK Vir.}
         \label{mod4}
   \end{figure*}    
   
\begin{figure*}
   \centering
   \includegraphics[angle=0,width=9cm]{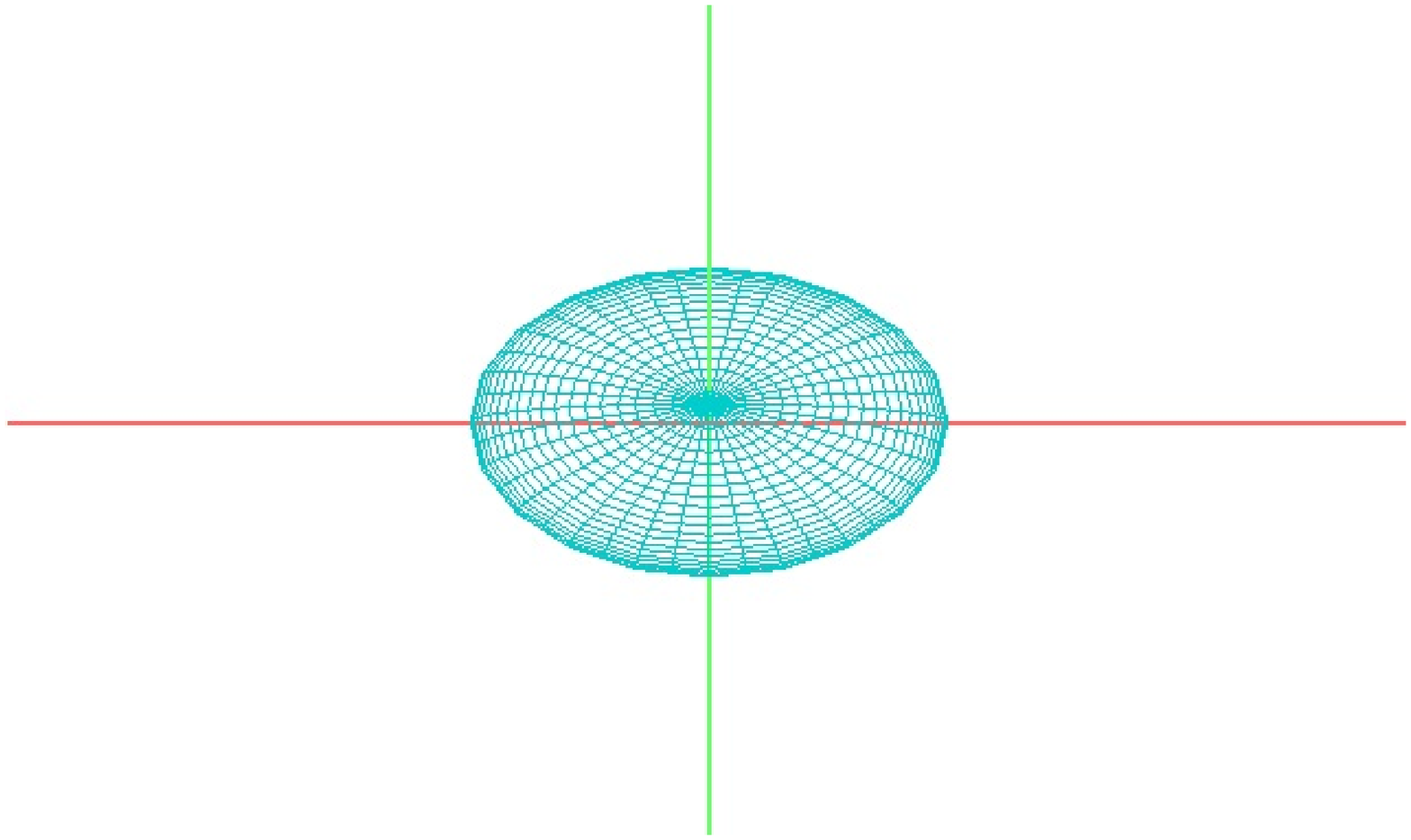}
   \includegraphics[angle=0,width=9cm]{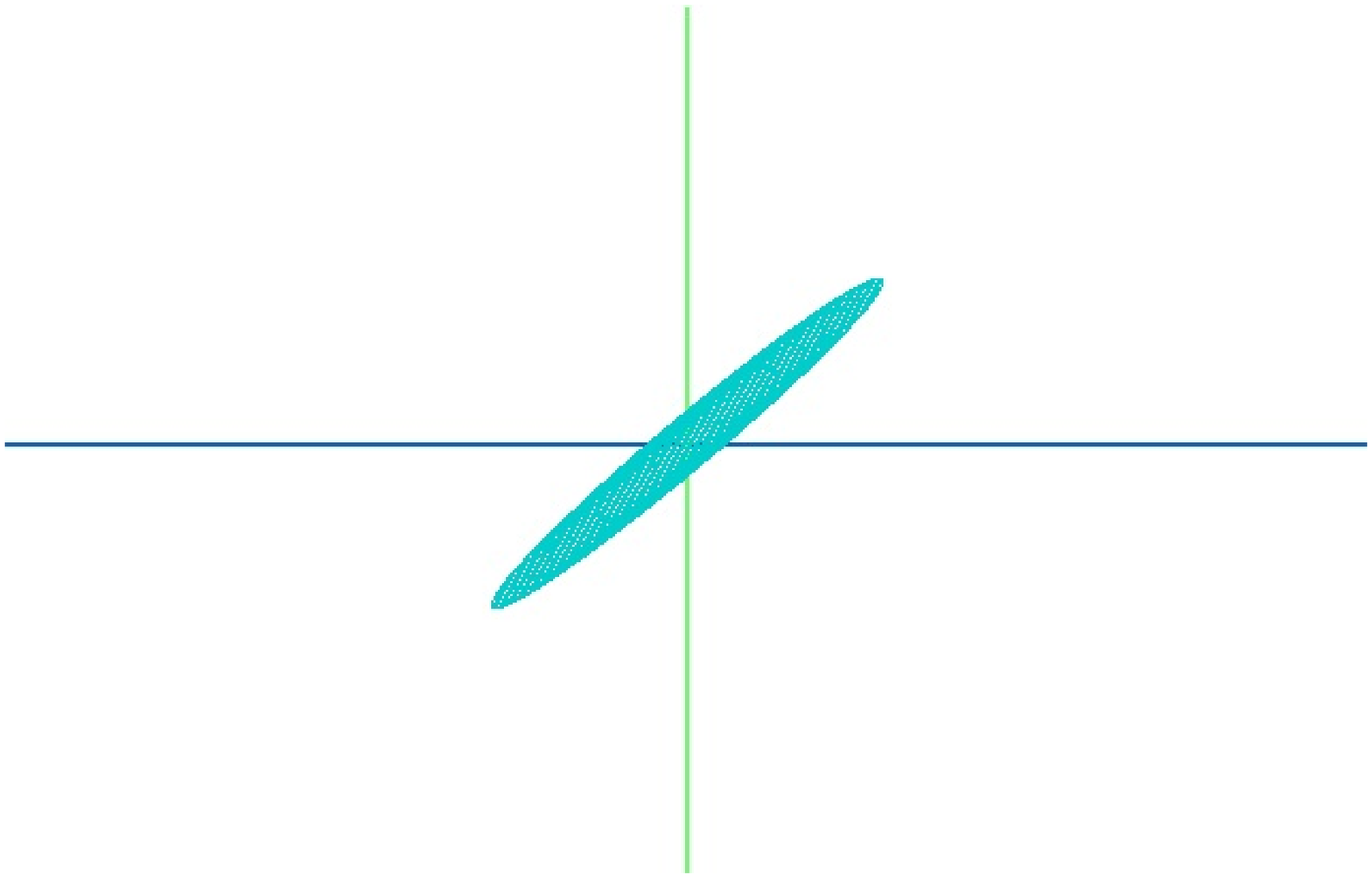}
      \caption{Three-dimensional mesh view of the model of the molecular envelope of RX Boo, which is defined by an oblate spheroid whose equatorial plane is inclined by $\sim$50$^{\circ}$ with respect to the plane of the sky. Left: View from Earth. Right: View from the direction defined by west in the plane of the sky (the observer is on the left). The X, Y, and Z axes are represented by red, green, and blue, respectively (positive directions are in the clockwise direction).}
         \label{mod5}
   \end{figure*}

\begin{figure*}
   \centering
   \includegraphics[angle=0,width=9cm]{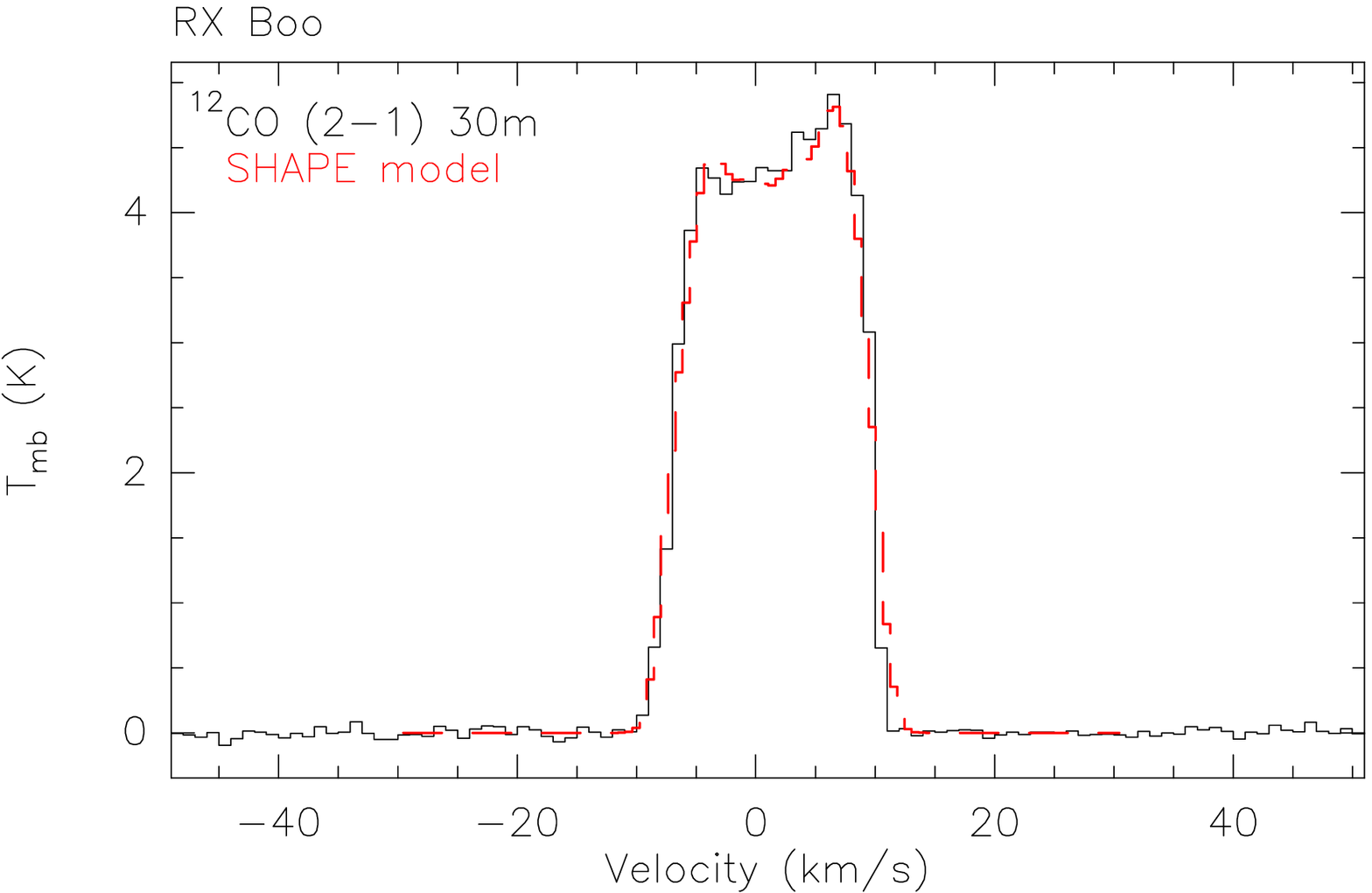}
      \caption{Resulting synthetic spectrum (red) and observation (black) for the $^{12}$CO \textit{J}=2-1 transition toward RX Boo.}
         \label{mod6}
   \end{figure*}       
   
\begin{figure*}
   \centering
   \includegraphics[angle=0,width=9cm]{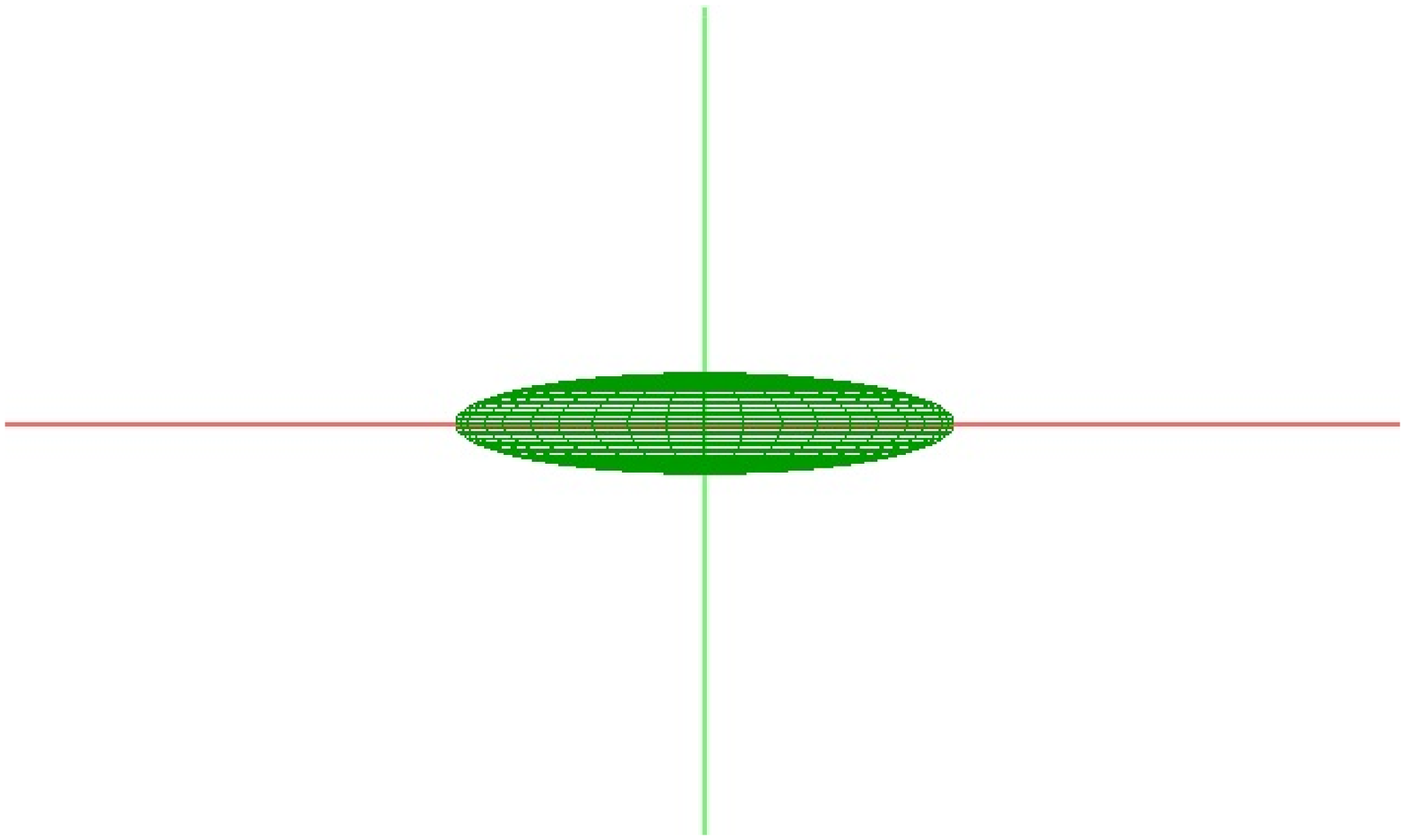}
   \includegraphics[angle=0,width=9cm]{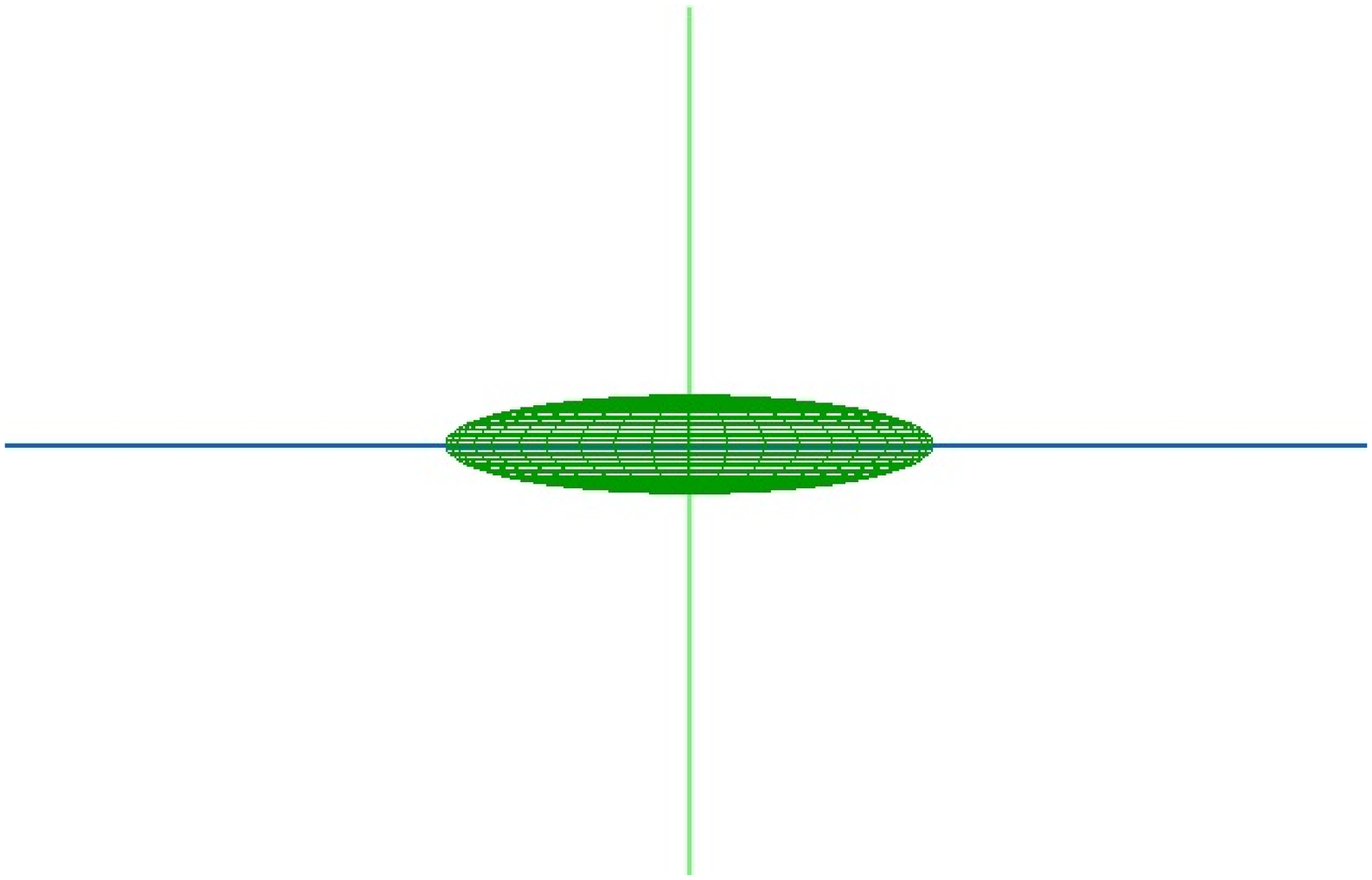}
      \caption{Three-dimensional mesh view of the model of the molecular envelope of $\tau$$^{4}$ Ser, which is defined by an edge-on oblate spheroid. Left: View from Earth. Right: View from the direction defined by west in the plane of the sky (the observer is on the left). The X, Y, and Z axes are represented by red, green, and blue, respectively (positive directions are in the clockwise direction).}
         \label{mod7}
   \end{figure*}

 \begin{figure*}
   \centering
   \includegraphics[angle=0,width=9cm]{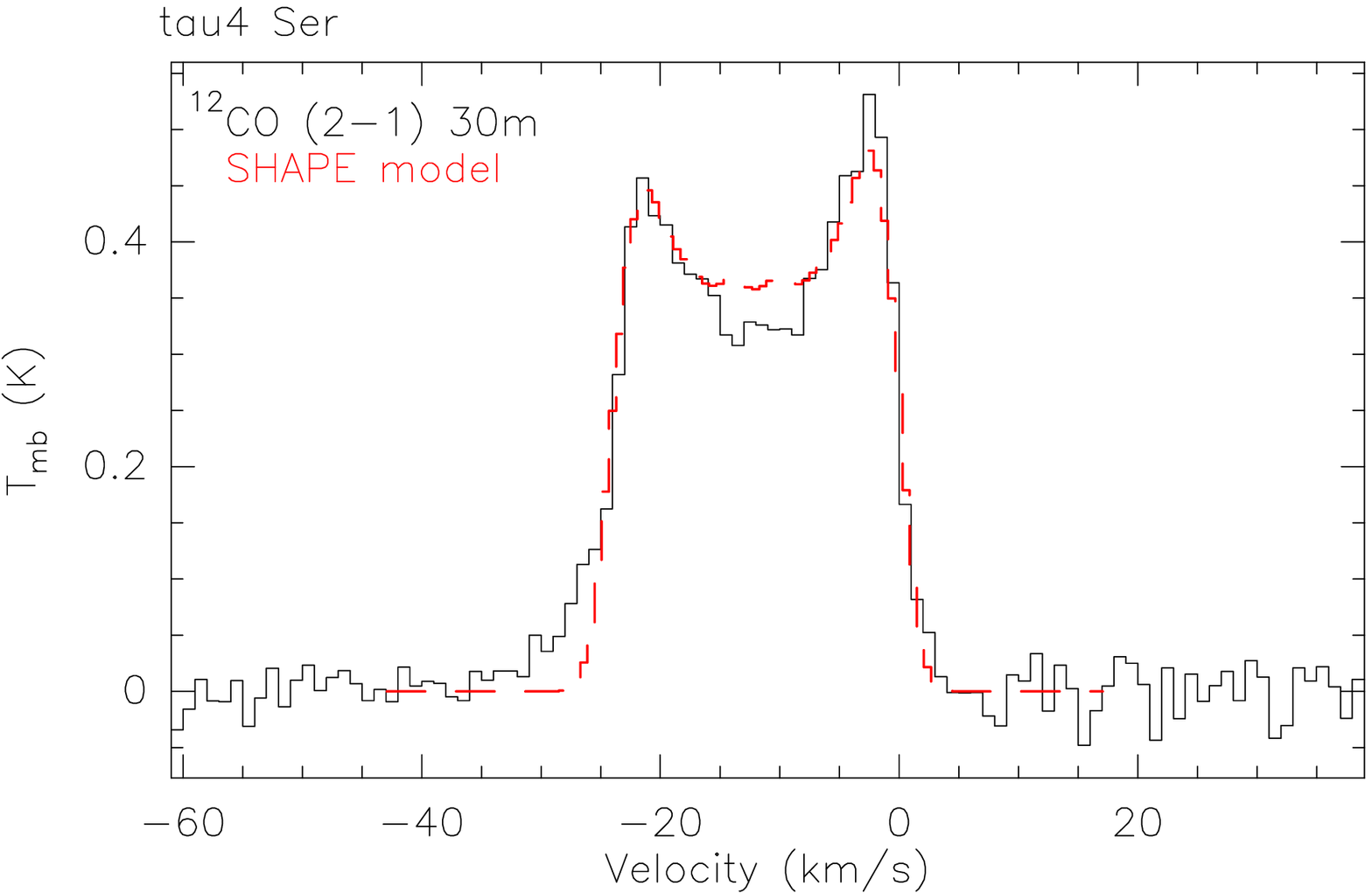}
      \caption{Resulting synthetic spectrum (red) and observation (black) for the $^{12}$CO \textit{J}=2-1 transition toward $\tau$$^{4}$ Ser.}
         \label{mod8}
   \end{figure*}      
   
\begin{figure*}
   \centering
   \includegraphics[angle=0,width=9cm]{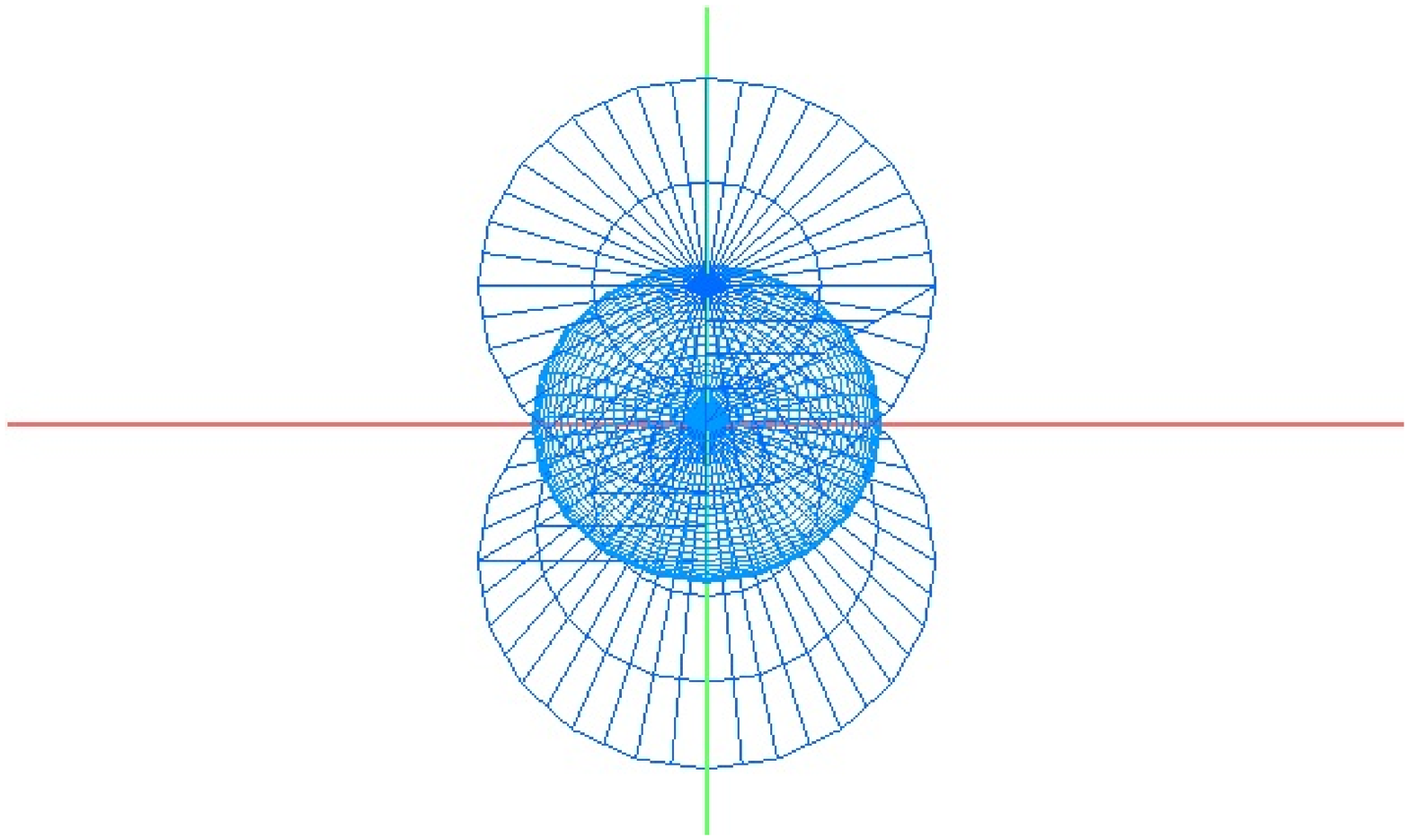}
   \includegraphics[angle=0,width=9cm]{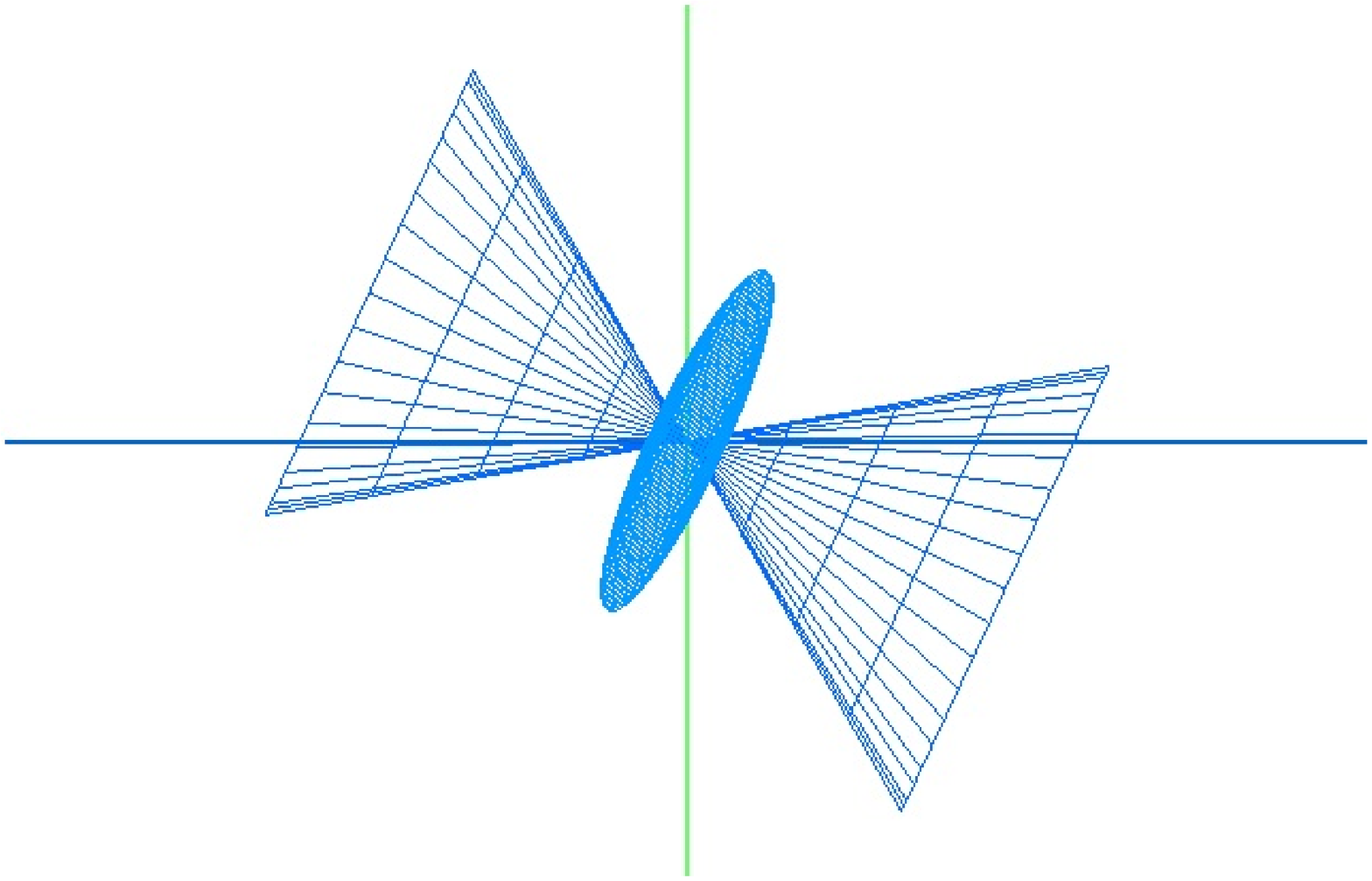}
      \caption{Three-dimensional mesh view of the model of the molecular envelope of V370 And, which is defined by an oblate spheroid whose equatorial plane is inclined by $\sim$25$^{\circ}$ with respect to the plane of the sky, attached to a biconical structure. Left: View from Earth. Right: View from the direction defined by west in the plane of the sky (the observer is on the left). The X, Y, and Z axes are represented by red, green, and blue, respectively (positive directions are in the clockwise direction). The symmetry axis of the envelope is perpendicular to the equatorial plane of the main structure (oblate spheroid).}
         \label{mod9}
   \end{figure*}
   
  \begin{figure*}
   \centering
   \includegraphics[angle=0,width=9cm]{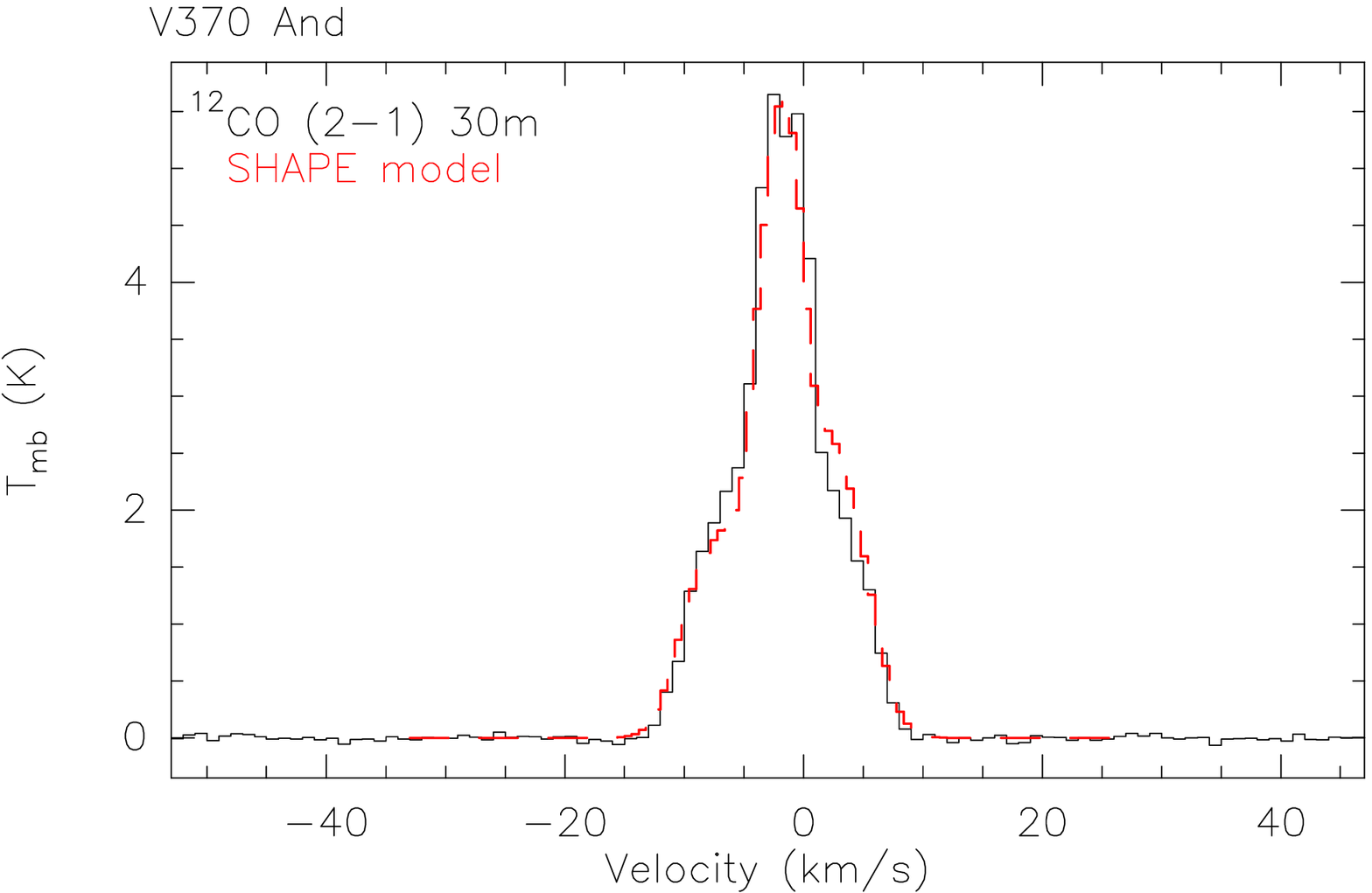}
      \caption{Resulting synthetic spectrum (red) and observation (black) for the $^{12}$CO \textit{J}=2-1 transition toward V370 And.}
         \label{mod10}
   \end{figure*} 
\clearpage  

\onecolumn

\newpage  
\section{Standard spherically symmetric envelope models}
   
In these models, we have considered spherically symmetric envelopes.   
   
\begin{table*}[htbp]
\caption{\label{tabmod} Physical conditions in the molecular envelopes of AQ Sgr, BK Vir, and V370 And derived from our model fitting of the $^{12}$CO data with SHAPE+shapemol.}
\centering
\scalebox{0.96}{
\begin{tabular}{lcccc}
\hline\hline
\\ [-1ex]
 Source                       & Gas density                             &  Temperature   &   Velocity     &  $\delta$$_{V}$   \\
                              & (cm$^{-3}$)                             & (K)            & (km\,s$^{-1}$) &  (km\,s$^{-1}$)   \\
\hline
\\ [-1ex] 
                              & \it{n(r)} = \it{n$_{o}$} \normalfont{$\times$ 1/(r$^{2}$+0.1)} & \it{T(r)} = \it{T$_{o}$} \normalfont{$\times$ 1/(r+0.1)}  & \it{V(r)} = \it{V$_{o}$}                              &    \\
                              & \it{n($\theta$)} = \it{n$_{o}$}                   & \it{T($\theta$)} = \it{T$_{o}$}              & \it{V($\theta$)} = \it{V$_{o}$}                       &    \\
                              & \it{n($\varphi$)} = \it{n$_{o}$}                  & \it{T($\varphi$)} = \it{T$_{o}$}             & \it{V($\varphi$)} = \it{V$_{o}$} $\times$ $\sqrt{r^2_2 + (r^2_1-r^2_2)\sin^{2}\left(\varphi/57.3\right)}$   &  \\
\hline
                              & \it{n$_{o}$}                                      & \it{T$_{o}$}                                 & \it{V$_{o}$}           &  \\
\hline
\\ [-1ex]  
  AQ Sgr                      & 4.94 $\times$ 10$^{4}$             & 50             & 3.5              &   2.9         \\
  BK Vir                      & 2.70 $\times$ 10$^{5}$             & 100            & 2.1              &   1.4         \\
\hline\hline
 V370 And                     &                                    &                                       &                         &  \\
\hline
\\ [-1ex] 
 Sphere                       & \it{n(r)} = \it{n$_{o}$} \normalfont{$\times$ 1/(r+0.1)}                  & \it{T(r)} = \it{T$_{o}$} \normalfont{$\times$ 1/(r+0.1)} &  \it{V(r)} = \it{V$_{o}$}        &  1.1   \\
                              & \it{n($\theta$)} = \it{n$_{o}$}                              & \it{T($\theta$)} = \it{T$_{o}$}             &  \it{V($\theta$)} = \it{V$_{o}$} &        \\
                              & \it{n($\varphi$)} = \it{n$_{o}$}                             & \it{T($\varphi$)} = \it{T$_{o}$}            &  \it{V($\varphi$)} = \it{V$_{o}$} $\times$ $\sqrt{r^2_2 + (r^2_1-r^2_2)\sin^{2}\left(\varphi/57.3\right)}$ &  \\    
                              & \it{n$_{o}$} = \normalfont{2.00 $\times$ 10$^{5}$} & \it{T$_{o}$} = \normalfont{100}                          &  \it{V$_{o}$} = \normalfont{0.8}              &        \\ 
\hline 
\\ [-1ex] 
 Biconical                    & \it{n(r)} = \it{n$_{o}$} \normalfont{$\times$ $\mid1/(r+0.1)\mid$} &  \it{T(r)} = \it{T$_{o}$} \normalfont{$\times$ 1/(r+0.1)}  &  \it{V(r)} = \it{V$_{o}$}        &  3.4 \\  
 structure                    & \it{n($\theta$)} = \it{n$_{o}$}                                   &  \it{T($\theta$)} = \it{T$_{o}$}              &  \it{V($\theta$)} = \it{V$_{o}$} &      \\
                              & \it{n}(\normalfont{z})~\tablefootmark{a} = \it{n$_{o}$}                        &  \it{T}(\normalfont{z}) = \it{T$_{o}$}                     &  \it{V}(\normalfont{z}) = \it{V$_{o}$}        &      \\
                              & \it{n$_{o}$} = \normalfont{1.30 $\times$ 10$^{5}$}      &  \it{T$_{o}$} = \normalfont{100}                           &  \it{V$_{o}$} = \normalfont{4.6}          &      \\    
\hline\hline                             
\end{tabular}}
\tablefoot{Description of the columns from left to right. (1) GCVS designation \citep{samus09}; (2-5) Gas density, gas temperature, gas velocity, and micro-turbulence velocity. We note that r, r$_1$, and r$_2$ are given in arcsec; r$_1$ = r$_2$ = \rco~(photodissociation radius taken from Table 2). $\varphi$ is given in degrees. \\
\tablefoottext{a}{We note that in this model we use cylindrical coordinates for defining the physical conditions in the biconical structure.}
}
\end{table*}     
   
   \begin{figure*}[htbp]
   \centering
   \includegraphics[angle=0,width=9cm]{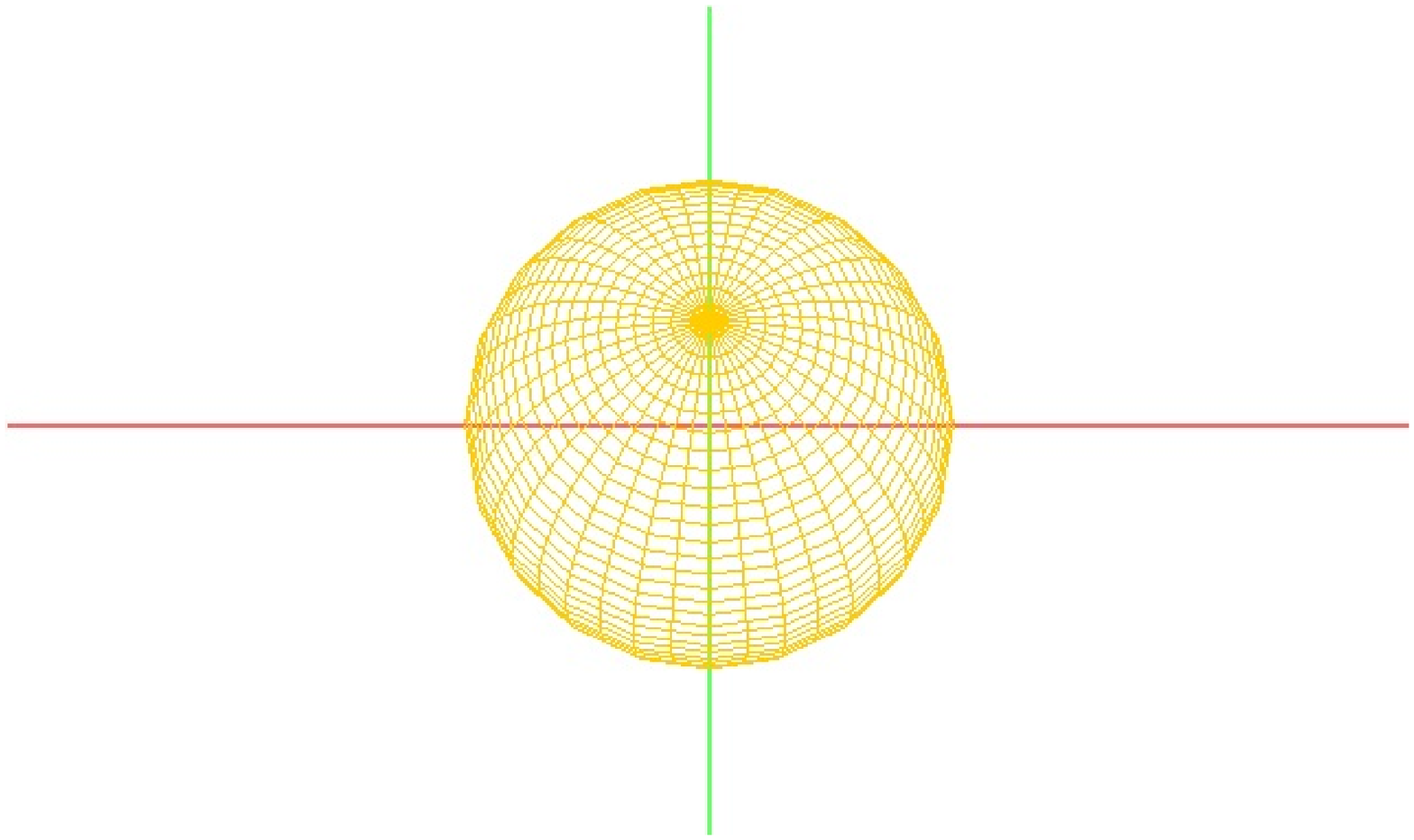}
   \includegraphics[angle=0,width=9cm]{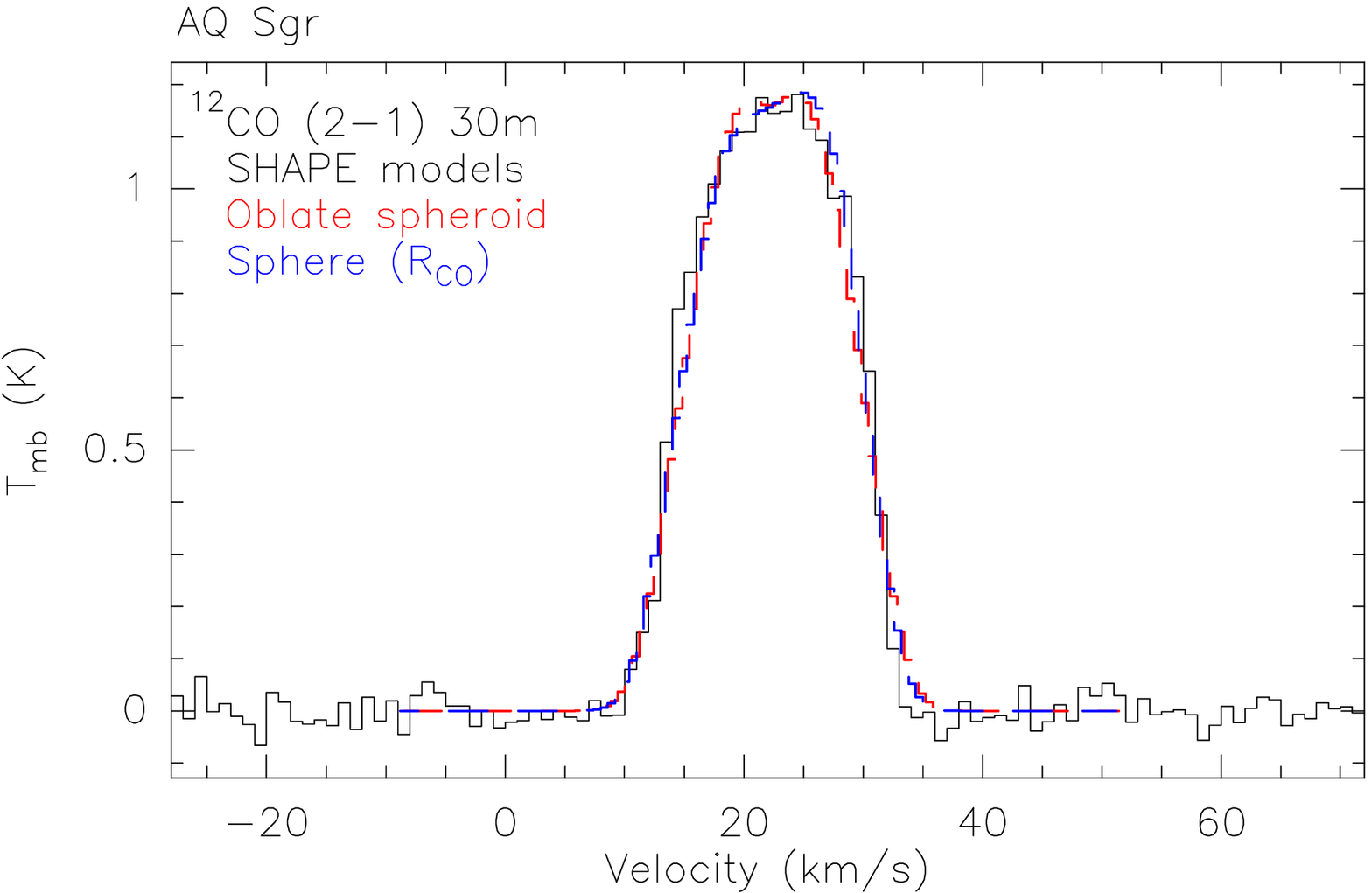}
      \caption{Three-dimensional mesh view of the model of the molecular envelope of AQ Sgr, which is defined by a sphere whose equatorial plane is inclined by $\sim$25$^{\circ}$ with respect to the plane of the sky. Left: View from Earth. X and Y axes are represented by red and green, respectively (positive directions are in the clockwise direction). Right: Resulting synthetic spectra for both models, the oblate spheroid (red) and the sphere (blue), and observation (black) for the $^{12}$CO \textit{J}=2-1 transition toward AQ Sgr. We note the similarity between the synthetic spectra for both models.}
      \label{mod1}
   \end{figure*}
   
   \begin{figure*}
   \centering
   \includegraphics[angle=0,width=9cm]{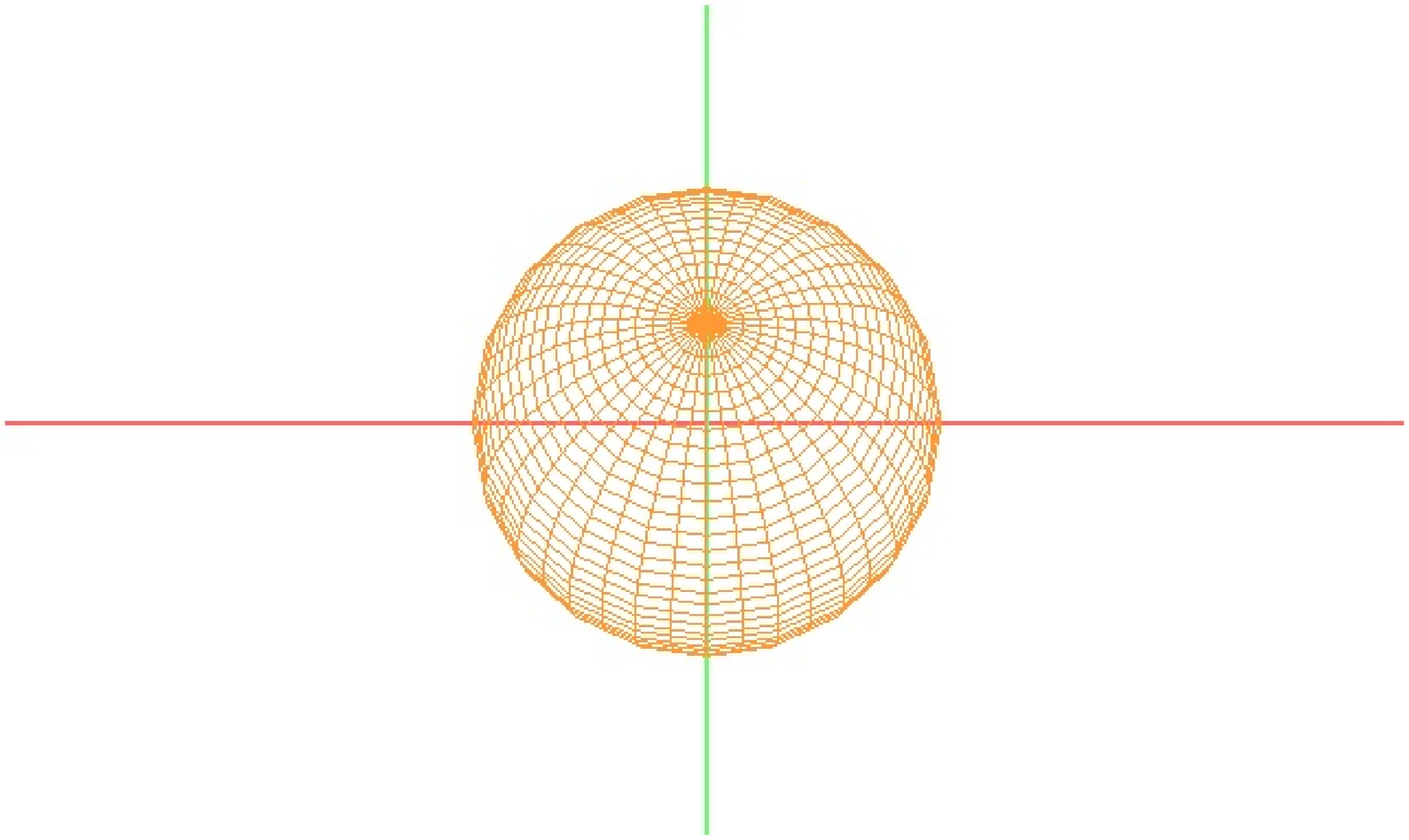}
   \includegraphics[angle=0,width=9cm]{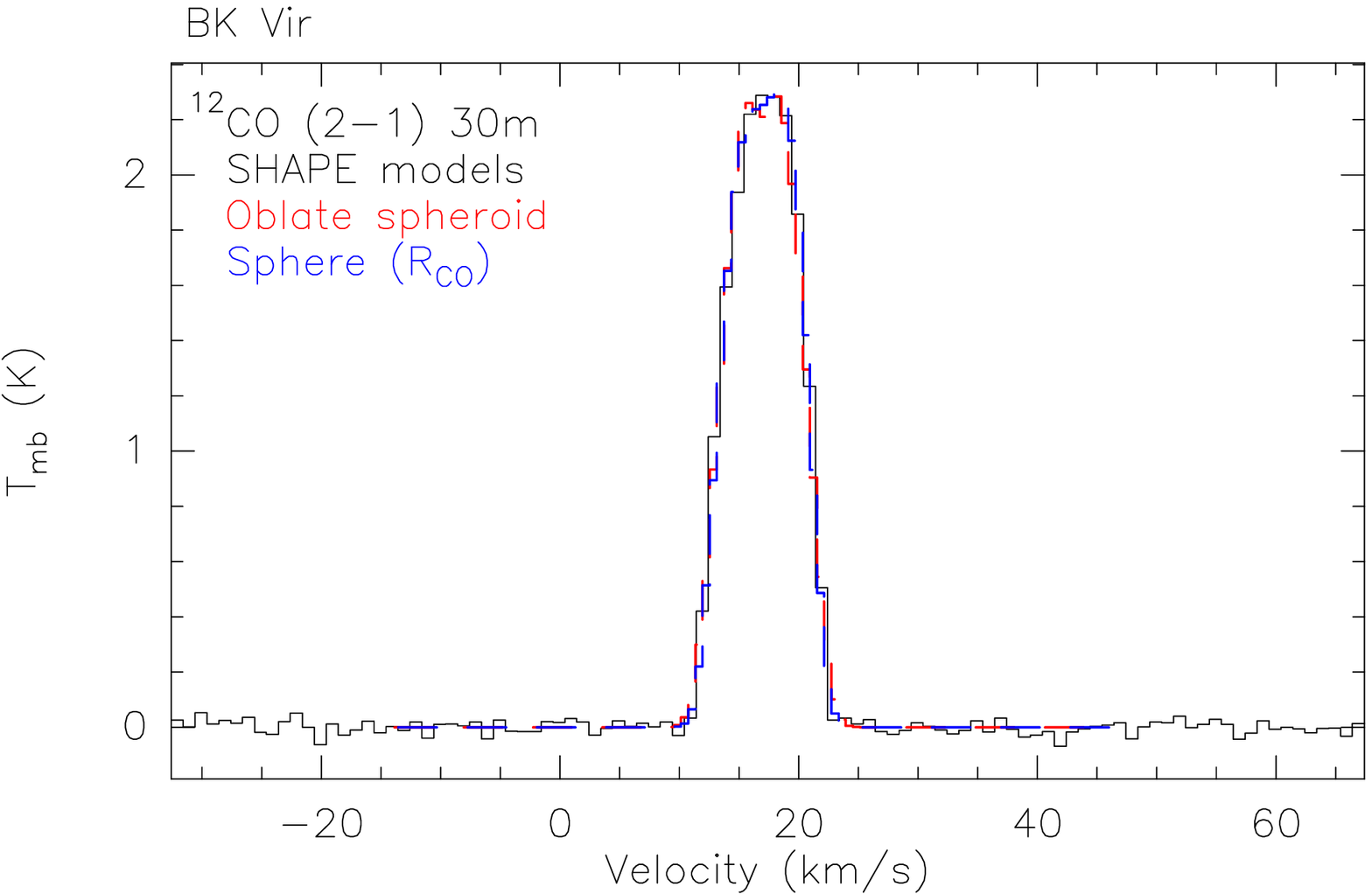}
      \caption{Three-dimensional mesh view of the model of the molecular envelope of BK Vir, which is defined by a sphere whose equatorial plane is inclined by $\sim$25$^{\circ}$ with respect to the plane of the sky. Left: View from Earth. The X and Y axes are represented by red and green, respectively (positive directions are in the clockwise direction). Right: Resulting synthetic spectra for both models, the oblate spheroid (red) and the sphere (blue), and observation (black) for the $^{12}$CO \textit{J}=2-1 transition toward BK Vir. We note the similarity between the synthetic spectra for both models.}
      \label{mod2}
   \end{figure*}
   
   \begin{figure*}
   \centering
   \includegraphics[angle=0,width=9cm]{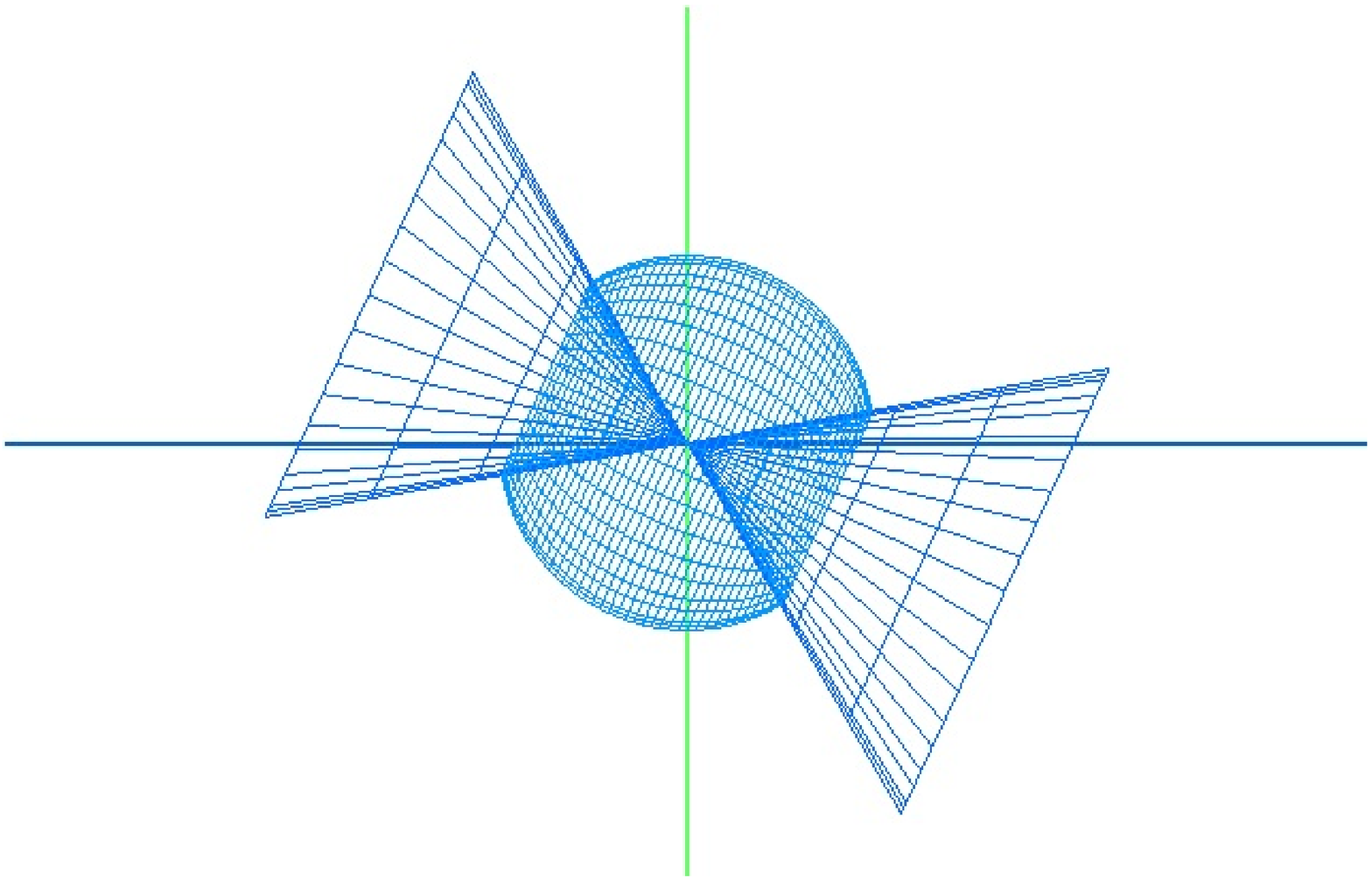}
   \includegraphics[angle=0,width=9cm]{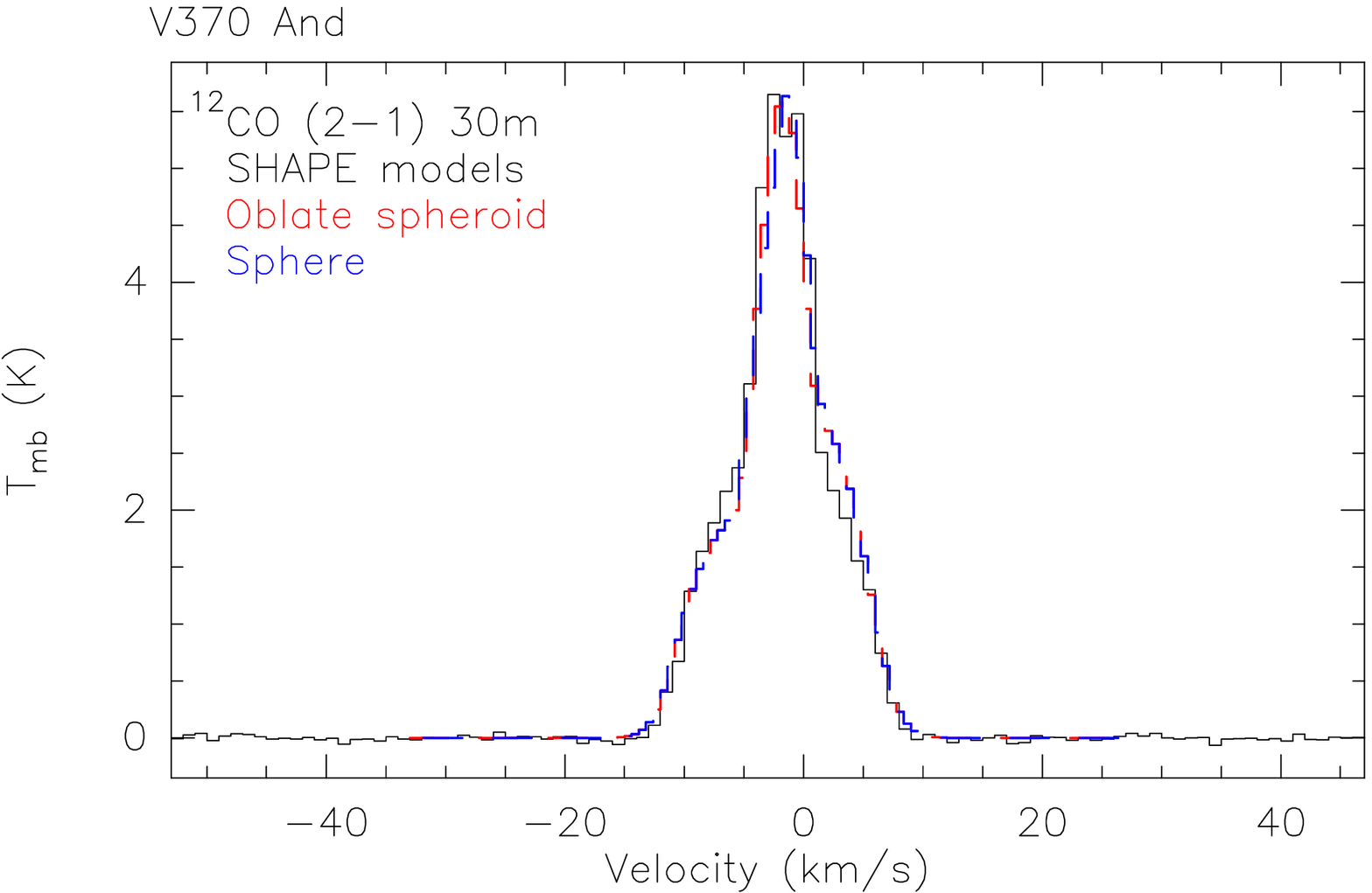}
      \caption{Three-dimensional mesh view of the model of the molecular envelope of V370 And, which is defined by a sphere whose equatorial plane is inclined by $\sim$25$^{\circ}$ with respect to the plane of the sky, attached to a biconical structure. Left: View from the direction defined by west in the plane of the sky (the observer is on the left). The Y and Z axes are represented by green and blue, respectively (positive directions are in the clockwise direction). Right: Resulting synthetic spectra for both models, the oblate spheroid (red) and the sphere (blue), both attached to the same biconical structure, and observation (black) for the $^{12}$CO \textit{J}=2-1 transition toward V370 And. We note the similarity between the synthetic spectra for both models.}
      \label{mod3}
   \end{figure*}
\clearpage 
\newpage  

\twocolumn
\section{Velocity field behavior for the different molecular envelopes}
   
In this section, we study the velocity field behavior for the different cases.
   
   \begin{figure}[hbt!]
   \centering
   \includegraphics[angle=0,width=8cm]{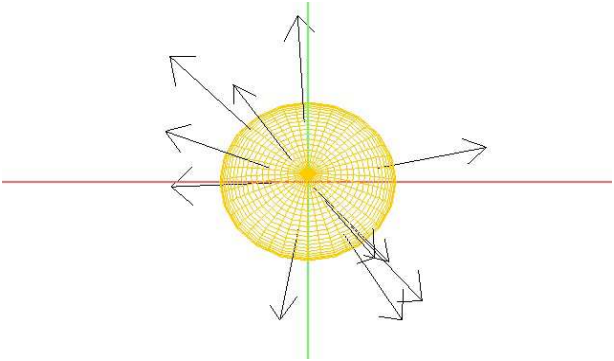}
      \caption{Velocity field vectors (arrows) indicating the expansion of the molecular envelope of AQ Sgr, which is defined by an oblate spheroid whose equatorial plane is inclined by $\sim$25$^{\circ}$ with respect to the plane of the sky. View from Earth. The X and Y axes are represented by red and green, respectively. The expansion increase from the poles to the equatorial plane of the structure. The equatorial expansion velocities are much higher than those in the polar region. }
      \label{mod4}
   \end{figure}
   
    \begin{figure}[hbt!]
   \centering
   \includegraphics[angle=0,width=8cm]{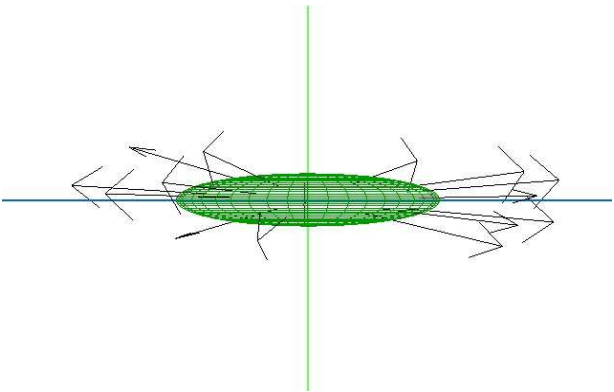}
      \caption{Velocity field vectors (arrows) indicating the expansion of the molecular envelope of $\tau$$^{4}$ Ser, which is defined by an edge-on oblate spheroid. View from the direction defined by west in the plane of the sky (the observer is on the left). The Y and Z axes are represented by green and blue, respectively. As in the case of AQ Sgr, the expansion is predominantly equatorial, with higher velocities in equatorial regions than those in the poles.}
      \label{mod5}
   \end{figure}
   
    \begin{figure}[hbt!]
   \centering
   \includegraphics[angle=0,width=8cm]{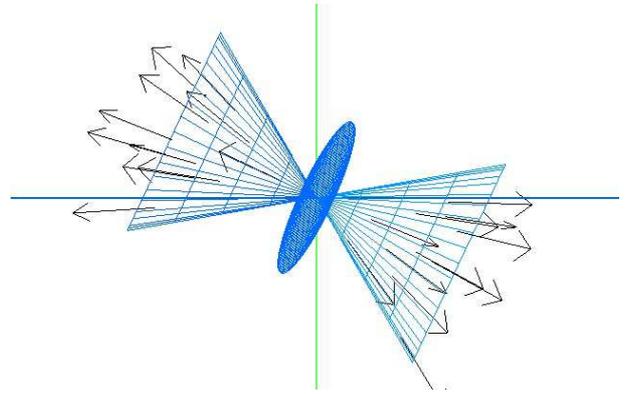}
      \caption{Velocity field vectors (arrows) indicating the expansion of the molecular envelope of V370 And, which is defined by an oblate spheroid whose equatorial plane is inclined by $\sim$25$^{\circ}$ with respect to the plane of the sky, attached to a biconical structure. A view from the direction defined by west in the plane of the sky is shown (the observer is on the left). The Y and Z axes are represented by green and blue, respectively. We note that the oblate spheroid (without arrows in order to avoid confusion) has a similar expansion to that of the previous examples. The biconical structure, defined by two cones, however, expands outwards from the central structure with constant velocity, indicating the presence of a bipolar outflow similar to that already found in EP Aqr (see above). The main symmetry axis of the molecular envelope (the same as the main axis of the cones), as well as the equatorial plane of the oblate spheroid, indicate a clear axial symmetry.}
      \label{mod6}
   \end{figure}
   
\end{appendix}

\end{document}